\documentclass[iop]{emulateapj}
\usepackage{apjfonts}
\usepackage{natbib}
\usepackage{subfigure}
\usepackage{booktabs}

\newcommand{\xmm}{\hbox{\it XMM-Newton\/}}
\newcommand{\chandra}{{\it Chandra\/}}
\newcommand{\nustar}{{\it NuSTAR\/}}
\newcommand{\flux}{{erg~cm$^{-2}$~s$^{-1}$}}
\newcommand{\mflux}{{erg~cm$^{-2}$~s$^{-1}$~Hz$^{-1}$}}
\newcommand{\lum}{{erg~s$^{-1}$}}
\newcommand{\mlum}{{erg~s$^{-1}$~Hz$^{-1}$}}
\newcommand\iona[2]{#1$\;${\scshape{#2}}}
\newcommand{\asca}{{\it ASCA\/}}
\usepackage{amsmath}

\begin{document}
\title{X-ray Insights into the Nature of 
PHL~1811 Analogs and Weak Emission-Line Quasars: Unification with a 
Geometrically Thick Accretion Disk?}

\author{
B.~Luo,\altaffilmark{1,2}
W.~N.~Brandt,\altaffilmark{1,2,3}
P.~B.~Hall,\altaffilmark{4}
Jianfeng~Wu,\altaffilmark{5}
S.~F.~Anderson,\altaffilmark{6}
G.~P.~Garmire,\altaffilmark{7}
R.~R.~Gibson,\altaffilmark{6}
R.~M.~Plotkin,\altaffilmark{8}
G.~T.~Richards,\altaffilmark{9}
D.~P.~Schneider,\altaffilmark{1,2}
O.~Shemmer,\altaffilmark{10}
and Yue~Shen\altaffilmark{11}
}
\altaffiltext{1}{Department of Astronomy \& Astrophysics, 525 Davey Lab,
The Pennsylvania State University, University Park, PA 16802, USA}
\altaffiltext{2}{Institute for Gravitation and the Cosmos,
The Pennsylvania State University, University Park, PA 16802, USA}
\altaffiltext{3}{Department of Physics, 104 Davey Lab,
The Pennsylvania State University, University Park, PA 16802, USA}
\altaffiltext{4}{Department of Physics \& Astronomy, York University, 4700 Keele Street, Toronto, ON M3J 1P3, Canada}
\altaffiltext{5}{Harvard-Smithsonian Center for Astrophysics, MS 6, 
60 Garden St, Cambridge, MA 02138, USA}
\altaffiltext{6}{Department of Astronomy, University of Washington, Box 351580, Seattle, WA 98195, USA} 
\altaffiltext{7}{Huntingdon Institute for X-ray Astronomy, LLC, 
10677 Franks Road, Huntingdon, PA 16652, USA}
\altaffiltext{8}{Department of Astronomy, University of Michigan, 1085 South University Ave, Ann Arbor, MI 48109, USA}
\altaffiltext{9}{Department of Physics, Drexel University, 3141 Chestnut Street, Philadelphia, PA 19104, USA}
\altaffiltext{10}{Department of Physics, University of North Texas, Denton, TX 76203, USA}
\altaffiltext{11}{Carnegie Observatories, 813 Santa Barbara Street, Pasadena, CA 91101, USA}

\begin{abstract}
We present an X-ray and multiwavelength study of
33 weak emission-line quasars (WLQs) 
and 18 quasars that are analogs of the extreme WLQ, PHL 1811, at 
$z\approx0.5$--2.9. New \chandra\ 1.5--9.5~ks exploratory
observations were obtained for 32 objects while the others have 
archival X-ray observations.
Significant fractions of these luminous type 1 quasars are distinctly 
X-ray weak
compared to typical quasars,
including 16 (48\%) of the WLQs and 17 (94\%) of the PHL 1811 analogs
with average X-ray weakness factors of 17 and 39, respectively.
We measure a relatively hard ($\Gamma=1.16_{-0.32}^{+0.37}$) effective 
power-law photon index for 
a stack of the X-ray weak subsample, suggesting X-ray
absorption, and 
spectral analysis of one PHL 1811 analog, J1521+5202, 
also indicates significant intrinsic
X-ray absorption.
We compare composite SDSS spectra for the X-ray weak and X-ray normal
populations and find several optical--UV tracers of X-ray weakness;
e.g., \iona{Fe}{ii} rest-frame equivalent width and relative color.
We describe how orientation effects under our previously proposed 
``shielding-gas'' scenario can likely
unify the
X-ray weak and X-ray normal populations.
We suggest that the shielding gas may naturally be understood as 
a geometrically thick inner
accretion disk that shields the broad line region
from the ionizing continuum. 
If WLQs and PHL~1811 analogs 
have very high Eddington ratios, the inner disk could be significantly
puffed up (e.g., a slim disk).
Shielding of the broad
emission-line region by a geometrically thick disk 
may have a significant role in setting
the broad distributions of \iona{C}{iv}
rest-frame equivalent width and blueshift for quasars more generally.
\end{abstract}

\keywords{accretion, accretion disks -- galaxies: active -- galaxies: nuclei --
quasars: emission lines -- X-rays: galaxies}

\section{INTRODUCTION}

Luminous X-ray emission is considered a universal property of 
active galactic nuclei (AGNs), and built upon this idea are
extragalactic X-ray surveys 
for finding AGNs efficiently throughout the universe \citep[e.g.,][and
references therein]{Brandt2014}.
For AGNs that are not radio loud (with a jet-linked X-ray enhancement)
or \hbox{X-ray} absorbed, the 
\hbox{X-ray-to-optical} power-law slope parameter ($\alpha_{\rm OX}$)\footnote{$\alpha_{\rm OX}$
is defined as
$\alpha_{\rm OX}=-0.3838\log(f_{2500~{\textup{\AA}}}/f_{2~{\rm keV}})$,
with $f_{2500~{\textup{\AA}}}$ and $f_{2~{\rm keV}}$
being the \hbox{rest-frame}
2500~\AA\ and 2~keV flux densities.} has
a highly significant correlation with 2500~\AA\ monochromatic 
luminosity ($L_{\rm 2500~{\textup{\AA}}}$) 
across $\approx5$ orders of magnitude in
UV luminosity \citep[e.g.,][]{Steffen2006,Just2007,Lusso2010}.
X-ray emission from AGNs is believed to originate from 
the accretion-disk ``corona'' via Comptonization of disk optical/UV/EUV photons
\citep[e.g.,][and references therein]{Turner2009}, although the details of this mechanism
remain mysterious.

Few AGNs are found to be intrinsically X-ray weak, i.e., producing
much less X-ray emission than expected from
the \hbox{$\alpha_{\rm OX}$--$L_{\rm 2500~{\textup{\AA}}}$}
relation \citep[e.g.,][]{Gibson2008}. A few candidates have been
suggested recently based on {\it Nuclear
Spectroscopic Telescope Array} (\nustar; \citealt{Harrison2013})
observations of significantly X-ray weak broad absorption line (BAL) quasars
\citep{Luo2013,Luo2014,Teng2014}.\footnote{BAL quasars are 
identified by their broad ($\ge2000$~km~s$^{-1}$ wide) 
blueshifted UV absorption lines \citep[e.g.,][]{Weymann1991}; 
e.g.,  
the \iona{C}{iv}~$\lambda1549$ line. BAL quasars are in general X-ray weak,
often due to absorption, but intrinsic X-ray weakness is a viable
explanation for a subset of BAL quasars.} 
The best-studied intrinsically X-ray weak
AGN is the type 1 quasar PHL~1811, 
a very bright ($B=13.9$) radio-quiet quasar
at $z=0.19$ \citep[e.g.,][]{Leighly2007b,Leighly2007}.
It is
\hbox{X-ray} weak by a factor of \hbox{$\approx30$--100} relative
to expectations from the \hbox{$\alpha_{\rm OX}$--$L_{\rm 2500~{\textup{\AA}}}$}
relation, and its X-ray weakness is likely intrinsic instead of being 
due to absorption given
its canonical X-ray spectrum (power-law photon index $\Gamma=2.3\pm0.1$),
lack of detectable photoelectric X-ray absorption, 
and 
short timescale X-ray variability \citep{Leighly2007}.
Interestingly, PHL~1811 also has an unusual UV spectrum
\citep{Leighly2007b}, which is dominated by strong \iona{Fe}{ii} 
and \iona{Fe}{iii} emission with very weak high-ionization lines.
The rest-frame equivalent width (REW) of the 
\iona{C}{iv}~$\lambda1549$ line ({6.6~\AA}) 
is a factor of $\approx5$ times smaller than
that measured from quasar composite spectra ({30~\AA}); 
this line is also blueshifted
(by $\approx1400$~km~s$^{-1}$)
and asymmetric, indicative of a wind component in the broad
emission-line region \citep[BELR; e.g.,][]{Richards2011}.
Based on photoionization modeling, \citet{Leighly2007b} suggested that many of
the unusual emission-line properties of PHL~1811 are a result of the soft
optical-to-X-ray
ionizing continuum caused by the intrinsic X-ray weakness.

The intriguing possible 
connection between the extreme emission-line and X-ray
properties of PHL~1811 prompted a search for more such X-ray weak quasars
using UV emission-line selection criteria (\citealt{Wu2011}, hereafter W11). 
A pilot sample of eight type 1, radio-quiet, non-BAL quasars with 
PHL~1811-like emission-line properties (including 
small \iona{C}{iv} REWs, large \iona{C}{iv} blueshifts, and strong
\iona{Fe}{ii}
and \iona{Fe}{iii} emission), termed PHL~1811 analogs, 
were selected for \hbox{X-ray} study. All of them turned out to be 
X-ray weak, by factors of $>4.8$ to $\ge34.5$, confirming the empirical
link between the \hbox{X-ray} weakness and unusual UV emission-line properties.
However, an \hbox{X-ray} stacking analysis revealed a hard spectrum on average
for this sample, albeit with a large uncertainty, suggesting that unlike
PHL~1811 itself that appears to be 
intrinsically X-ray weak, these PHL~1811 analogs 
may often be X-ray absorbed. {More PHL~1811 analogs selected in addition to
the eight W11 pilot objects
are clearly required
to constrain better the nature of these extreme quasars.}

There is another small population of type 1 quasars that have \hbox{X-ray} and 
UV emission-line properties overlapping with those of the PHL~1811 analogs:
radio-quiet
weak emission-line quasars (WLQs). Strong broad emission lines 
in the optical and UV are a characteristic feature of radio-quiet 
quasars.\footnote{In radio-loud systems, the line emission can sometimes be
diluted by the synchrotron emission from a relativistic jet, as typically
seen in BL Lac objects.} It was thus surprising when \citet{McDowell1995} 
discovered the 
first WLQ PG~1407+265, with unusually weak
Ly$\alpha$, \iona{C}{iv}~$\lambda1549$, \iona{C}{iii}]~$\lambda1909$, 
and H$\beta$ lines.
With the large spectroscopic quasar sample provided by the 
Sloan Digital Sky Survey (SDSS; \citealt{York2000}), more WLQs were
discovered. These were 
originally at $z>2.2$ where Ly$\alpha$ coverage is available
\citep[$\approx90$ WLQs; e.g.,][]{Fan1999,Anderson2001,Plotkin2008,Plotkin2010,Diamond2009}, 
and later extended to lower redshifts \citep[$\approx100$ WLQs; e.g.,][]{Collinge2005,Hryniewicz2010,Plotkin2010b,Plotkin2010,Nikolajuk2012,Meusinger2014} 
requiring weak \iona{C}{iv} and/or other lines 
at longer wavelengths. The fraction of X-ray weak quasars among either
the high-redshift or lower-redshift WLQs is high ($\approx50\%$; 
\citealt{Shemmer2009}; \citealt{Wu2012}, hereafter W12), again suggesting
a link between the weak UV line emission and X-ray weakness.

Based on the overall similarities between the PHL~1811 analogs and X-ray
weak WLQs, W11 argued that PHL~1811 analogs are a subset of WLQs,
despite small technical differences in their UV line REW selection 
criteria.\footnote{PHL~1811 analogs were required to have \iona{C}{iv}
REW $<10$~\AA,
while the WLQs in the W11 study were from the 
\citet{Plotkin2010} catalog which requires REW $\la5$~\AA\ for all UV 
emission features (e.g., \iona{C}{iv}, \iona{C}{iii]}, and \iona{Mg}{ii}).
No apparent difference was found between the PHL~1811 analogs
having $<5$~\AA\ \iona{C}{iv} REWs and those having 5--10~\AA\ \iona{C}{iv} 
REWs. Therefore,
the different REW criteria were 
considered a technical selection effect and the WLQ criterion could be relaxed 
to \iona{C}{iv} REW $<10$~\AA, which is the lower 3$\sigma$ limit of the
log-normal \iona{C}{iv} REW distribution (\citealt{Diamond2009}; W12). \label{footnote-sel}}
{WLQs contain both X-ray normal and X-ray weak quasars (we adopt an
X-ray weakness factor of 3.3 as the 
dividing threshold between X-ray weak and X-ray normal quasars; 
see 
Section~\ref{sec-aox} below), while 
PHL~1811 analogs are likely \hbox{X-ray} weak WLQs 
due to the
additional selection criteria 
of strong UV Fe emission and large \iona{C}{iv} blueshift
(W11; W12). A larger sample of PHL~1811 analogs than the pilot sample of 
eight would help examine further the above suggested connection.}

Except for the unusual UV emission-line and X-ray properties, 
the PHL~1811 analogs and WLQs appear to be typical quasars in terms of other observable
multiwavelength
properties (e.g., \citealt{Lane2011}; W11; W12). 
Various explanations have been proposed for the nature of the PHL~1811 analogs
and WLQs, such
as an anemic BELR where there is a significant deficit of line-emitting gas
in the BELR \citep[e.g.,][]{Shemmer2010}, a brief evolutionary stage
where the BELR is not fully developed \citep[e.g.,][]{Hryniewicz2010}, 
a soft ionizing spectral energy distribution (SED) 
produced by the cold accretion disk of a very massive
black hole \citep[e.g.,][]{Laor2011}, a soft ionizing continuum 
due to intrinsic \hbox{X-ray} weakness \citep[e.g.,][]{Leighly2007b}, and
a soft ionizing continuum due to small-scale absorption (e.g., W11; W12).
However, for the general population of PHL~1811 analogs and WLQs,
the absorption-induced soft ionizing continuum appears the most likely
scenario, based on systematic studies, albeit using small samples,
of their X-ray and multiwavelength
properties (e.g., W11; W12).

A small-scale ``shielding gas''
scenario was proposed in W11 to explain and unify
PHL~1811 analogs and WLQs, which was broadly 
motivated by the shielding gas
generally required in the disk-wind model for BAL quasars
\citep[e.g.,][]{Murray1995,Proga2000}.
In the W11 scenario, some shielding gas
interior to the BELR shields
all or most of the BELR from the nuclear ionizing continuum, resulting
in the observed weak UV emission lines. If our line of sight
intersects the X-ray absorbing shielding gas, a PHL~1811 analog
or an X-ray weak WLQ is observed; 
if not, an X-ray normal WLQ is observed. 
Since BAL quasars were excluded in the selection of PHL~1811 analogs and WLQs,
the inclination angle (with respect to disk normal) should probably 
still be relatively small so that the line of
sight does not intercept the (often equatorial) 
disk wind which would produce BALs in the observed
spectra.
The reason this shielding gas is unusually effective
at screening the BELR in the PHL~1811 analogs and WLQs
remains uncertain,
but it should be a rare occurrence given the small numbers of 
PHL~1811 analogs and WLQs discovered.

The fractions of PHL~1811 analogs and WLQs among typical quasars are small,
$\la1\textrm{--}2\%$ (W11). 
However, studies of rare and extreme objects often 
clearly reveal phenomena that are more generally applicable,
as such effects are more difficult to identify in the overall population
\citep[cf.][]{Eddington1922}. We shall indeed argue the case
for such generality later in this paper (see Section~6.3 below).
With the pilot studies of W11 and W12 examining systematically the 
X-ray and emission-line properties of PHL~1811 analogs and WLQs, 
progress has been made
toward understanding the nature of these extreme
quasars (e.g., the W11 shielding-gas scenario). 
However, the small X-ray sample available
in their work, with only eight PHL~1811 analogs and 11 radio-quiet WLQs
(which are also divided into X-ray weak and X-ray normal categories for 
the case of WLQs),
limited further investigations.
A larger sample is critically needed to reduce the uncertainties of stacking
and joint spectral analyses, examine correlations between 
the degree of X-ray weakness 
and emission-line properties, assess why PHL~1811 analogs are preferentially
X-ray weak, and further explore the shielding-gas scenario.

As an extension of the W11 and W12 work, we present here 
an X-ray and multiwavelength study of 18 PHL~1811 analogs
and 33 WLQs, including 33 objects with new \chandra\ observations. 
We describe the sample selection and \chandra\ 
data analysis in Sections 2 and 3, respectively.
The multiwavelength properties, including the $\alpha_{\rm OX}$ parameters,
continuum SEDs, and radio properties, are presented in Section~4.
In Section~5, we perform X-ray stacking and joint spectral analyses,
estimate Eddington ratios,
construct composite SDSS spectra, and identify spectral indicators for the 
X-ray weak subsample. In Section~6, we 
discuss the unification of the X-ray weak 
and X-ray normal PHL~1811 analogs and WLQs 
under the W11 shielding-gas scenario, and 
we propose that a geometrically thick inner
accretion disk may act as
the shielding gas in the general population of PHL~1811 analogs
and WLQs. We summarize in Section~7.

We caution that, consistent with the exploratory nature of this work, 
the PHL~1811 analogs
and WLQs could be a heterogeneous population.
The discussions and conclusions are mainly applicable to 
the general population of PHL~1811 analogs
and WLQs, 
and we acknowledge that other explanations are possible
for a fraction of our sample.
We also stress that PHL~1811 analogs were selected based on 
their PHL~1811-like UV emission-line properties, and they
are not necessarily intrinsically X-ray weak like PHL~1811 itself.
In fact, the X-ray weakness of PHL~1811 analogs is likely caused by
absorption, based on the studies of W11 and our work here.
Throughout this paper,
we use J2000 coordinates and a cosmology with
$H_0=65.1$~km~s$^{-1}$~Mpc$^{-1}$, $\Omega_{\rm M}=0.329$,
and $\Omega_{\Lambda}=0.671$ \citep[e.g.,][]{Ade2014}.
Full J2000 names of the targets are listed in the tables while abbreviated
names are used in the text.
We quote uncertainties at a 1$\sigma$ confidence level
and upper and lower limits at
a 90\% confidence level.

\section{SAMPLE SELECTION AND {{\em CHANDRA}} OBSERVATIONS}

\subsection{New X-ray Sample of PHL~1811 Analogs} \label{sec-sel11}
Our new sample of PHL~1811 analogs was selected from the SDSS 
Data Release 7 (DR7; \citealt{Abazajian2009})
quasar catalog \citep{Schneider2010}
following 
similar procedures to those in W11, with the major difference being 
a relaxation in the redshift requirement. 
Specifically, we require redshift
$z>1.7$, SDSS $r$-band magnitude $m_{r}<18.85$,
\iona{C}{iv}~$\lambda1549$ REW $<10$~\AA, 
\iona{C}{iv} blueshift $>1000$~km~s$^{-1}$,
strong UV
\iona{Fe}{iii} (UV48 $\lambda2080$) and/or 
\iona{Fe}{ii} (2250--2650~\AA) emission,\footnote{The
strong UV \iona{Fe}{ii} and \iona{Fe}{iii} 
emission relative to the other weak UV lines
in PHL~1811 analogs can be explained in the scenario of
a soft ionizing continuum; see \citet{Leighly2007b} for details.} 
no detection of BALs or mini-BALs (absorption troughs 
\hbox{500--2000~km~s$^{-1}$} wide),
and radio-loudness parameter $R<10$ (radio quiet).\footnote{The radio-loudness parameter 
is defined as
$R=f_{5~{\rm GHz}}/f_{\rm 4400~{\textup{\AA}}}$
\citep[e.g.,][]{Kellermann1989}, where $f_{5~{\rm GHz}}$ and 
$f_{\rm 4400~{\textup{\AA}}}$ are the flux densities at
rest-frame 5~GHz and 4400~\AA, respectively.}
The relaxed redshift requirement allows selection of a larger 
sample than in W11 which includes more optically bright objects for 
efficient \chandra\ observations.
The choice of a radio-quiet sample is to avoid any contamination from
jet-linked X-ray emission that might confuse the results.
Some basic quasar properties (e.g., redshift, \iona{C}{iv} REW) 
were initially adopted or 
derived from the catalogs of \citet{Hewett2010} and \citet{Shen2011}. 
The \iona{C}{iv} blueshifts were measured based on the adopted redshifts.
We later performed our own measurements of 
the redshifts, emission-line properties 
(Section~\ref{sec-linemeasure} below), 
and radio-loudness parameters 
(Section~\ref{sec-radio} below) for our \chandra\ targets.
The UV \iona{Fe}{ii} and \iona{Fe}{iii} line strength was assessed
visually among an initially selected sample satisfying the other
criteria, and quasars with stronger \iona{Fe}{ii} and \iona{Fe}{iii} emission
(relative to the other weak-lined quasars)
were favored.

Compared to the redshift requirement in W11, $2.125\le z\le2.385$,
our $z>1.7$ criterion no longer requires coverage of the Ly$\alpha$ and UV
\iona{Fe}{ii} emission, although the visual inspection stage of the
\iona{Fe}{ii} line strength generally selected targets with $z<3$.
An initial sample of 66 quasars was selected, including six of the
eight radio-quiet PHL~1811 analogs in W11 (the other two 
have smaller \iona{C}{iv} blueshifts than our criterion here).
We ranked these objects according to the REW and blueshift
of the \iona{C}{iv} line, and selected 10 bright 
($m_{r}<18.2$)
targets with the lowest
\iona{C}{iv} REWs and highest \iona{C}{iv} blueshifts for \chandra\
observations. The selected quasars span a redshift range of 1.7--2.9.
They were observed in \chandra\ Cycle 14 with
3.7--9.5~ks exposures (Table~1) using the S3 CCD of
the Advanced CCD Imaging
Spectrometer (ACIS; \citealt{Garmire2003}).\footnote{We also 
observed PHL~1811 itself
with a 2~ks \chandra\ exposure. It was detected with flux and effective
power-law photon index consistent
with previous X-ray observations and did not show unexpected variability
(e.g., a return to a nominal level of X-ray emission).}

In addition, we obtained a 40~ks \chandra\ exposure of
J1521+5202 that had a previous 4~ks
\chandra\ exposure \citep{Just2007} and was studied as a PHL~1811 analog 
in W11. This remarkable source is one of the most luminous quasars in
the SDSS catalog \citep[$M_{i}=-30.2$; e.g.,][]{Just2007}, and we aimed to acquire reliable
basic spectroscopic information with this longer \chandra\ observation.
X-ray properties derived from this longer observation are used in the relevant
analyses of this study.

\subsection{New X-ray Sample of WLQs} \label{sec-selwlq}

Our WLQ targets were mainly selected from the \citet{Plotkin2010} catalog
of radio-quiet WLQs which have REW $\la5$~\AA\ for
all emission features. {There was no \iona{C}{iv} blueshift requirement 
for the \citet{Plotkin2010} WLQs.}
We chose bright 
($i$-band magnitude $m_{i}<18.6$) WLQs, excluding objects that are
identified as stars based on their proper motions, have 
potential absorption features (e.g., 
intervening absorption, mini-BALs, extremely red continuum), or have
already been studied in W11 or \citet{Shemmer2009}. There were
21 such WLQs selected from the \citet{Plotkin2010} catalog.
In addition, we include in our WLQ sample two bright radio-quiet 
quasars found in the literature
that have similarly weak emission-line features, HE 0141$-$3932 
(\citealt{Reimers2005}; REW $\la15$~\AA\ for \iona{C}{iv} and 
\iona{Mg}{ii}~$\lambda2799$)
and 2QZ~J2154$-$3056 (\citealt{Londish2004}; weak [\iona{O}{iii}]~$\lambda5007$
and H$\beta$ emission).
These 23 WLQs span a redshift range of 0.5--2.5,
and they were observed in \chandra\ Cycle 14 with
1.5--3.5~ks exposures using the ACIS-S3 CCD (Table~1).

However, of these 23 WLQ targets,
J1332+0347 was later found to show a 
\iona{C}{iv} BAL in its VLT/X-shooter spectrum 
(\citealt{Plotkin2015}). It 
is also a lensed quasar \citep{Morokuma2007} which may have
its emission-line REWs affected by gravitational lensing amplification
\cite[e.g.,][]{Shemmer2009}. Therefore, we will not include
J1332+0347 in the figures or our analyses below, since it may not be a 
bona-fide WLQ. However, 
we do present the basic properties of this object
in Tables~1--3 for completeness.

We also note that HE 0141$-$3932 and 2QZ~J2154$-$3056 
were selected differently from the other WLQs. However,
the inclusion of these
two objects does not bias our results, as they do not have SDSS spectra
and are not included in the
majority of the statistical analyses below.
Our final new X-ray sample of WLQs 
includes 22 objects.

\subsection{Redshift, UV Emission-Line, and Continuum Measurements} \label{sec-linemeasure}
We measured redshifts and UV emission-line properties 
for our new X-ray samples of PHL~1811 analogs
and WLQs the same way as in Section~2.2 of W11. 
Briefly, the SDSS Catalog Archive Server (CAS) redshifts were initially 
adopted. These redshifts may not be precise in some cases 
due to the weakness and sometimes significant blueshifts of the emission lines. 
We examined the spectra and made small
adjustments for some sources based on strong line features (e.g., 
\iona{Mg}{ii} emission, narrow absorption features).
We adopted $z=2.238$ for J1521+5202 based on its best-fit H$\beta$ line
(W11).
The emission-line properties, including the 
REWs of the \iona{C}{iv}~$\lambda1549$,
\iona{Si}{iv}~$\lambda1397$, $\lambda1900$ complex
(mainly \iona{C}{iii}]~$\lambda1909$), 
\iona{Fe}{iii} UV48 $\lambda2080$,
and \iona{Mg}{ii}~$\lambda2799$ emission features, 
and the
blueshift and FWHM of the \iona{C}{iv} line, were
measured 
interactively. 
The wavelength region for each line was chosen based on the 
upper and lower wavelength limits given in Table 2 of \citet{Vandenberk2001},
and a power-law local continuum was
fitted between the lower and upper 10\% of the wavelength region
for the line measurements.

In addition, we measured the UV \iona{Fe}{ii} REW that was not included 
in the previous studies of W11 and W12;
it was measured between
2250~\AA\ and 2650~\AA\ following the same approach above.
For three objects that do not have
full spectral coverage of this wavelength range, a lower limit was derived;
for the two cases where the covered fraction of this line feature 
is $>60\%$ based on the SDSS quasar composite
spectrum \citep{Vandenberk2001}, an estimated \iona{Fe}{ii} REW 
was also derived from this fractional coverage.
The uncertainty of this UV \iona{Fe}{ii} REW is dominated by the 
continuum fitting, 
and a 10\% error is assumed. 

\begin{figure}
\centerline{
\includegraphics[scale=0.5]{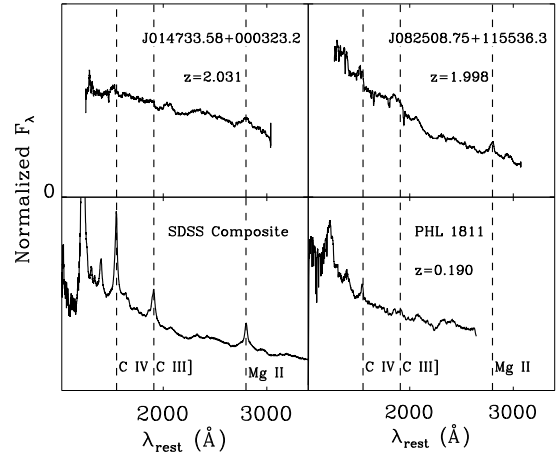}
}
\caption{
SDSS spectra for the PHL~1811 analogs and WLQs in the \chandra\ Cycle 14
sample, {following the order in Table~1}
(an extended version of this figure showing all the spectra
is available in the online journal).
The $y$-axis is the flux density in arbitrary linear units.
Each spectrum has been corrected for Galactic extinction and
smoothed using a boxcar of width 20 pixels,
We do not show the spectrum of J1521+5202, which was already included in
W11.
The composite spectrum of SDSS quasars from
\citet{Vandenberk2001} and the {\it HST} spectrum of PHL~1811
\citep{Leighly2007b} are shown for comparison.
}
\label{fig-sdssspectra}
\end{figure}

The adopted redshifts are listed in Table~1,
and the UV emission-line properties are shown in Table~2.
We also derived relative SDSS $g-i$ colors, $\Delta(g-i)$, for our targets,
which are defined as
the colors referenced to the median color at a given redshift
\citep[e.g.,][]{Richards2001}. The $g-i$ colors were taken from
\citet{Schneider2010}, and the median color was computed within a
redshift bin of $\Delta z=0.1$. The relative $g-i$ colors are listed
in Table~3, with positive values representing relatively red spectra.
We plot the individual SDSS spectra for the PHL~1811 analogs 
and WLQs in Figure~\ref{fig-sdssspectra}, with the composite 
spectrum of SDSS quasars from
\citet{Vandenberk2001} and the {\it Hubble Space Telescope} ({\it HST}\/) 
spectrum of PHL~1811
\citep{Leighly2007b} shown for comparison.

\subsection{Previous Samples Used in this Study}
In order to improve the statistics of our study, 
we also include previous samples 
of radio-quiet PHL~1811 analogs and WLQs from W11 and W12.
The basic X-ray and multiwavelength properties of these objects were
adopted from W11 and W12.
There are also X-ray samples of high-redshift ($z\approx3$--6) WLQs studied
previously \citep[e.g.,][]{Shemmer2006,Shemmer2009}. We do not include 
those WLQs here due to the significant difference in redshift (and subsequently
different rest-frame
SDSS spectral coverage). The general comparison between the high-redshift
WLQs and lower-redshift WLQs was presented in Section~5 
of W12, and their overall
properties are consistent with each other.

We include 7 additional radio-quiet PHL~1811 analogs from W11
(the other radio-quiet PHL~1811 analog in W11 is J1521+5202; 
see Section~\ref{sec-sel11})
and 10 radio-quiet
WLQs from W12. There is one object,
J0903+0708, in W11 that does not satisfy the criteria for being a
PHL~1811 analog but should be considered a WLQ. For simplicity of 
the presentation in this paper, we added this object to the W12 WLQ sample.

\subsection{The Full Sample and Its Luminosity, Redshift, and Emission-Line 
Properties}

\begin{figure}
\centerline{
\includegraphics[scale=0.5]{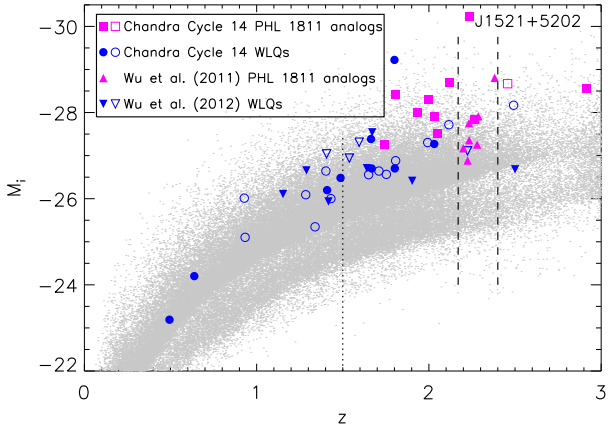}
}
\caption{
Redshift vs.\ absolute $i$-band magnitude
for the samples of
PHL~1811 analogs and WLQs. 
The underlying gray dots represent objects from the SDSS DR7 quasar catalog. 
The vertical dashed lines illustrate
the narrow redshift range for
the PHL~1811 analogs in W11; our newly selected PHL~1811 analogs
have a broader redshift range and generally higher fluxes than
the W11 objects. The vertical dotted line marks the redshift ($z=1.5$)
where our sample objects start to have \iona{C}{iv} coverage and be
included in the \iona{C}{iv}
subsample. Solid symbols represent X-ray weak objects, while
open symbols represent X-ray normal objects.}
\label{fig-lz}
\end{figure}

In the full sample, 
we have in total 18 ($11+7$) radio-quiet PHL~1811 analogs and 33 
($22+10+1$) radio-quiet WLQs.
We show in Figure~\ref{fig-lz} the locations of these objects in
the redshift vs.\ absolute $i$-band magnitude plane, which
highlights the broader redshift range and brighter optical fluxes
of the PHL~1811 analogs
in our \chandra\ Cycle 14 sample. 
There is relatively 
little overlap between PHL~1811 analogs and WLQs in this plot,
purely due to the differing selection approaches (Sections~\ref{sec-sel11}
and \ref{sec-selwlq});
physically, they are likely similar types of object (W11).

\begin{figure*}
\centerline{
\includegraphics[scale=0.5]{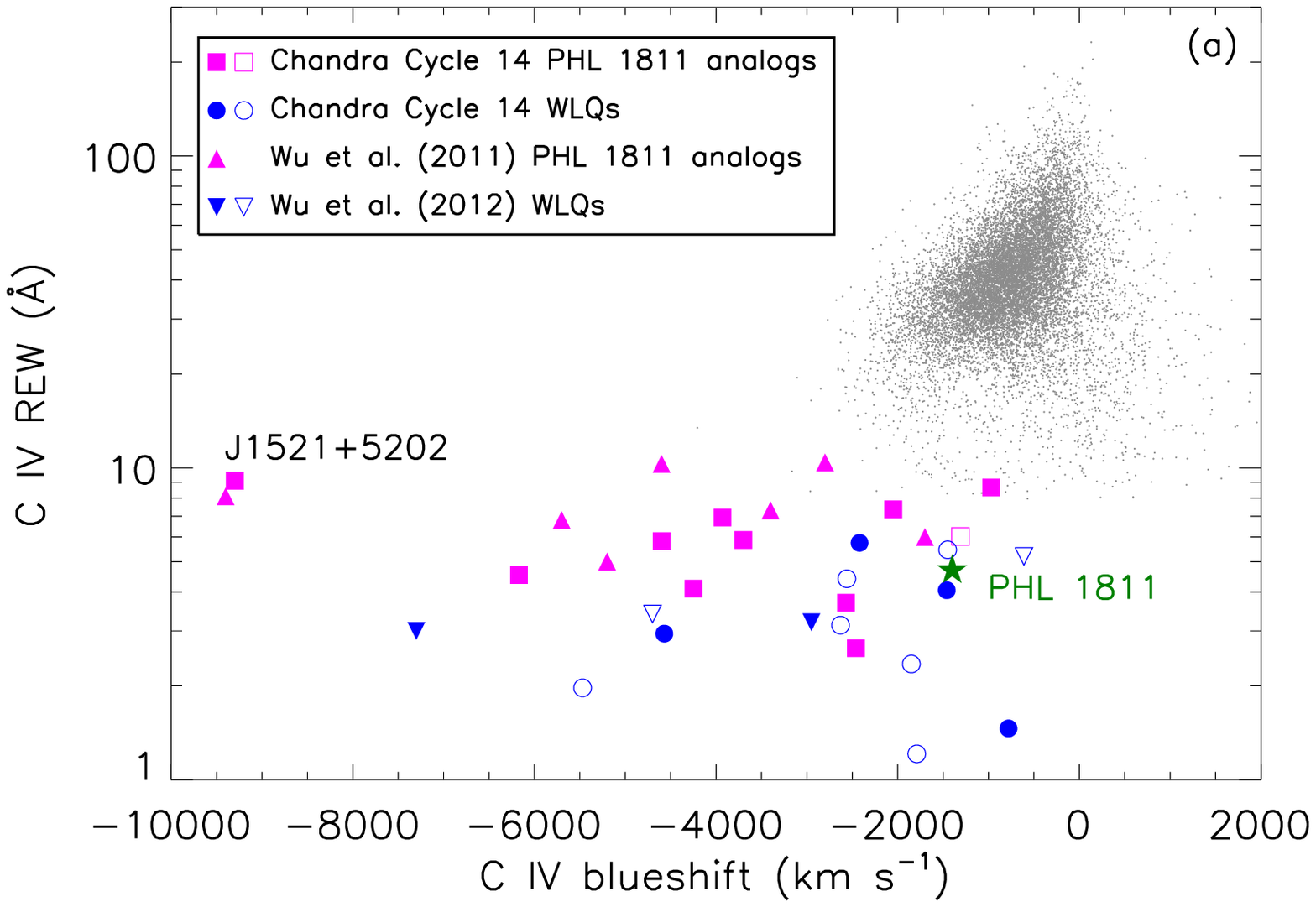}
\includegraphics[scale=0.5]{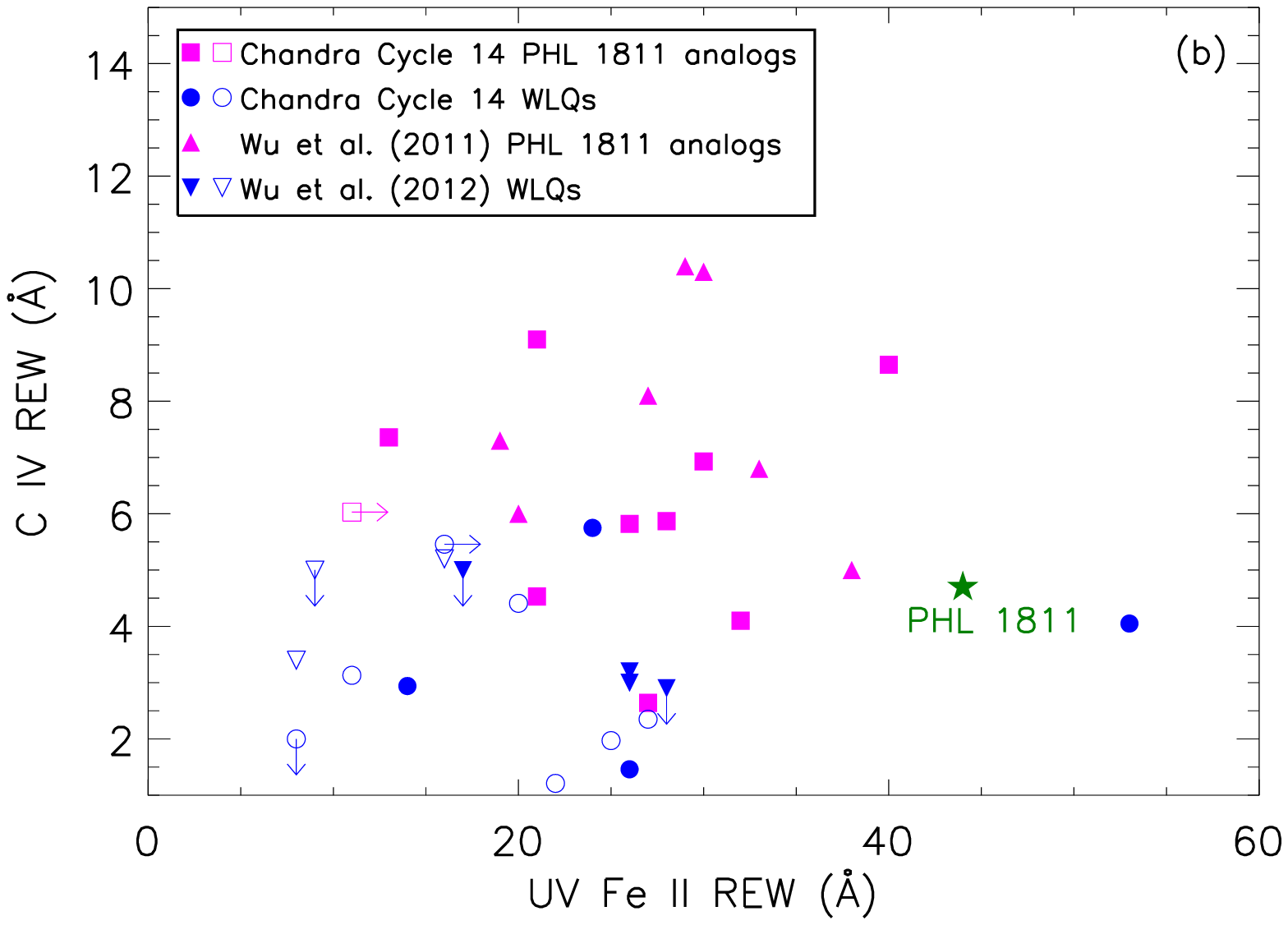}
}
\caption{
(a) \iona{C}{iv} REW
vs.\ \iona{C}{iv} blueshift for the samples of
PHL~1811 analogs and WLQs. Only objects having
both \iona{C}{iv} REW and blueshift measurements are shown
here.
The underlying gray dots are
the 13\,582 typical radio-quiet quasars in \citet{Richards2011}.
The PHL~1811 analogs and WLQs have remarkably weak
(as expected from the selection criteria detailed in
Sections~\ref{sec-sel11}
and \ref{sec-selwlq})
and usually
strongly blueshifted
\iona{C}{iv} emission. 
(b) \iona{C}{iv} REW
vs.\ \iona{Fe}{ii} REW for the samples of
PHL~1811 analogs and WLQs. Arrows represent upper limits (for 
\iona{C}{iv} REWs) or lower limits (for \iona{Fe}{ii} REWs).
The PHL~1811 analogs have in general larger \iona{Fe}{ii} REWs than 
the WLQs by 
selection.
In both panels, solid symbols represent X-ray weak objects, while
open symbols represent X-ray normal objects; the 
green star represents PHL~1811.
}
\label{fig-vc4}
\end{figure*}

In Figure~\ref{fig-vc4}a we compare
the \iona{C}{iv} REWs and blueshifts of our sample objects 
(not all the sources in the full sample have \iona{C}{iv} coverage) to those
of typical radio-quiet quasars \citep{Richards2011}.
The PHL~1811 analogs and WLQs have remarkably weak and strongly blueshifted
\iona{C}{iv} emission, although \iona{C}{iv} blueshift was not 
a selection criterion for the WLQs. 
For comparison, only $0.48\%$ of the radio-quiet quasars
in \citet{Richards2011} have \iona{C}{iv} ${\rm REW}<10$~\AA, and
$30\%$ ($1.6\%$) have \iona{C}{iv} blueshift $>1000$~km~s$^{-1}$ 
($>2000$~km~s$^{-1}$). Due to the differing selection approaches
(e.g., Footnote~\ref{footnote-sel}), the PHL~1811 analogs have in general
somewhat
larger \iona{C}{iv} REWs than the WLQs, 
but we do not believe this indicates any fundamental difference between 
them; both groups have strikingly weak \iona{C}{iv}
emission relative to the general quasar population.
In Figure~\ref{fig-vc4}b we compare the 
\iona{Fe}{ii} REWs for the PHL~1811 analogs and WLQs.
By our sample selection,
the PHL~1811 analogs also have generally larger \iona{Fe}{ii} REWs than 
the WLQs, although there is significant overlap between the two groups.

Since some of our WLQs are at relatively low redshifts ($z\le1.5$) with no
\iona{C}{iv} coverage in the SDSS spectra, the full sample might
have contamination from objects having stronger \iona{C}{iv} lines 
(e.g., \citealt{Plotkin2015})
or \iona{C}{iv} BALs (e.g., similar to J1332+0347 noted
above). Therefore, in some of the following analyses, we present results
based on the subsample of objects at $z>1.5$ 
with \iona{C}{iv} coverage (the \iona{C}{iv}
subsample) to avoid such contamination. All the 18 
PHL~1811 analogs are in the \iona{C}{iv}
subsample, and 18 of the 33 WLQs 
({11 of the 22 
in the newly observed X-ray sample of WLQs})
are in this subsample.\footnote{The 
\iona{C}{iv}
subsample also includes J1013+4927,
J1604+4326, and J2115+0001 in W12 that do not have \iona{C}{iv}
measurements in Table 2 of W12 but actually have upper limits on the
\iona{C}{iv} REWs.}

\section{X-RAY DATA ANALYSIS}
\subsection{X-ray Counts and Photometric Properties} \label{sec-xprop}
We reduced and analyzed the \chandra\ Cycle 14 data
using mainly the \chandra\ Interactive Analysis
of Observations (CIAO) tools.\footnote{See
\url{http://cxc.harvard.edu/ciao/} for details on CIAO.}
For each source, we reprocessed the data
using the {\sc chandra\_repro} script to apply the latest calibration.
The background light curve of each observation was inspected
and background flares were removed using the {\sc deflare}
script with an iterative 3$\sigma$ clipping algorithm.
The cleaned exposure times are listed in
Table~1.

For each source, we created images in the 0.5--8 keV (full),
0.5--2~keV (soft), and 2--8~keV (hard) bands 
from the cleaned event file using the standard
\asca\ grade set (\asca\ grades 0, 2, 3, 4, and 6).
We then ran {\sc wavdetect} \citep{Freeman2002}
on the images to search for X-ray sources
with a ``$\sqrt{2}$~sequence'' of wavelet scales (i.e.,\ 1, 1.414, 2,
2.828, and 4 pixels)
and a false-positive probability threshold of 10$^{-6}$.
If a source is detected in at least one band, we adopt the 
{\sc wavdetect} position that is closest to its SDSS position 
as the X-ray position. 
Seven of the 11 PHL 1811 analogs are detected, and
15 of the 22 WLQs are detected.
The X-ray-to-optical positional offsets
for our targets are small,
ranging from 0.1\arcsec\ to 0.9\arcsec\ with a mean value
of $0.43\pm0.06\arcsec$. We also verified that there is no 
confusion with the X-ray source identification (e.g., no close pairs). 
If a source is not detected by {\sc wavdetect}, the SDSS position
is adopted as the X-ray position.

We performed aperture photometry to 
assess the detection significance and 
extract source counts in 
each of the three energy bands. Source counts were extracted using a
2\arcsec-radius circular aperture centered on the \hbox{X-ray} position, 
which corresponds to 
encircled-energy fractions (EEFs) of 0.939, 0.959, and 0.907 in 
the full, soft, and hard bands, respectively.
Background counts were extracted
from an annular region centered on the X-ray position
with inner radius 10\arcsec\ and outer radius
40\arcsec. 
For each source in each band,
we computed a binomial no-source probability, $P_{\rm B}$,
to assess the significance of the source
signal \citep[e.g.,][]{Broos2007,Xue2011,Luo2013},
which is defined as
\begin{equation}
P_{\rm B}(X\ge S)=\sum_{X=S}^{N}\frac{N!}{X!(N-X)!}p^X(1-p)^{N-X}~.
\end{equation}
In this equation, $S$ is the total number of counts in the
source-extraction region; $N=S+B$, where $B$
is the total number of counts in the background extraction 
region; $p=1/(1+BACKSCAL)$,
where $BACKSCAL$ is {the ratio of the area of the background region to
that of the source region.}
$P_{\rm B}$ represents the
probability of observing the source counts by chance 
under the assumption that there is no real source at the relevant location.
If the $P_{\rm B}$ value
in a band is smaller than 0.01 ({corresponding to a 
$\ge2.6\sigma$ detection}),
we considered the source detected in this band and calculated
the 1$\sigma$ errors on the net counts,
which were derived from the 1$\sigma$ errors
on the extracted source and
background counts \citep{Gehrels1986} following the numerical
method in Section 1.7.3 of \citet{Lyons1991}.
{We note that this $P_{\rm B}$ threshold is appropriate in our 
case where the position of interest is precisely specified in advance.}
If the $P_{\rm B}$ value is larger than 0.01, we considered the
source undetected and derived
an upper limit on the source counts
using the Bayesian approach of \citet{Kraft1991}
for a 90\% confidence level. 
In Table~1, we list
the source counts and upper limits in the soft and hard bands.

We derived a 0.5--8 keV effective power-law photon index 
($\Gamma_{\rm eff}$) for each source from the
band ratio between the hard and soft band counts, calibrated using 
the Portable, Interactive, Multi-Mission Simulator
(PIMMS)\footnote{\url{http://cxc.harvard.edu/toolkit/pimms.jsp}.} with 
the assumption of a power-law spectrum modified with
Galactic absorption \citep{Dickey1990}.
{$\Gamma_{\rm eff}$ represents the real power-law photon index
if the underlying spectrum is truly a power law with Galactic
absorption.}
The uncertainties of (or limits on)
the band ratios (and subsequently
$\Gamma_{\rm eff}$) were
derived using the
Bayesian code {\sc behr} \citep{Park2006}.
For sources undetected in both the soft and hard bands,
$\Gamma_{\rm eff}$ cannot be constrained and a value of 1.9
was adopted for flux or flux density computations later 
(Section~\ref{sec-aox} below),
which is typical for 
luminous radio-quiet quasars
\citep[e.g.,][]{Reeves1997,Just2007,Scott2011}.
The band ratios and $\Gamma_{\rm eff}$ values are listed in Table~1.

\subsection{Spectral Analysis of J1521+5202} \label{sec-J1521}

J1521+5202 is an exceptionally luminous quasar (e.g., Figure~\ref{fig-lz}).
However,  it is remarkably X-ray weak (by a factor of $\approx35$), as first reported by
\citet{Just2007} using a 4 ks \chandra\ observation. As a PHL~1811 analog,
J1521+5202 was studied in W11, which also presented 
a near-infrared (NIR) spectrum showing very weak [\iona{O}{iii}] emission and 
a higher optical \iona{Fe}{ii}/H$\beta$ ratio than those of typical quasars.
The source was only weakly detected in
the 4~ks \chandra\ observation with 3 full-band counts, 
preventing further constraints on the 
band ratio and effective photon index. With our much longer 
\chandra\ observation (37.1~ks cleaned exposure) of J1521+5202,
we derived a band ratio of $1.1_{-0.2}^{+0.3}$ and 
an effective power-law photon index of $0.6\pm0.2$ (Table~1). This flat
effective photon index, compared to the typical
value of $\Gamma\approx1.9$ for luminous radio-quiet quasars
\citep[e.g.,][]{Reeves1997,Just2007,Scott2011},
suggests that the significant X-ray weakness of 
this source is likely due to strong absorption.

\begin{figure}
\centerline{
\includegraphics[scale=0.5]{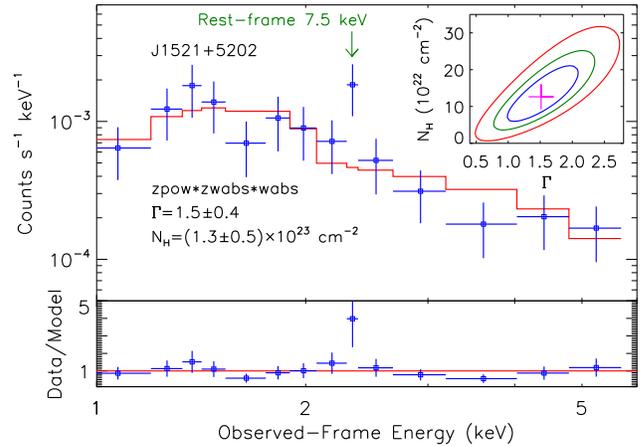}
}
\caption{
\chandra\ spectrum of J1521+5202 (blue)
overlaid with the best-fit model (red). 
The 0.3--8~keV spectrum was
fitted with an absorbed power-law model modified with Galactic absorption
using cstat in XSPEC. For display purposes, 
the data are
grouped with a minimum of six counts per bin.
The inset shows the best-fit $\Gamma$ and $N_{\rm H}$ values (magenta plus),
and their
1$\sigma$ (blue), 2$\sigma$ (green), and 3$\sigma$ (red) 
contours.
The bottom panel shows the ratio of the data to the best-fit model.
The best-fit model suggests strong intrinsic X-ray absorption,
with $N_{\rm H}=(1.3\pm0.5)\times10^{23}$~cm$^{-2}$.}
\label{fig-j1521spec}
\end{figure}

We then performed basic spectral analysis for J1521+5202. The 0.3--8~keV
(rest-frame 1.0--25.5 keV)  
spectrum was extracted using the CIAO
{\sc specextract} script within a circular aperture of 4\arcsec\
radius centered on the X-ray position. There are $\approx90$ source counts
extracted including $<1$ background count. The background spectrum was
extracted from an annular region with inner radius
6\arcsec\ and outer radius 15\arcsec, which is free of X-ray source
contamination. Given the limited source counts, we 
fitted the spectrum using XSPEC
(version 12.8.1; \citealt{Arnaud1996}) with 
a simple model consisting of a power law modified by
both intrinsic absorption (at $z=2.24$)
and Galactic absorption. The $C$ statistic (cstat) was used due to the 
limited photon counts \citep{Cash1979}.
The best-fit photon index is
$\Gamma=1.51_{-0.35}^{+0.38}$, and the intrinsic absorption column density is 
$N_{\rm H}=(1.31_{-0.48}^{+0.56})\times10^{23}$~cm$^{-2}$ ($C=88$ for 
75 degrees of freedom). The spectrum and the best-fit model are displayed
in Figure~\ref{fig-j1521spec}. There is a possible emission-line feature
at rest-frame 7.5~keV, which would likely correspond to a blueshifted 
Fe line if confirmed \citep[e.g.,][]{Reeves2004}. More data are 
required to constrain this feature. 

The spectral analysis also suggests significant X-ray absorption in J1521+5202,
in agreement with the simple flat $\Gamma_{\rm eff}$ argument above. 
We note that the absorption derived from our simple spectral model
cannot entirely account for the observed X-ray weakness;
after correction for this absorption J1521+5202 
is still a factor of $\approx7$ times
X-ray weak (a $\approx2.5\sigma$ deviation 
from the \hbox{$\alpha_{\rm OX}$--$L_{\rm 2500~{\textup{\AA}}}$}
relation). This result suggests that our simple model probably did not recover
the real (larger) absorption column density associated with a more
complex absorber, as is commonly seen in highly obscured
AGNs.\footnote{Therefore, the best-fit
$\Gamma$ value is likely not representative of the intrinsic
power-law spectral shape, and we cannot obtain a reliable 
estimate of the Eddington ratio 
based on this X-ray photon
index (see Section~\ref{sec-ledd} for further discussion).} 
Alternatively,
J1521+5202 may also be 
intrinsically X-ray weak by a factor of a few
besides the strong X-ray absorption.

\section{MULTIWAVELENGTH PROPERTIES}

\subsection{Distribution of the X-ray-to-Optical Power-Law Slopes} \label{sec-aox}

We measured the $\alpha_{\rm OX}$ parameters for our \chandra\ Cycle 14 sample objects.
The \hbox{rest-frame} 
2500~\AA\ flux densities
were adopted from the \citet{Shen2011} SDSS quasar catalog; for three
objects that were not included in this catalog, the flux densities
were interpolated/extrapolated from their optical photometric data.
The \hbox{rest-frame} 2~keV flux densities were derived from the 
soft-band net counts (Table~1) using PIMMS assuming a power-law 
spectrum modified with Galactic absorption; the observed soft band (0.5--2 keV)
covers rest-frame 2~keV for our sources.
If a source is not detected in the soft band, 
an upper limit 
on $f_{2~{\rm keV}}$ was calculated.
There are two sources, J0147+0003 and J1133+1142, that are not detected in
the soft band but are detected in the hard band. We still present
upper limits on $f_{2~{\rm keV}}$ (and $\alpha_{\rm OX}$) for these two
sources, although the $f_{2~{\rm keV}}$ flux density could be estimated
via extrapolation of the hard-band flux assuming a power-law photon index.
This approach does not affect the analyses in this study. The $\alpha_{\rm OX}$
values, along with other relevant parameters, are listed in Table~3.
The level of \hbox{X-ray} weakness is usually measured by the 
$\Delta\alpha_{\rm OX}$ parameter, which is 
defined as the difference
between the observed $\alpha_{\rm OX}$ and the one expected from
the \hbox{$\alpha_{\rm OX}$--$L_{\rm 2500~{\textup{\AA}}}$} relation 
($\Delta\alpha_{\rm OX}=\alpha_{\rm OX,obs}-\alpha_{\rm OX,exp}$).
From $\Delta\alpha_{\rm OX}$, we can derive the factor of X-ray weakness 
($f_{\rm weak}=403^{-\Delta\alpha_{\rm OX}}$) with respect to the expected X-ray flux.
These parameters are also listed in Table~3.

\begin{figure}
\centerline{
\includegraphics[scale=0.5]{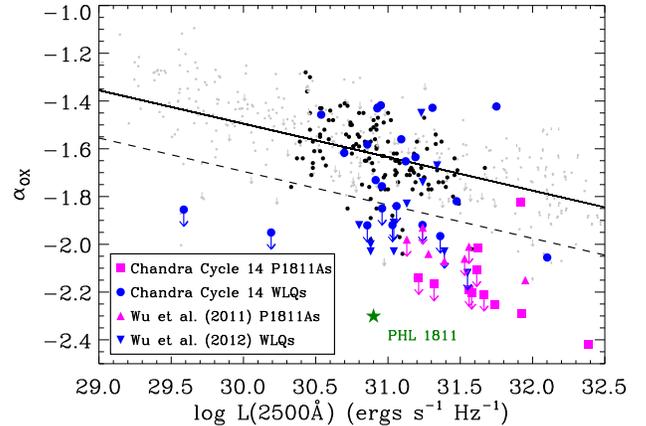}
}
\caption{
X-ray-to-optical power-law slope ($\alpha_{\rm OX}$) vs.\ 2500 \AA\
monochromatic luminosity
for the samples of
PHL~1811 analogs and WLQs.
For comparison, the 132 radio-quiet, non-BAL quasars in the improved version
of \citet{Gibson2008} Sample B (see Footnote 16 of W11)
are represented by the black dots, and PHL~1811 is shown as the green star.
The small grey dots and downward arrows (upper limits)
are from the samples of \citet{Steffen2006} and \citet{Just2007},
and the solid
line represents the \citet{Just2007} 
$\alpha_{\rm OX}$--$L_{\rm 2500~{\textup{\AA}}}$ relation.
The dashed line ($\Delta\alpha_{\rm OX}=-0.2$) marks the 
division between X-ray weak and X-ray normal quasars adopted in this study.
}
\label{fig-aox}
\end{figure}

The $\alpha_{\rm OX}$ vs.\ $L_{\rm 2500~{\textup{\AA}}}$ distribution 
for our full sample is displayed in Figure~\ref{fig-aox}; also shown 
for comparison are PHL 1811 and some typical AGN samples 
\citep{Steffen2006,Just2007,Gibson2008}. 
We adopt $\Delta\alpha_{\rm OX}=-0.2$ ($f_{\rm weak}=3.3$) to be 
the dividing threshold between X-ray weak and X-ray normal quasars, 
which corresponds to a $\approx1.3\sigma$
($\approx90\%$ single-sided confidence level)
offset given the $\alpha_{\rm OX}$ rms scatter in Table~5 of
\citet{Steffen2006}. A remarkable fraction of our sample sources are 
X-ray weak, similar to PHL~1811. Specifically, 
of the 18 PHL~1811 analogs, 17 ($94_{-23}^{+6}\%$)
are X-ray weak, and 16 ($48_{-12}^{+15}\%$) of the 33 WLQs
are X-ray weak. In the \iona{C}{iv} subsample, 8 ($44_{-15}^{+22}\%$) 
of the 18 WLQs
are X-ray weak. For comparison, the fraction of X-ray weak
objects in the improved version of 
\citet{Gibson2008} Sample B 
(see Footnote 16 of W11) is only 7.6\% (10/132).

\begin{figure*}
\centerline{
\includegraphics[scale=0.5]{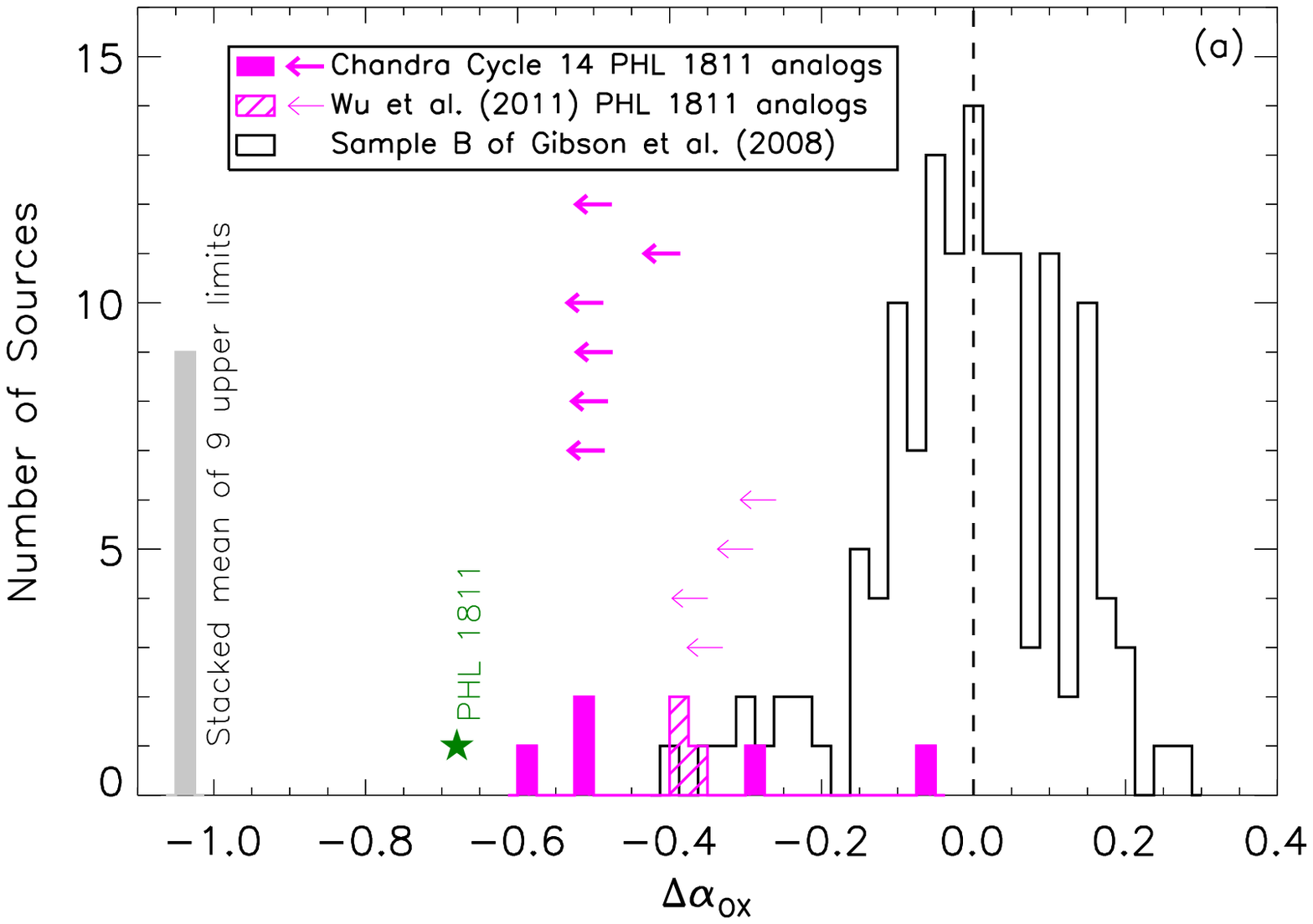}
\includegraphics[scale=0.5]{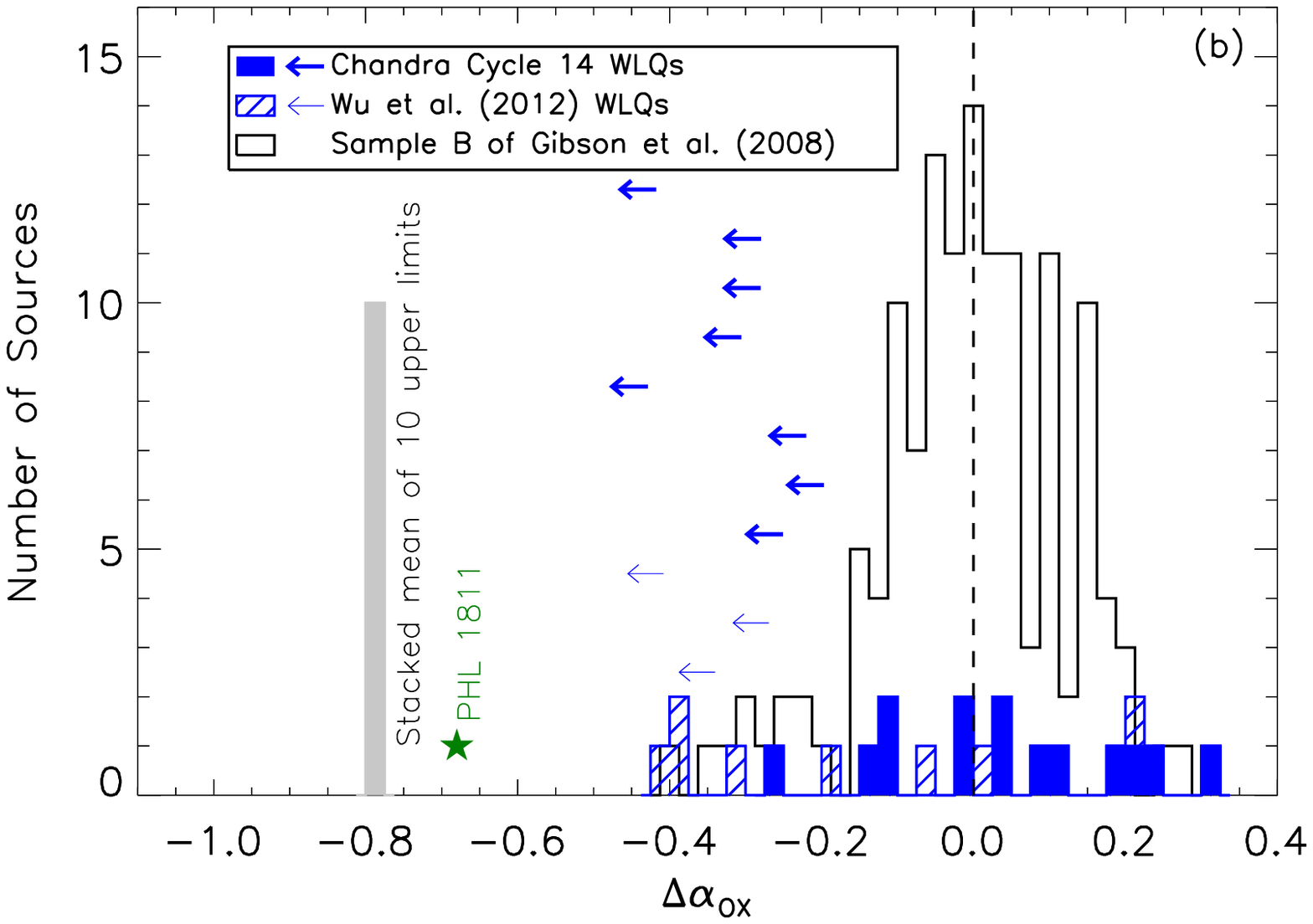}
}
\caption{Distribution of the $\Delta\alpha_{\rm OX}$ values for the
(a) PHL~1811 analogs and (b) WLQs in our sample. The solid
and hatched shaded histograms (thick and thin 
leftward arrows) represent X-ray detected
(undetected) sources in the \chandra\ Cycle 14 sample and the previous
W11 or W12 sample, respectively. In both panels, the gray bar shows
the average $\Delta\alpha_{\rm OX}$ value for the undetected sources
estimated via stacking analyses (see Section~\ref{sec-stacking} below);
there is one \xmm\ undetected source that was not included in the stacking
in each case. For comparison, the $\Delta\alpha_{\rm OX}$ distribution
for the 132 \citet{Gibson2008} Sample B quasars
is presented as the unshaded histogram, and the $\Delta\alpha_{\rm OX}$
value of PHL~1811 is shown as the green star. A significant fraction of the
PHL~1811 analogs and WLQs are X-ray weak compared to typical quasars.}
\label{fig-aoxhist}
\end{figure*}

The significantly different fractions of X-ray weak objects among typical 
quasars and our 
samples of PHL~1811 analogs and WLQs 
are also evident in the $\Delta\alpha_{\rm OX}$ distributions
(Figure~\ref{fig-aoxhist}). 
As in W11 and W12,
we used the Peto-Prentice test 
implemented in the Astronomy Survival Analysis
package (ASURV; e.g., \citealt{Feigelson1985,Lavalley1992}) 
to assess whether two samples follow the
same distribution.
For PHL~1811 analogs, the probability of their
$\Delta\alpha_{\rm OX}$ values being drawn from the same parent
population as those of typical quasars (improved version of
\citealt{Gibson2008} Sample B) is $\approx10^{-23}$ (10.4$\sigma$).
For WLQs,
this probability is $3\times10^{-5}$ (4.1$\sigma$).

Since most of our X-ray weak 
PHL~1811 analogs and WLQs are not X-ray
detected and only upper limits on $\Delta\alpha_{\rm OX}$ are available,
we also performed X-ray stacking analyses (see Section~\ref{sec-stacking} below)
to obtain the average level of \hbox{X-ray} weakness.
The stacking results are listed
in Table~4 as subsamples ``P1811A SB undet'' and ``WLQ undet''
for the undetected sources (excluding two \xmm\ undetected
sources), 
and the stacked $\Delta\alpha_{\rm OX}$
values are plotted as gray bars in Figure~\ref{fig-aoxhist}. 
The corresponding average X-ray weakness factors for the undetected
PHL~1811 analogs and WLQs are 513 and 108, respectively.
As noted above, there are two PHL~1811 analogs in the stacked sample, 
J0147+0003 and J1133+1142, that are not detected in
the soft band but are detected in the hard band. Excluding these two 
sources from the stacking still yields a large negative stacked 
$\Delta\alpha_{\rm OX}$
value ($-0.87$), as shown in the results for subsample ``P1811A undet'' 
in Table~4.

Including detected objects, the stacked $\Delta\alpha_{\rm OX}$ values 
are $-0.61$ and $-0.47$ for the X-ray weak PHL~1811 analogs and WLQs, 
corresponding to average X-ray weakness factors of 39 and 17, respectively.
The X-ray weak WLQs in the \iona{C}{iv} subsample have a
similar stacked $\Delta\alpha_{\rm OX}$
value ($-0.62$; average X-ray weakness factor of $\approx41$)
to the X-ray weak PHL~1811 analogs.
In an absorption scenario, an X-ray weakness factor of 40 corresponds to 
$N_{\rm H}\approx9\times10^{23}$~cm$^{-2}$
for $z=2$ and $\Gamma=1.9$ in the {\sc MYTorus} model \citep{Murphy2009}.
Using the Kaplan--Meier estimator in ASURV, we also derived the mean
$\Delta\alpha_{\rm OX}$ values for the X-ray weak PHL~1811 analogs and WLQs,
which are $-0.49\pm0.03$ and $-0.38\pm0.02$, respectively.
Note that the Kaplan--Meier estimator assumes that the censored data have an 
identical distribution as the measured data \citep{Feigelson1985}, which
may not be applicable to our $\Delta\alpha_{\rm OX}$ values here;
also, the mean and stacked values differ in the sense that the mean
$\Delta\alpha_{\rm OX}$ estimates the geometric mean of the X-ray weakness 
factors while the stacked $\Delta\alpha_{\rm OX}$ estimates the arithmetic 
mean.

\subsection{IR-to-X-ray Spectral Energy Distributions} \label{sec-sed}

\begin{figure}
\centerline{
\includegraphics[scale=0.5]{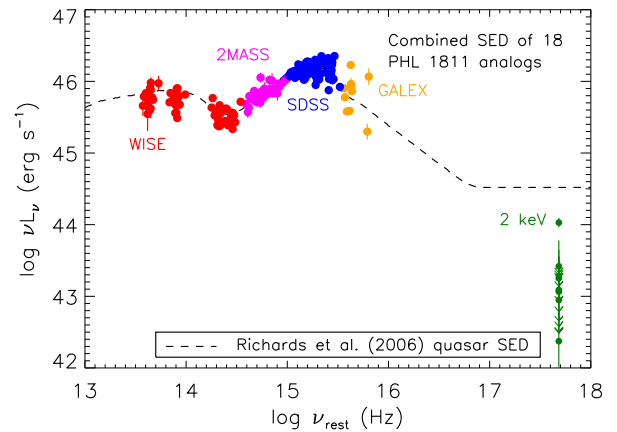}
}
\caption{
Combined SED for the PHL~1811 analogs.
The IR-to-UV SED data were collected from the
{\it WISE} (red), 2MASS (magenta), SDSS (blue), and {\it GALEX} (orange)
catalogs. These data have been corrected for Galactic extinction
following the dereddening
approach of \citet{Cardelli1989} and \citet{Odonnell1994}.
The green data points and arrows show the 2~keV
luminosities and upper limits.
The SED for each object
was scaled to the composite quasar SED of optically
luminous quasars \citep{Richards2006} at rest-frame 3000~\AA,
and then combined.
Some of the {\it GALEX} data points affected by the
Lyman break are removed in the combined SED for display purposes.
The combined IR-to-UV SED of PHL~1811 analogs is similar to the 
composite quasar SED.
We do not separate the X-ray normal PHL~1811 analog (J1537+2716) from
the other X-ray weak objects as it also has a typical quasar SED.
}
\label{fig-sed1811}
\end{figure}

\begin{figure*}
\centerline{
\includegraphics[scale=0.5]{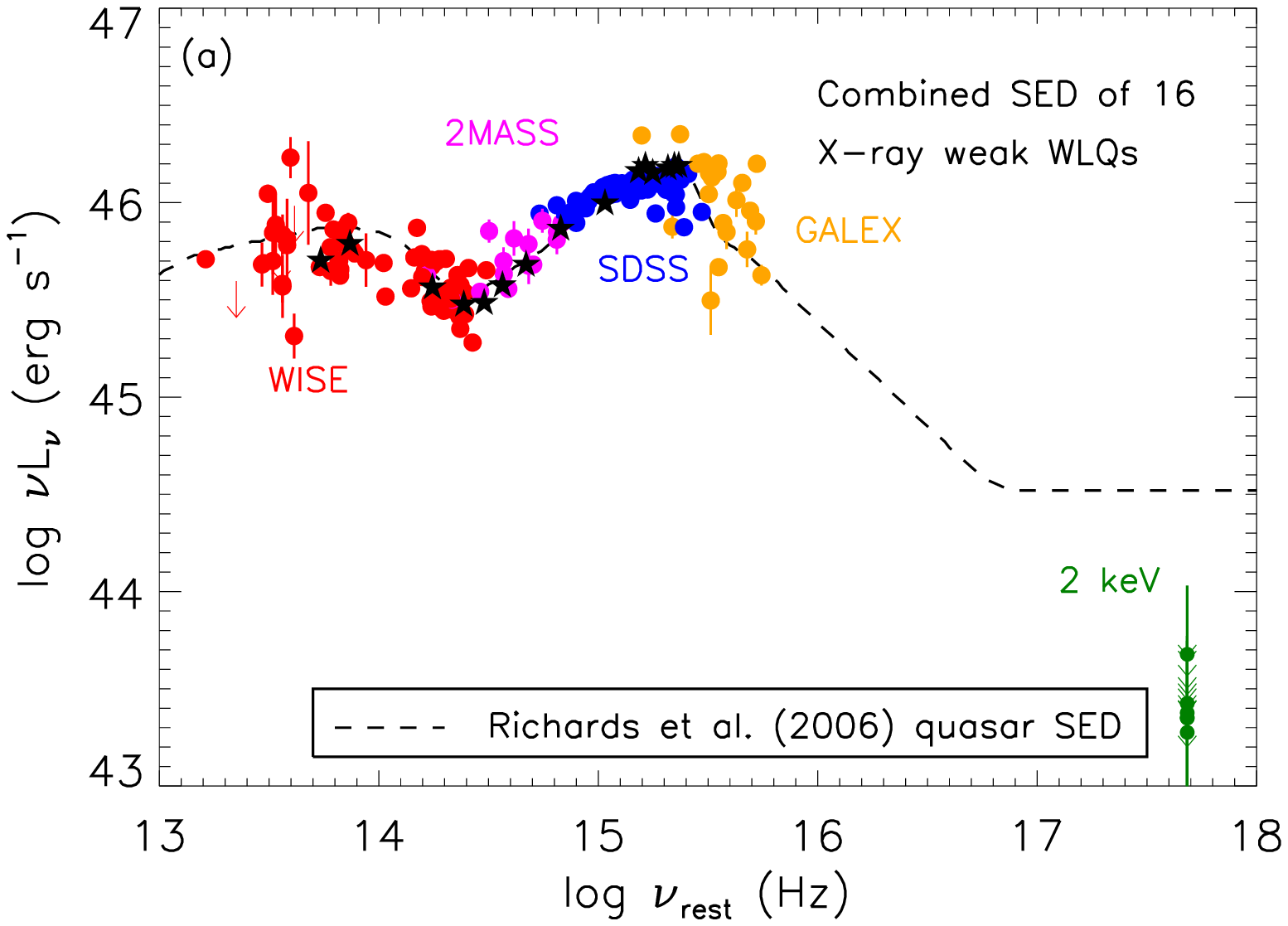}
\includegraphics[scale=0.5]{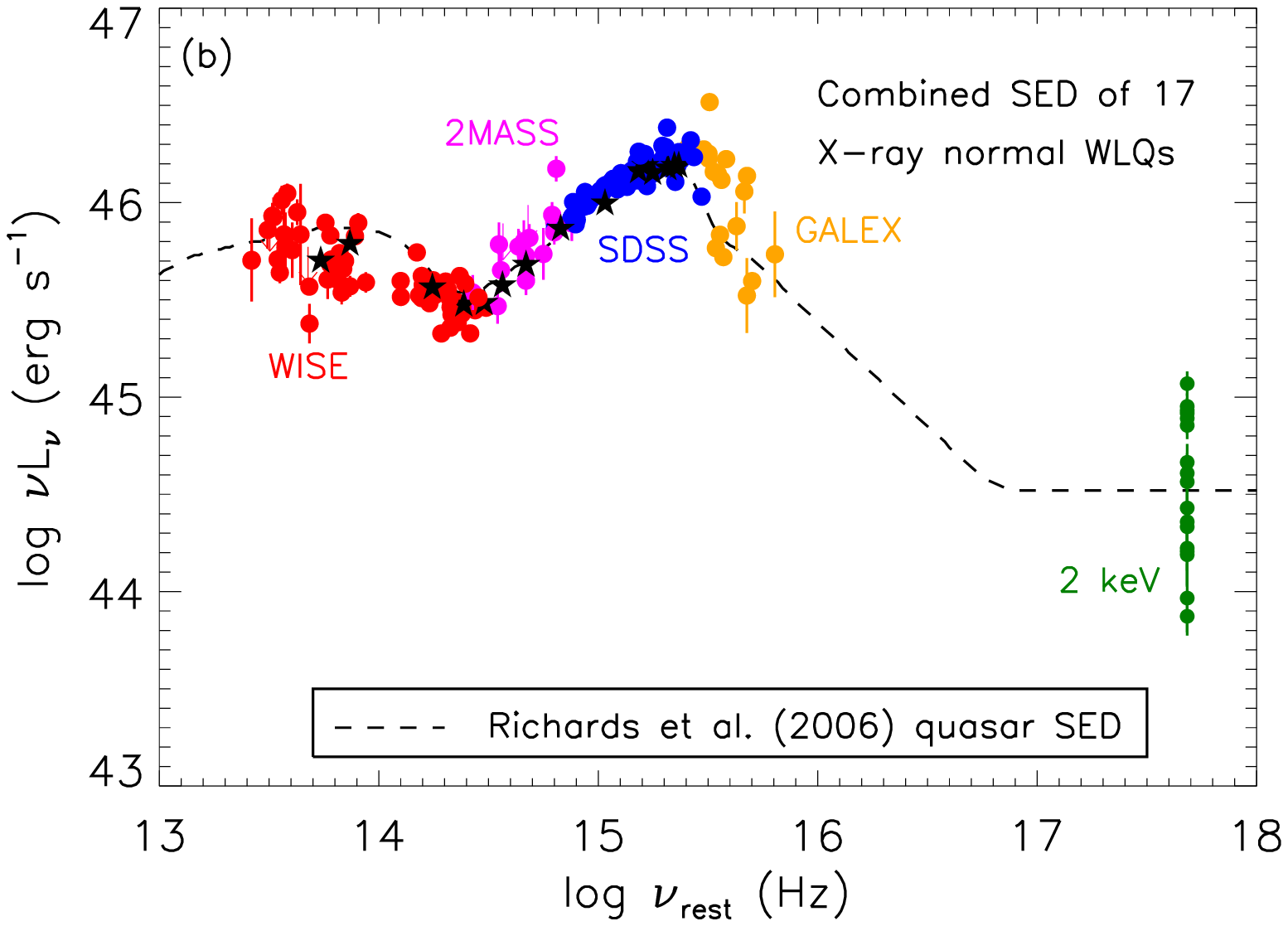}
}
\caption{
Same as Figure~\ref{fig-sed1811} but for the 
(a)
X-ray weak and (b) \hbox{X-ray} normal WLQs.
The black stars represent the composite SED of 18 high-redshift 
(\hbox{$z=2.7$--5.9}) WLQs in \citet{Lane2011}.
The combined SEDs of the X-ray weak and X-ray normal WLQs are similar
from the IR to UV, and they agree well with the composite quasar SED.
}
\label{fig-sedwlq}
\end{figure*}

We constructed infrared (IR) to X-ray continuum
SEDs for our full sample of
PHL~1811 analogs and WLQs, using photometric
data collected from the {\it Wide-field Infrared Survey Explorer}
({\it WISE}; \citealt{Wright2010}), Two Micron All Sky Survey (2MASS;
\citealt{Skrutskie2006}), SDSS, and/or {\it Galaxy Evolution Explorer}
({\it GALEX}; \citealt{Martin2005}) catalogs. 
The optical and UV data have been corrected for Galactic extinction
following the dereddening
approach of \citet{Cardelli1989} and \citet{Odonnell1994}.
The combined SEDs for the PHL~1811 analogs and WLQs 
are displayed in Figures~\ref{fig-sed1811} and \ref{fig-sedwlq}, respectively.
The mean SED of optically luminous SDSS quasars from \citet{Richards2006}
is shown for comparison. 
Because of our sample selection criteria, 
the PHL~1811 analogs are about
a factor of $\approx3$ times 
more optically luminous on average than the \citet{Richards2006}
optically luminous
SDSS quasars, while the WLQs have comparable luminosities to the SDSS
quasars. Regardless of whether they are X-ray weak or X-ray normal, 
our PHL~1811 analogs and WLQs in general have IR-to-UV SEDs 
similar to those of typical quasars. 
Such SEDs differ significantly from those of BL Lac objects,
indicating that our sample is not contaminated by any radio-quiet
BL Lac candidates (e.g., see discussion on J1109+3736 in W12).
Studies of a sample of high-redshift 
($z=2.7$--5.9) WLQs also found typical quasar SEDs \citep{Lane2011}.

We estimated bolometric luminosities ($L_{\rm Bol}$) for our sample objects
by integrating
the \citet{Richards2006} composite SED ({between $10^{12.5}$--$10^{19}$~Hz})
scaled to their rest-frame 3000~\AA\ luminosities.\footnote{The 0.5--10~keV
luminosity only constitutes 4.5\% of the bolometric luminosity
in the \citet{Richards2006} composite SED. Therefore, assuming
a nominal level of X-ray emission for our sample objects when estimating
bolometric luminosities does not affect the results materially.}
The resulting bolometric luminosities for the full sample
are listed in Table~5, ranging from
\hbox{$\approx2\times10^{45}$~\lum} to $\approx2\times10^{48}$~\lum\
with a median value of $\approx8\times10^{46}$~\lum.
We compared these bolometric luminosities to those estimated via integrating
the SEDs directly. The ratios of the two luminosities have a median value 
of $\approx1$, and an rms scatter of $\approx3$, suggesting that 
the typical uncertainty associated with the estimated $L_{\rm Bol}$ values is 
approximately a factor of three. There may also be additional 
systematic uncertainty if the EUV properties of these objects 
differ from those of typical quasars (e.g., see Sections~\ref{sec-shieldingscenario}
and \ref{sec-phy}; \citealt{Krawczyk2013}).

\subsection{Radio Properties} \label{sec-radio}

Only radio-quiet ($R<10$) quasars are included in our sample.
The radio-loudness parameter was computed as
$R=f_{5~{\rm GHz}}/f_{\rm 4400~{\textup{\AA}}}$
\citep[e.g.,][]{Kellermann1989}, where the
rest-frame 5~GHz and 4400~\AA\ flux densities were converted from the
observed 1.4 GHz and rest-frame 2500~\AA\ flux densities (or their
upper limits)
assuming a radio
power-law slope of $\alpha_{\rm r}=-0.8$
($f_\nu\propto \nu^{\alpha}$; e.g., \citealt{Falcke1996}; \citealt{Barvainis2005})
and an optical power-law slope of $\alpha_{\rm o}=-0.5$
(e.g., \citealt{Vandenberk2001}).
We obtained radio flux information at 1.4~GHz via cross-matching to the
Faint Images of the Radio Sky at
Twenty-Centimeters (FIRST) survey catalog \citep{Becker1995,White1997}.
For sources not detected by the FIRST survey, the
upper limits on the radio fluxes were set to $0.25+3\sigma_{\rm rms}$ mJy,
where $\sigma_{\rm rms}$ is the rms noise of the FIRST survey at the
source position and 0.25 mJy is to account for the CLEAN bias
\citep{White1997}. The $R$ values for the \chandra\ Cycle 14 targets
are listed in Table~3.

Given the radio-quiet selection criterion,
all our sample objects have optical-to-radio power-law slopes
$\alpha_{\rm RO}>-0.21$ (e.g., Equation~4 of W12) and are in the radio-quiet
region of the $\alpha_{\rm OX}$ vs. $\alpha_{\rm RO}$ plot (Figure 7 of W12).
However, the two objects with the highest $R$ values in our sample
(J0844+1245 with $R=8.0$ and J1156+1848 with $R=7.4$; see Table~3)
are considered
radio-intermediate in \citet{Shen2011} with $R_{\rm Shen}=11.2$ and
$R_{\rm Shen}=10.6$, respectively. The radio-loudness parameter
was defined differently in \citet{Shen2011}
with $R_{\rm Shen}=f_{5~{\rm GHz}}/f_{\rm 2500~{\textup{\AA}}}$, which
leads to the above small discrepancy. Both quasars are X-ray normal, and
we cannot exclude entirely
the effect of jet contamination in these two objects. In
fact, J0844+1245 is an outlier in the diagnostic
plot for X-ray weak quasars (Figure~\ref{fig-colorfe2} below), which might
be related to its relatively high radio loudness. Since both objects
are still radio-quiet under our definition, and we have also verified that
the inclusion of the two does not affect our statistical analyses
significantly,
we do not remove them from our study.

\section{SAMPLE ANALYSES}

\subsection{X-ray Spectral Stacking Analyses of the X-ray Weak Objects}  \label{sec-stacking}

The X-ray spectral analysis of J1521+5202 
(Section~\ref{sec-J1521}) suggested significant X-ray absorption. For the other
X-ray weak objects in our sample, there are few X-ray 
photons detected for spectral analyses.
The derived effective photon indices (Table~1) 
also have large uncertainties and cannot
constrain whether a source has a flat X-ray spectrum that is 
indicative of absorption. In this case, X-ray stacking analyses
are often used to assess the average spectral properties of the sample.
With this approach, W11 and W12 found relatively flat/hard stacked
effective photon indices for their samples, although with large error bars,
suggestive of X-ray absorption. With the addition of our new X-ray samples,
we can improve the statistics of the stacking results and obtain tighter
spectral constraints.

We stacked the X-ray photometry by 
combining the extracted source and background
counts of individual sources and following 
the same procedure as in Section~\ref{sec-xprop} to derive 
photometric properties of the stacked source. 
Since our objects are generally at high redshifts (a mean value of
$z=1.85$ for the full X-ray weak sample), we are probing
the highly
penetrating X-ray emission in
the $\approx2$--23~keV rest-frame band that is more reliable for assessing 
the presence of heavy absorption than lower-energy observations.
Stacking analyses were 
performed for several subsamples of X-ray weak objects; the results 
are given in Table~4, which include the number of sources used in the
stacking, mean redshift, total stacked exposure time, 
stacked soft- and hard-band counts,
stacked effective photon index, and stacked $\alpha_{\rm OX}$ and 
$\Delta\alpha_{\rm OX}$. 
For the full X-ray weak sample, the stacked 
effective photon index is relatively hard 
($\Gamma_{\rm eff}=1.37_{-0.23}^{+0.25}$) compared to the typical
value of $\Gamma\approx1.9$ for luminous radio-quiet quasars 
\citep[e.g.,][]{Reeves1997,Just2007,Scott2011}. The hard average spectral
shape is more evident in the subsample of X-ray weak 
PHL~1811 analogs, with $\Gamma_{\rm eff}=1.16_{-0.32}^{+0.37}$.
The stacking results suggest that, unlike PHL~1811 itself that appears
intrinsically X-ray weak (at least in the 0.5--8 keV band), these X-ray weak
PHL~1811 analogs and WLQs on average are \hbox{X-ray} absorbed, consistent
with the conclusion from the spectral
analysis of J1521+5202 in Section~\ref{sec-J1521}. However, given 
the uncertainties associated with the stacking procedure (e.g.,  
individual sources contribute differently to the stacked signal due to
the different fluxes, exposure times, and rest-frame bands probed), 
we cannot exclude the possibility
that a fraction of the sources are intrinsically X-ray weak like PHL~1811
(e.g., those sources with no or poor constraints on $\Gamma_{\rm eff}$).

\subsection{Joint Spectral Fitting of the X-ray Normal Objects}  \label{sec-jointfit}

The X-ray stacking analyses of the X-ray weak objects likely did not provide
constraints on their intrinsic X-ray spectral shapes (due to the
likely modification of their spectra by X-ray absorption), and these
objects have too few X-ray photons for spectral fitting 
(the stacked number of counts for all the X-ray weak objects is only 87).
Therefore, to obtain a relatively tight constraint on the intrinsic 
X-ray power-law photon indices of PHL~1811 analogs and WLQs,
we
performed joint spectral fitting of the 
one \hbox{X-ray} normal
PHL~1811 analog and 17 \hbox{X-ray} normal WLQs.
The individual source and background spectra were extracted following the 
same procedure as in Section~\ref{sec-J1521}, and all eighteen 
spectra in the rest-frame $>2$~keV band were fitted jointly 
with an absorbed power-law model using cstat in XSPEC, allowing each source to have its own
redshift and Galactic column density. 
There are $\approx610$ total spectral counts for the fitting including 
only $\approx2$ background counts, and these spectral counts are not
dominated by the counts from one or a few objects.

\begin{figure}
\centerline{
\includegraphics[scale=0.5]{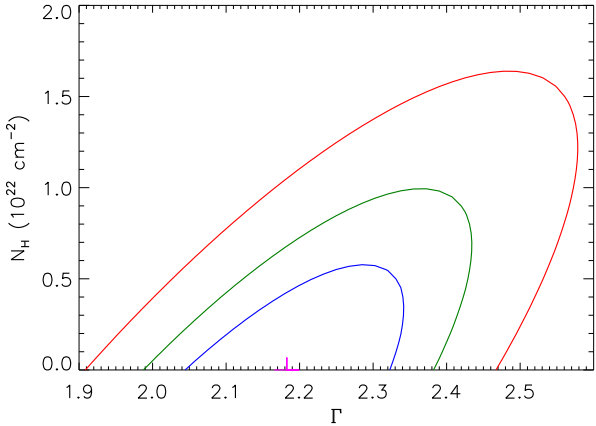}
}
\figcaption{
Contour plot for the photon index and intrinsic absorption column density from
the joint spectral analysis for the X-ray normal PHL~1811 analogs and WLQs.
The blue, green, and red curves are the
1$\sigma$, 2$\sigma$, 3$\sigma$ contours, respectively. The magenta plus
sign indicates the best-fit $\Gamma$ and $N_{\rm H}$ values.
\label{fig-con}}
\end{figure}

The best-fit photon index is
$\Gamma=2.18\pm0.09$ ($C=449$ for 487 degrees of freedom),
and no intrinsic absorption is required with an upper limit
of $N_{\rm H}<2.9\times10^{21}$~cm$^{-2}$. The 1--3$\sigma$
contours for the best-fit $\Gamma$ and $N_{\rm H}$ values are shown in
Figure~\ref{fig-con}. The same analysis was also performed for the 
\iona{C}{iv} subsample, which includes one \hbox{X-ray} normal
PHL~1811 analog and nine X-ray normal WLQs. The resulting photon index is
slightly higher, $\Gamma=2.26\pm0.11$ ($C=283$ for 330
degrees of freedom), and no intrinsic absorption is required
($N_{\rm H}<1.4\times10^{21}$~cm$^{-2}$).
Similar joint fitting was performed for 
seven high-redshift radio-quiet WLQs in \citet{Shemmer2009},
and their resulting $\Gamma$ constraint is consistent with our
$\Gamma$ values within the errors, although
our uncertainties are factors of $\approx5$ times smaller
owing to the substantially larger number of 
spectral counts available.

\subsection{Eddington-Ratio Estimates} \label{sec-ledd}
The extreme emission-line properties of
PHL~1811 analogs and WLQs
could perhaps be related to rapid or even super-Eddington
accretion (e.g., \citealt{Leighly2007b,Leighly2007,Shemmer2010}; W11),
and thus we estimated their Eddington
ratios ($L_{\rm Bol}/L_{\rm Edd}$) based on their estimated bolometric
luminosities and supermassive black hole (SMBH) masses.
Most of our PHL~1811 analogs and WLQs
have their SMBH masses
estimated in \citet{Shen2011} using the single-epoch virial mass approach,
with 25 objects having \iona{Mg}{ii}-based estimates and 21 objects having
\iona{C}{iv}-based estimates.
The \iona{C}{iv} virial mass approach is likely not
applicable for our objects
as the prominent \iona{C}{iv} blueshifts indicate
a strong wind component for the \iona{C}{iv} BELR 
\citep[e.g.,][]{Baskin2005,Richards2011,Trakhtenbrot2012,Shen2013}, and 
\iona{C}{iv}-based masses may be an order 
of magnitude higher than the real values for quasars with weak \iona{C}{iv}
lines
\citep[e.g.,][]{Kratzer2015}.

For the 25 objects with \iona{Mg}{ii}-based SMBH masses, the median
$L_{\rm Bol}/L_{\rm Edd}$ value is 0.27
with an interquartile range of 0.17 to 0.38 (the mean $L_{\rm Bol}/L_{\rm Edd}$
value is 0.39).
Even the \iona{Mg}{ii}-based SMBH masses
are perhaps not reliable for our objects that
have weak \iona{Mg}{ii} line emission, although not as
abnormally
weak as the \iona{C}{iv} line emission 
(e.g., \citealt{Plotkin2015}).\footnote{Being a high-ionization
line, the \iona{C}{iv} line of PHL~1811 analogs and 
WLQs is expected to be
more significantly affected by a soft ionizing SED compared 
to low-ionization lines such as \iona{Mg}{ii} \citep[e.g.,
Section~4.1.4 of][]{Leighly2007b}.} 
A recent VLT/X-shooter study of WLQs \citep{Plotkin2015}
also suggests that,
at least sometimes, the \iona{Mg}{ii} BELR is complex and
may not be virialized in these exceptional objects.

Four of the objects
with \iona{Mg}{ii}-based masses 
(J0945+1009, J1411+1402, J1417+0733, and
J1447$-$0203)
recently 
have had their masses estimated using the H$\beta$ line profiles
(\citealt{Plotkin2015}), 
and the masses are on average smaller by
a factor of $\approx3$. The H$\beta$ lines of WLQs are 
generally weak at a similar level to their \iona{Mg}{ii} lines
(\citealt{Plotkin2015}), but the H$\beta$ mass estimator
is considered the most reliable one
among all the single-epoch virial mass approaches 
\citep[e.g.,][and references therein]{Shen2013}.
The estimated Eddington ratios for these four objects 
plus J1521+5202 which also has an H$\beta$-based mass
(W11) range from 0.9 to 2.8 with a median value of 1.4 (a mean value of 1.7).\footnote{These Eddington ratios differ
from those in W11 and \citet{Plotkin2015} as the 
bolometric luminosities were estimated via different 
approaches.} The available SMBH mass and Eddington-ratio
estimates for our full sample are listed in Table~5.
{The uncertainties on these SMBH masses include measurement errors
($\approx0.15$~dex; \citealt{Shen2011}), systematic errors of 
the virial-mass approach ($\approx0.4$--0.5~dex; e.g., \citealt{Shen2013}),
and additional systematic errors (unknown but likely large) 
related to the likely unusual BELRs
of our extreme quasars.}
Given the substantial uncertainties associated with the 
SMBH masses and likely also the estimated bolometric luminosities (see 
Section~\ref{sec-sed}),
the estimated Eddington ratios do not provide strong constraints
on the accretion status of the PHL~1811 analogs and WLQs.

We also make basic
estimates of $L_{\rm Bol}/L_{\rm Edd}$
from the hard \hbox{X-ray} power-law photon indices obtained from
our joint spectral analyses (Section~\ref{sec-jointfit}), which have been demonstrated to be a
useful {\it direct} probe of the Eddington ratio
\citep[e.g.,][]{Shemmer2008,Risaliti2009,Brightman2013}.
Quasars with high Eddington ratios generally have very soft X-ray
photon indices. For example,
PHL~1811 has $\Gamma=2.3\pm0.1$ and $L_{\rm Bol}/L_{\rm Edd}\approx1.6$
\citep{Leighly2007}.
Based on the empirical $\Gamma\textrm{--}L_{\rm Bol}/L_{\rm Edd}$ relations
\citep{Shemmer2008,Risaliti2009,Brightman2013},
the best-fit hard \hbox{X-ray} photon index for our \hbox{X-ray} normal
objects, $\Gamma\approx2.2$,
corresponds to $L_{\rm Bol}/L_{\rm Edd}\approx1$. Therefore,
the joint spectral fitting results suggest high
Eddington ratios for our \hbox{X-ray} normal objects, or
our PHL~1811 analogs and WLQs in general considering that
the X-ray weak and \hbox{X-ray} normal objects
could be unified (W11; see also Section~6 below).
There is substantial object-to-object intrinsic scatter in the
$\Gamma\textrm{--}L_{\rm Bol}/L_{\rm Edd}$ relations.
Thus, by using the average $\Gamma$ from the joint spectral fitting
of 10--18 objects,
our constraint on the typical $L_{\rm Bol}/L_{\rm Edd}$
should be more reliable than that from the $\Gamma$ values
of a few individual
objects.

{In summary, the Eddington-ratio estimates derived from the 
virial SMBH masses
are highly uncertain and perhaps systematically
in error. We caution that individual measurements should
not be over-interpreted and, at best, one can look at the
group properties for some average indication.
The group properties from the H$\beta$-based virial masses
suggest the Eddington ratio may be high.
Meanwhile, the large hard \hbox{X-ray} photon index 
from the joint spectral fitting suggests a high
Eddington ratio for our objects as a group. We consider it
plausible that our PHL~1811 analogs and WLQs are accreting at high
or even super-Eddington rates.}

\subsection{Composite SDSS Spectra} \label{sec-compspec}

We constructed composite SDSS spectra for our PHL~1811 analogs and WLQs, 
to compare their average UV spectral properties to those of PHL~1811 and typical
quasars and also to make comparisons between the X-ray weak and X-ray normal
populations within our sample.
Median composite spectra were created following \citet{Vandenberk2001}.
Basically, each individual spectrum was corrected for Galactic extinction,
smoothed using a boxcar of width 10 pixels,
and normalized at rest-frame 2240~\AA.
The continuum at rest-frame 2240~\AA\ is not significantly contaminated
by \iona{Fe}{ii} or other lines \citep[e.g.,][]{Rafiee2011}.
The composite spectra were derived by computing the
median flux density at each wavelength where at least
five objects have spectral coverage.

\begin{figure*}
\centerline{
\includegraphics[scale=0.5]{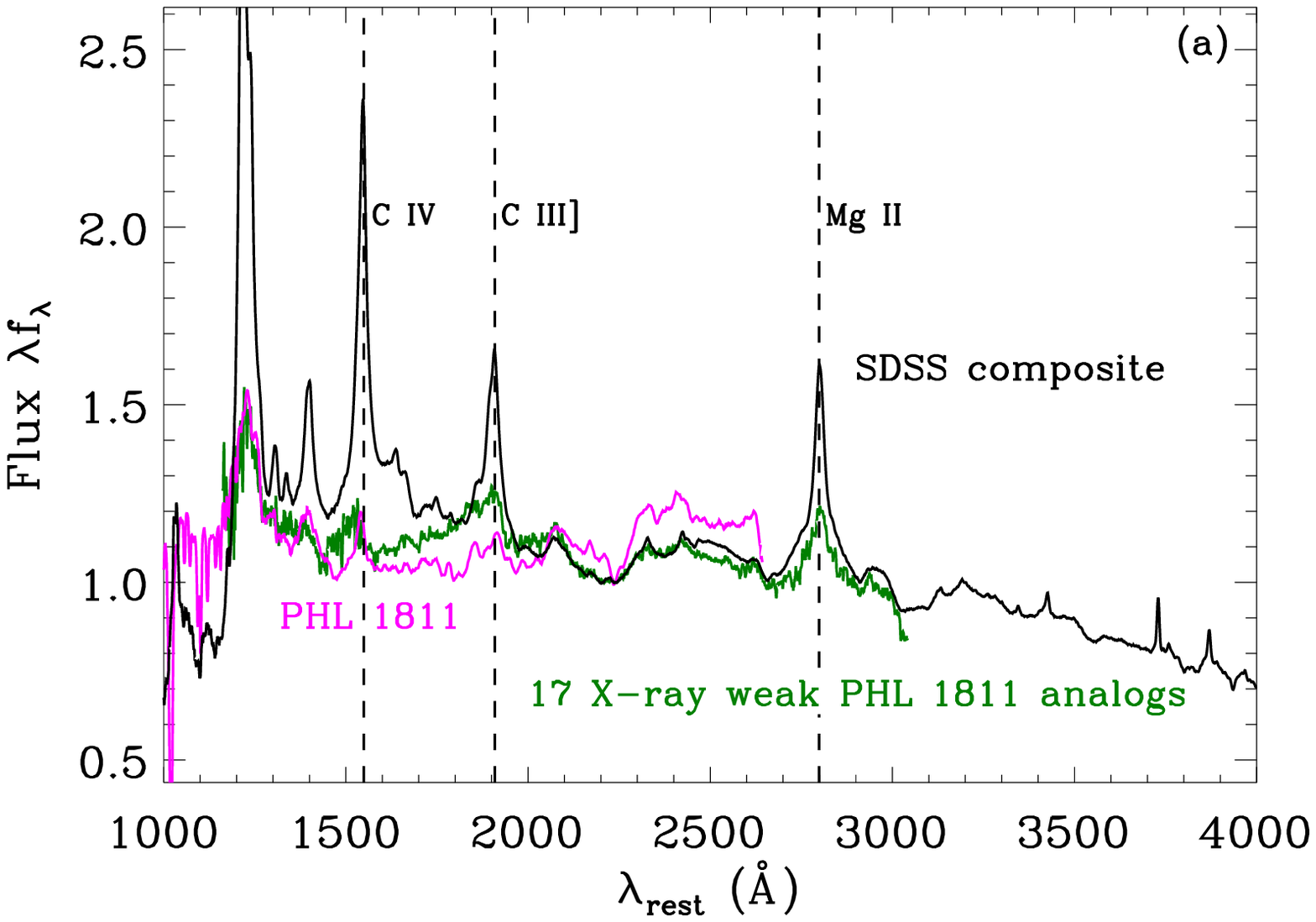}
\includegraphics[scale=0.5]{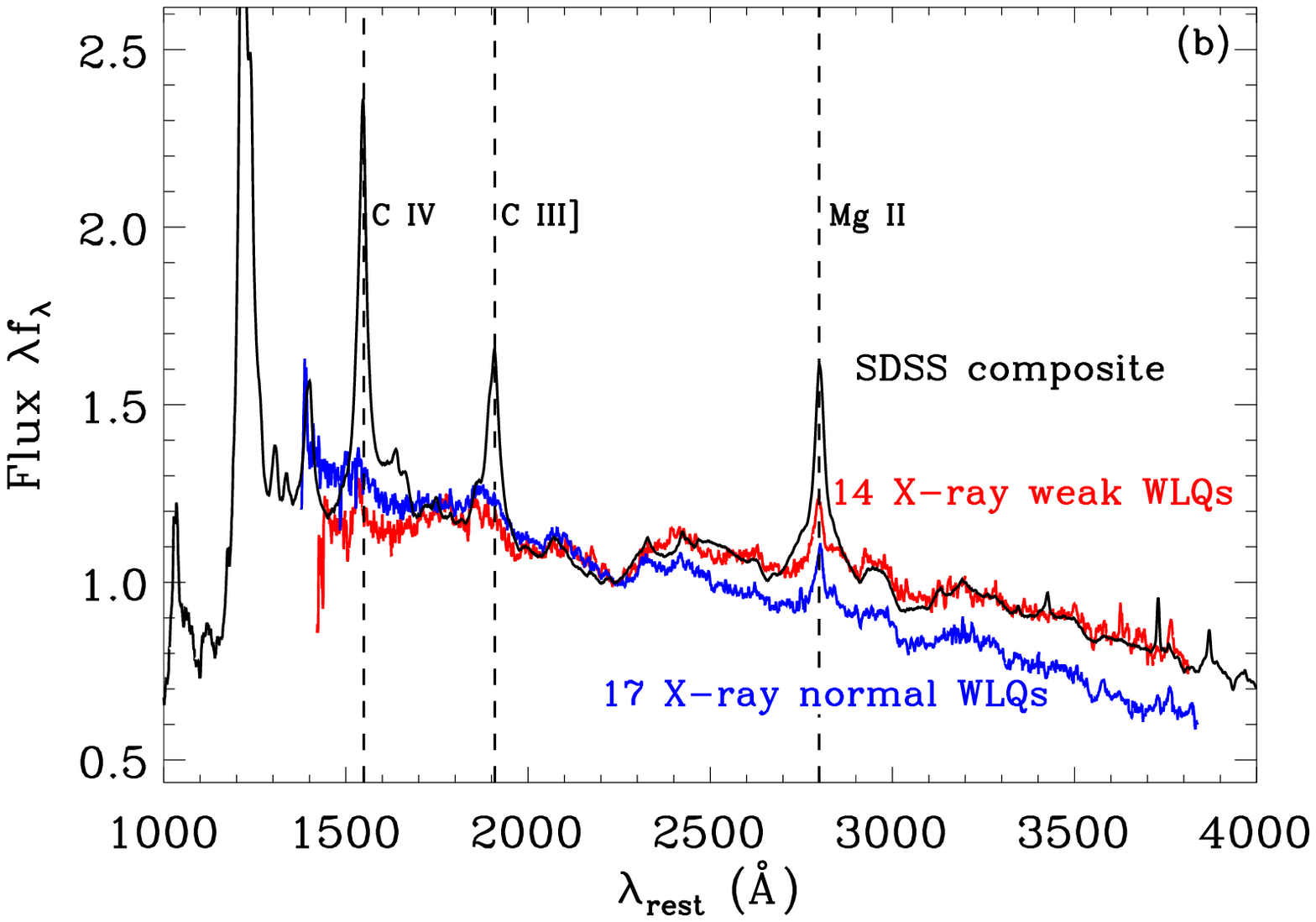}
}
\caption{
Comparison of the composite SDSS spectra for the
(a) 17 X-ray weak PHL~1811 analogs and
PHL 1811 and (b) 14
X-ray weak and 17 \hbox{X-ray} normal WLQs,
with the
$y$-axis being the flux in arbitrary linear units.
In both panels, the SDSS composite spectrum in \citet{Vandenberk2001}
is also shown for comparison.
PHL~1811 and its analogs have similar redder continua
than the SDSS composite spectrum, with
PHL~1811 having stronger UV \iona{Fe}{ii} (2250--2650~\AA) emission.
The X-ray weak WLQs have generally
redder continua than the SDSS composite spectrum,
while the X-ray normal WLQ composite spectrum is bluer and also
exhibits weaker UV \iona{Fe}{ii} emission.
}
\label{fig-compspecall}
\end{figure*}

The composite SDSS spectrum for the 17 X-ray weak PHL~1811 analogs is 
shown in Figure~\ref{fig-compspecall}a;
the \citet{Vandenberk2001} composite spectrum of SDSS quasars
and the {\it HST} spectrum of PHL~1811 \citep{Leighly2007b} are
included for comparison.
The weak emission lines, including \iona{C}{iv}, \iona{C}{iii}], 
and \iona{Mg}{ii},
are evident in PHL~1811 and its analogs. Meanwhile, 
the PHL~1811 analogs have similar \iona{Fe}{iii} UV48 $\lambda2080$
and UV \iona{Fe}{ii} (2250--2650~\AA) emission to the SDSS composite spectrum, 
while
PHL~1811 has even stronger UV \iona{Fe}{iii} and \iona{Fe}{ii} emission.
On average, PHL~1811 and its analogs have similar continua that are 
redder
than the SDSS composite spectrum between 
$\approx1500$--2700~\AA.\footnote{The PHL~1811 spectrum is bluer
than the SDSS composite spectrum at short wavelengths ($\la1300$~\AA; \citealt{Leighly2007b}) where most of the PHL~1811 analogs do not have spectral coverage.}
Using the SDSS filter information\footnote{\url{http://classic.sdss.org/dr1/instruments/imager/index.html\#filters}.} and the median redshift of $z=2.2$,
we measured a relative $g-i$ color (in the observed frame) 
of $\Delta(g-i)=0.24$
for the
PHL~1811 analog composite spectrum. 

\begin{figure*}
\centerline{
\includegraphics[scale=0.5]{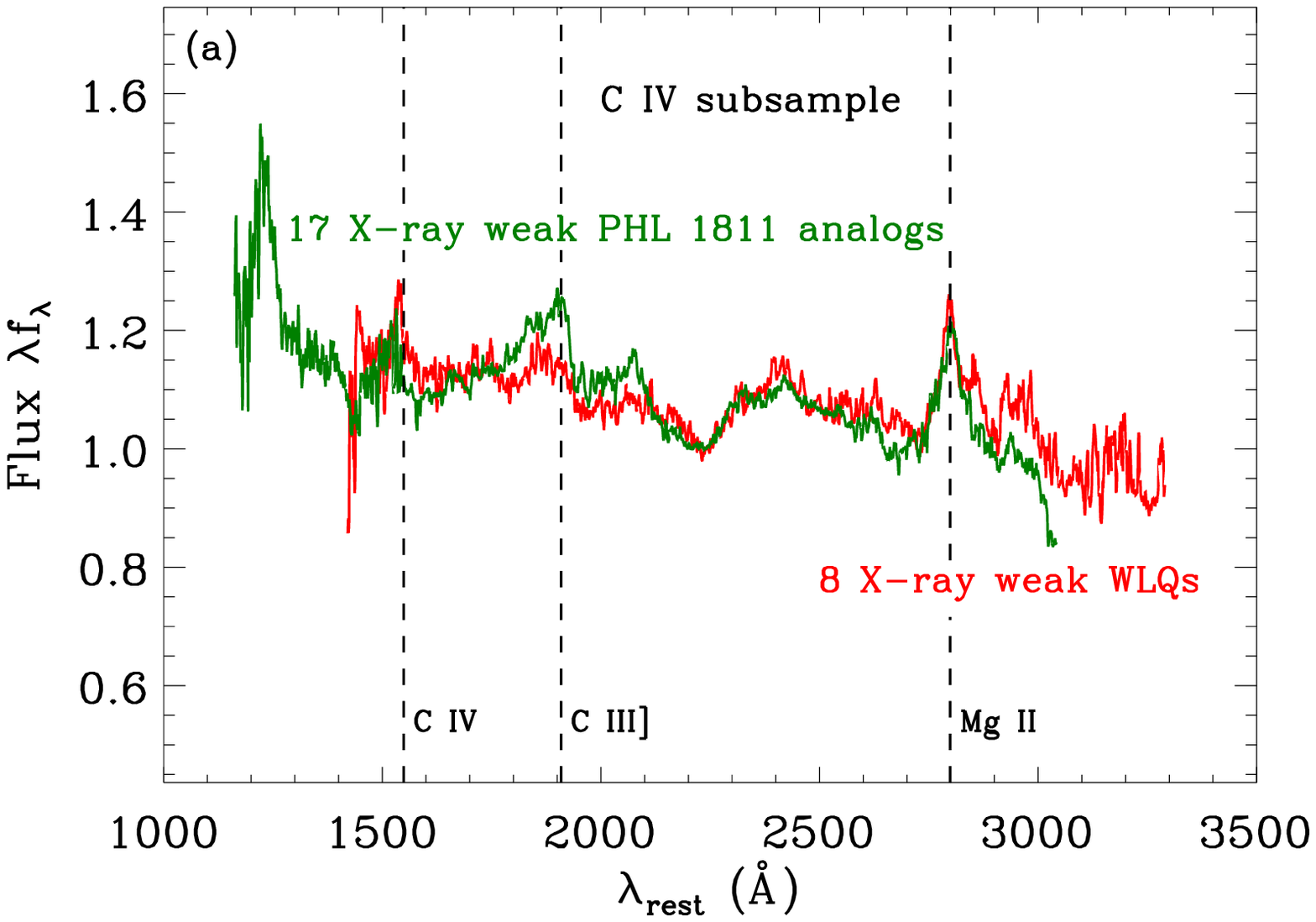}
\includegraphics[scale=0.5]{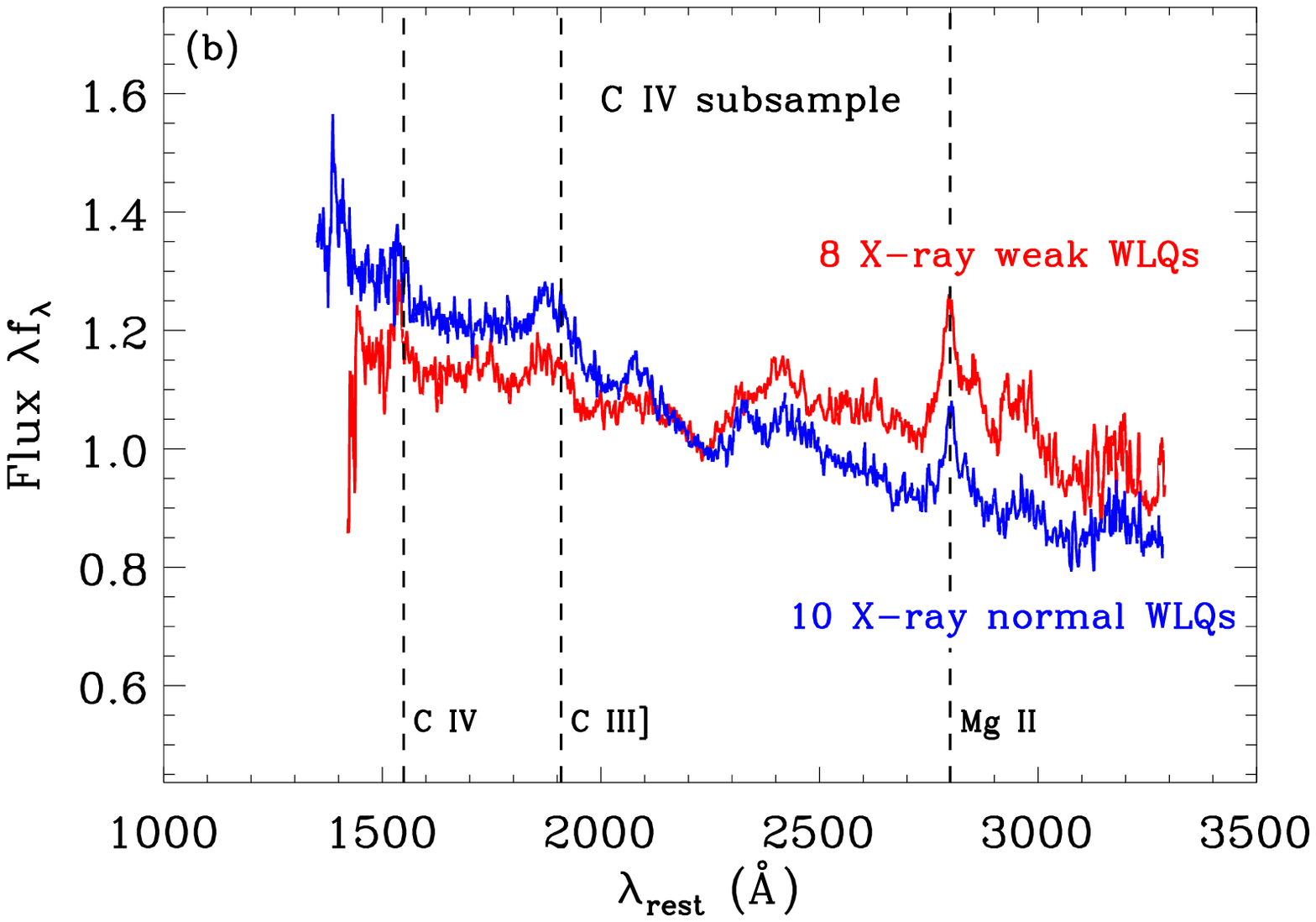}
}
\caption{
Comparison of the composite SDSS spectra for the
(a) 17 X-ray weak PHL~1811 analogs and
8 X-ray weak WLQs and (b) 8
X-ray weak and 10 \hbox{X-ray} normal WLQs in the \iona{C}{iv} subsample.
The composite spectra for the X-ray weak PHL~1811 analogs and
WLQs are similar to each other.
The X-ray weak WLQs have redder continua and stronger
UV \iona{Fe}{ii} (2250--2650~\AA) emission than the X-ray normal WLQs.
}
\label{fig-compspec}
\end{figure*}

To check the reliability of our approach for
creating composite spectra, we 
made a composite spectrum for the
7251 SDSS quasars satisfying
the redshift, magnitude,
radio-loudness, and non-BAL criteria of PHL~1811 analogs
(Section~\ref{sec-sel11}), and it agrees very well
with the \citet{Vandenberk2001} SDSS quasar
composite spectrum in the \hbox{1500--3500~\AA} range.
In addition, to check whether the redder color 
[$\Delta(g-i)=0.24$] of the 17 X-ray weak 
PHL~1811 analogs is an outcome of small sample bias,
we randomly
picked 17 SDSS quasars 
from the above sample and measured
the $\Delta(g-i)$ value of their composite spectrum.
This practice was repeated 10\,000 times and the 
resulting $\Delta(g-i)$ distribution has an rms scatter
of 0.037. Therefore, the redder color of the PHL~1811 analogs
is a highly significant (6.4$\sigma$) result.

The composite SDSS spectra for the 14
X-ray weak and 17 \hbox{X-ray} normal WLQs are shown in
Figure~\ref{fig-compspecall}b, with comparison 
to the SDSS quasar composite spectrum.
Similar to the PHL~1811 analogs, the X-ray weak WLQs have weak \iona{C}{iv}, \iona{C}{iii}], and \iona{Mg}{ii} emission lines but normal UV \iona{Fe}{iii}
and \iona{Fe}{ii} emission, compared to the SDSS composite spectrum.
The X-ray weak WLQs have generally
redder continua than the SDSS composite spectrum,
with $\Delta(g-i)=0.11$.
The \hbox{X-ray} normal WLQ composite spectrum is bluer
than the SDSS composite spectrum, with $\Delta(g-i)=-0.11$.\footnote{Recently,
\citet{Meusinger2014} found that their
sample of SDSS quasars with weak emission lines have generally bluer
continua than typical quasars. However, their sample was selected
differently from our WLQs, and their method to generate
the composite spectrum also differs from ours. Therefore, the
\citet{Meusinger2014} results cannot be directly compared to our
findings here.}
It also has weaker UV \iona{Fe}{ii} emission than the X-ray weak WLQ
or SDSS composite spectra
(more evident in Figures~\ref{fig-findavc4} and \ref{fig-fe2rew} below).

\begin{figure}
\centerline{
\includegraphics[scale=0.5]{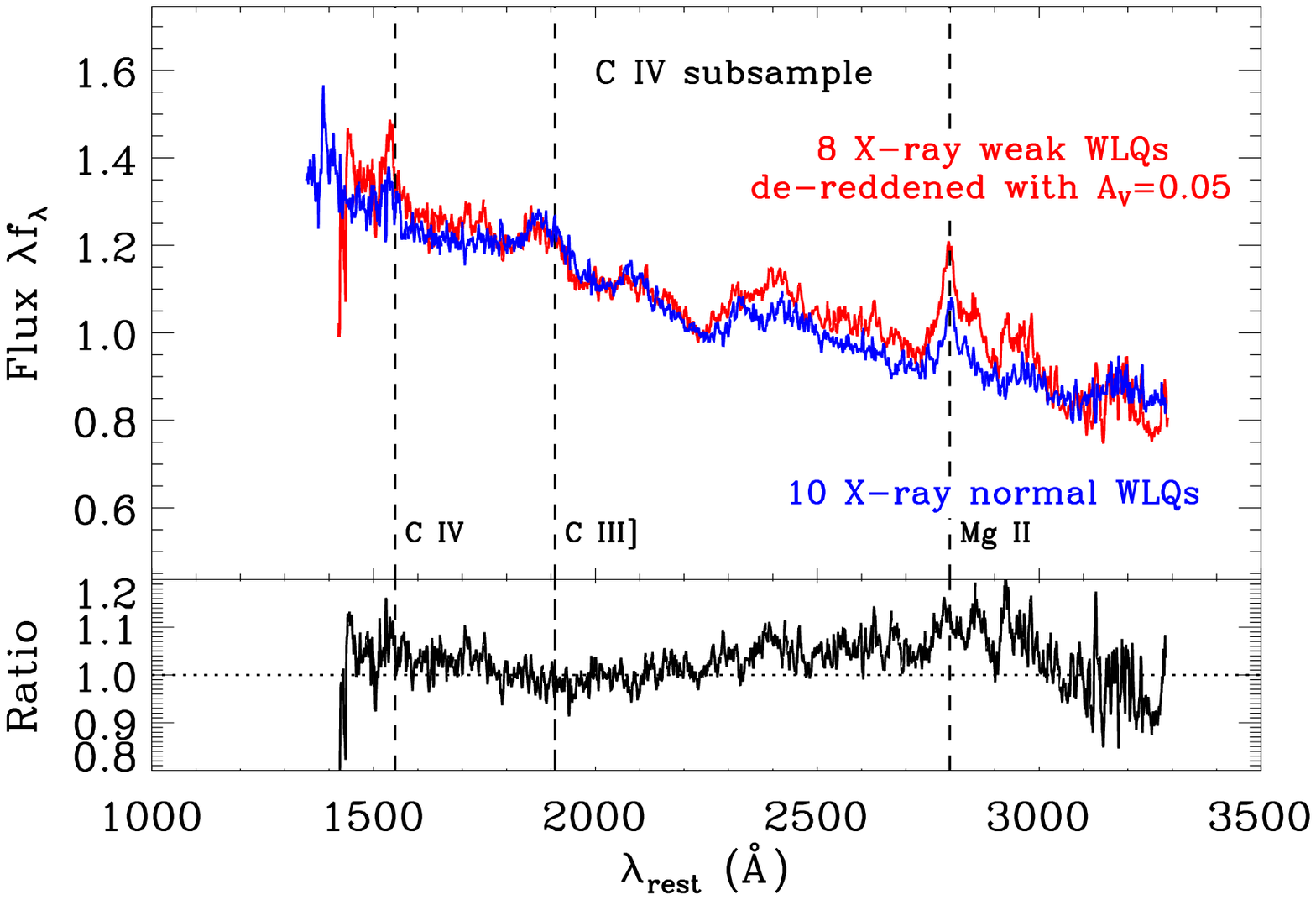}
}
\caption{
Composite SDSS spectrum for the eight X-ray weak WLQs de-reddened with
$A_{V}=0.05$~mag compared to the composite spectrum for the 10
\hbox{X-ray} normal WLQs in the \iona{C}{iv} subsample.
The bottom panel shows the ratio of the two spectra.
The de-reddening was performed assuming an
SMC extinction law.
The
de-reddened X-ray weak composite spectrum matches well with the
X-ray normal composite spectrum, except for some excess
UV \iona{Fe}{ii} and \iona{Mg}{ii} emission.
}
\label{fig-findavc4}
\end{figure}

Composite spectra were also made for the \iona{C}{iv} subsample.
In Figure~\ref{fig-compspec}, we compare the composite spectra for 
the X-ray weak PHL~1811 analogs, X-ray weak WLQs, and X-ray normal WLQs
in the \iona{C}{iv} subsample. These spectra are similar to 
the corresponding composite spectra for the full sample.
The composite spectra for the X-ray weak PHL~1811 analogs and
WLQs are similar to each other.\footnote{The SDSS spectrum of the
one X-ray normal PHL~1811 analog (J1537+2716) is more similar to the
spectra of X-ray normal WLQs; it is bluer and has weaker UV \iona{Fe}{ii}
emission than the composite spectrum of the X-ray weak WLQs (e.g., see Figure~\ref{fig-colorfe2} below).}
The X-ray weak WLQs have redder continua and stronger
UV \iona{Fe}{ii} (2250--2650~\AA) emission than the X-ray normal WLQs.

We de-reddened the composite spectrum for the X-ray weak WLQs with various
$A_V$ values to test whether the difference between the
X-ray weak and X-ray normal composite spectra can be explained by
dust extinction. A Small Magellanic Cloud
(SMC) extinction law (\citealt{Gordon2003}; $R_V=2.74$) was adopted which is
usually used to describe intrinsic quasar reddening 
\citep[e.g.,][]{Hopkins2004,Glikman2012}. As shown in Figure~\ref{fig-findavc4},
mild dust extinction of $A_V=0.05$~mag could describe well the 
redder composite spectrum for the \hbox{X-ray} weak WLQs, with the 
remaining residuals mainly in the UV \iona{Fe}{ii} and \iona{Mg}{ii} emission.
This small amount of dust reddening corresponds to a gas column density
of $N_{\rm H}<10^{20}$~cm$^{-2}$ assuming a Galactic gas-to-dust
ratio \citep[e.g.,][]{Guver2009}, and thus the extra
dust for the reddening 
cannot explain the observed X-ray weakness which requires 
$N_{\rm H}\approx10^{24}$~cm$^{-2}$ (Section~\ref{sec-aox}).
This simple testing of reddening cannot, of course,
exclude the possibility of an intrinsically redder spectrum for the 
\hbox{X-ray} weak WLQs.

\subsection{Spectral Diagnostics of X-ray Weak Quasars} \label{sec-weakdiag}

Studies of the composite SDSS spectra revealed distinct
characteristics of our X-ray weak PHL~1811 analogs and WLQs 
with respect to the X-ray normal
population, including redder continua and stronger UV \iona{Fe}{ii} emission.
We thus performed a statistical analysis of various
emission-line and continuum properties, searching for possible spectral
diagnostics of X-ray weak quasars that would help reveal their nature.

We ran Peto-Prentice tests to assess whether
the distributions of the optical--UV spectral properties
of the X-ray weak objects differ from those of
the X-ray normal objects. {We combined PHL~1811 analogs and WLQs 
in these tests considering that they can be unified 
(W11; see also Section~6 below).}
We examined the
\iona{C}{iv} REW, \iona{C}{iv} blueshift,
\iona{C}{iv} FWHM, REWs of the $\lambda1900$, \iona{Fe}{ii},
\iona{Fe}{iii}, and \iona{Mg}{ii} emission features, 
and also the relative
SDSS color $\Delta(g-i)$. 
The tests were performed for both the full sample and the \iona{C}{iv}
subsample; the results are shown in Table~6.

We found that statistically, the X-ray weak PHL~1811 analogs and WLQs
have larger \iona{Fe}{ii} REWs
and redder
$\Delta(g-i)$ colors than
the X-ray normal objects, both at the more than 
3$\sigma$ significance level (3.8$\sigma$ and 4.6$\sigma$). The 
significance levels of the differences drop slightly (to
2.9$\sigma$ and 3.7$\sigma$) for the \iona{C}{iv}
subsample, probably due to the smaller sample size. For the other spectral
properties in either the full sample or the \iona{C}{iv}
subsample, discrepancies are found at the 1.2--2.7$\sigma$ level:
the X-ray weak population has in general stronger \iona{C}{iv} blueshifts
(1.7$\sigma$),
larger \iona{C}{iv} FWHMs (2.2$\sigma$), 
and higher UV emission-line REWs (e.g., \iona{C}{iv} REW at 2.3$\sigma$) 
than 
the X-ray normal population.

\begin{figure*}
\centerline{
\includegraphics[scale=0.5]{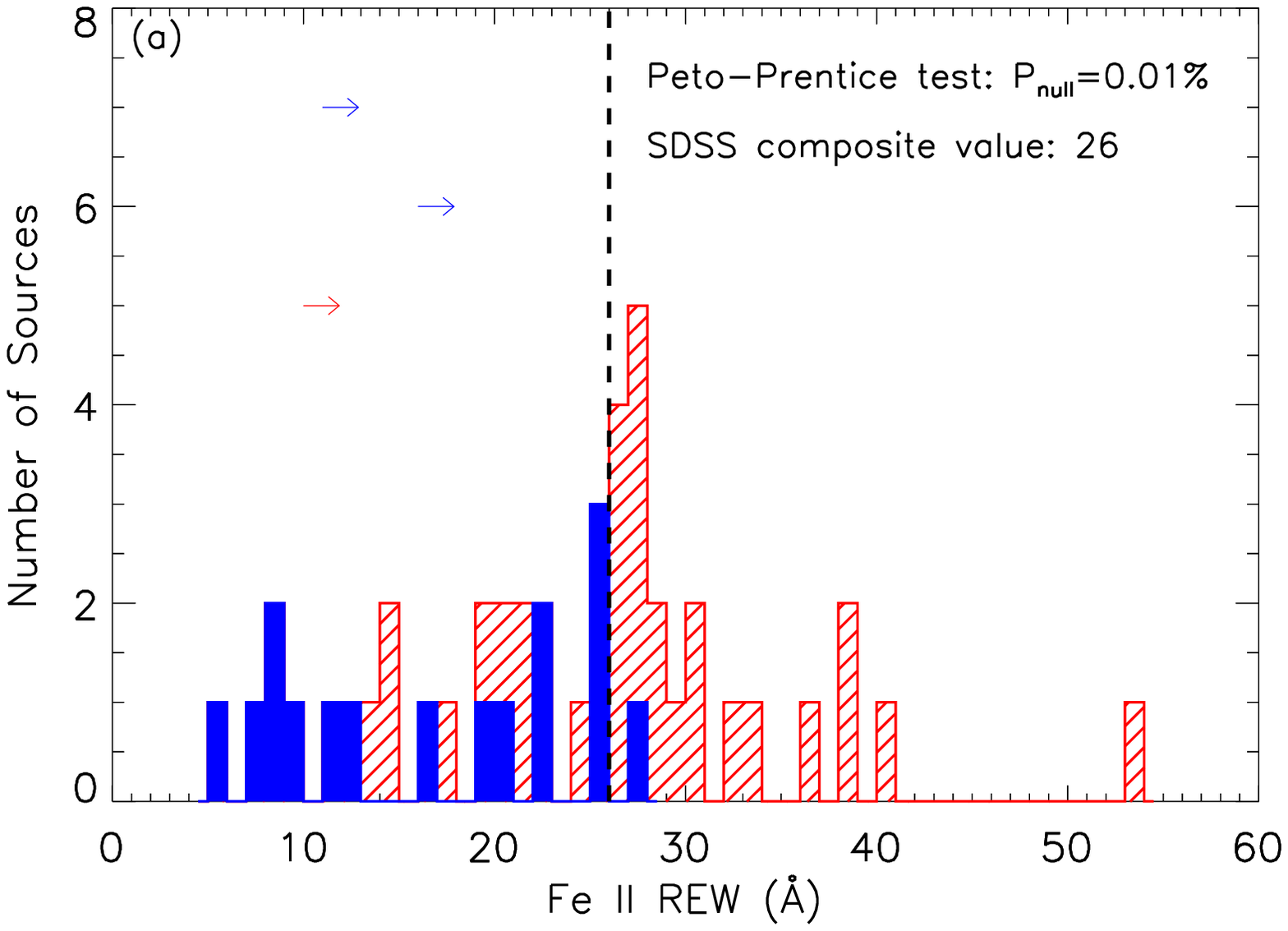}
\includegraphics[scale=0.5]{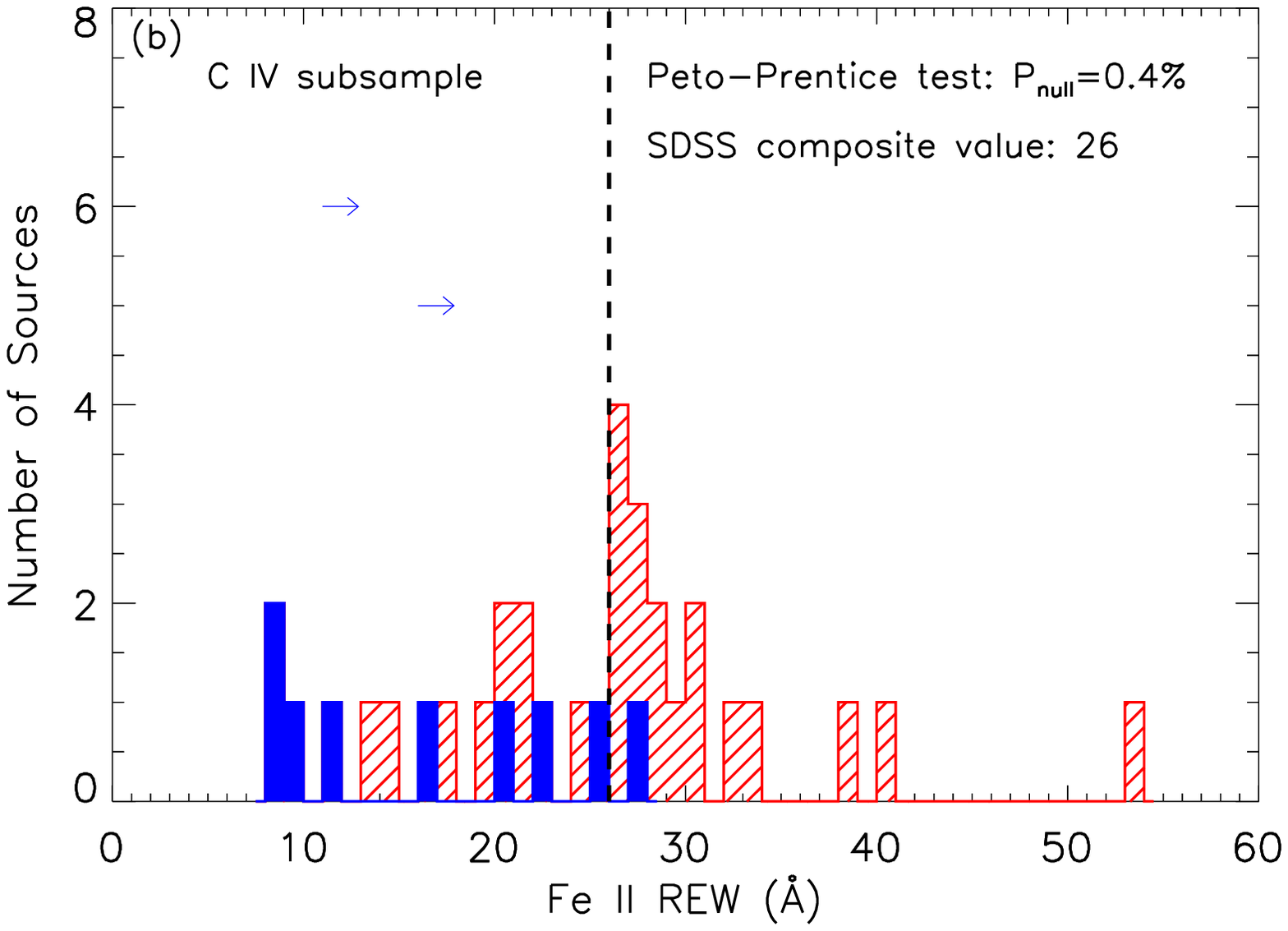}
}
\caption{
Distributions of the \iona{Fe}{ii} REWs
for the (a) full sample and (b) \iona{C}{iv} subsample.
The hatched red and solid
blue shaded histograms and arrows (representing limits) are for the
X-ray weak and X-ray normal objects, respectively.
We list the probability of the two distributions
being drawn from the same parent population based on the Peto-Prentice test.
We also list,
and plot as the vertical dashed line, the \iona{Fe}{ii}
REW measured from the SDSS composite spectrum in \citet{Vandenberk2001}.
The X-ray weak objects have on average
larger
\iona{Fe}{ii} REWs than the X-ray normal objects.
}
\label{fig-fe2rew}
\end{figure*}

\begin{figure}
\centerline{
\includegraphics[scale=0.5]{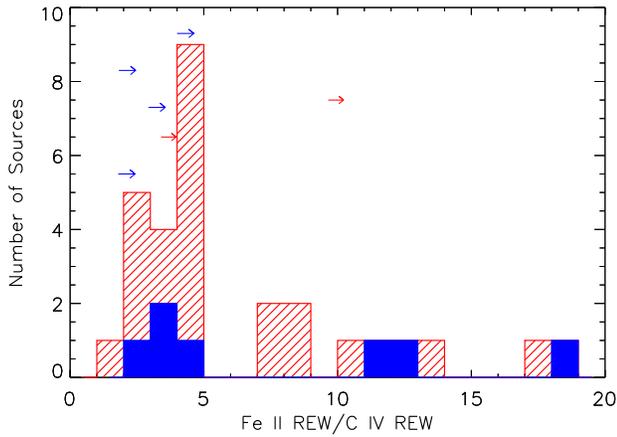}
}
\caption{
Distribution of the \iona{Fe}{ii} REW to \iona{C}{iv} REW ratios
for the X-ray weak (red) and X-ray normal (blue) objects.
The X-ray weak and normal objects
have similar distributions of these line ratios,
and the Peto-Prentice test also suggests no significant difference
($P_{\rm null}=0.5$).
}
\label{fig-fe2c4ratio}
\end{figure}

\begin{figure*}
\centerline{
\includegraphics[scale=0.5]{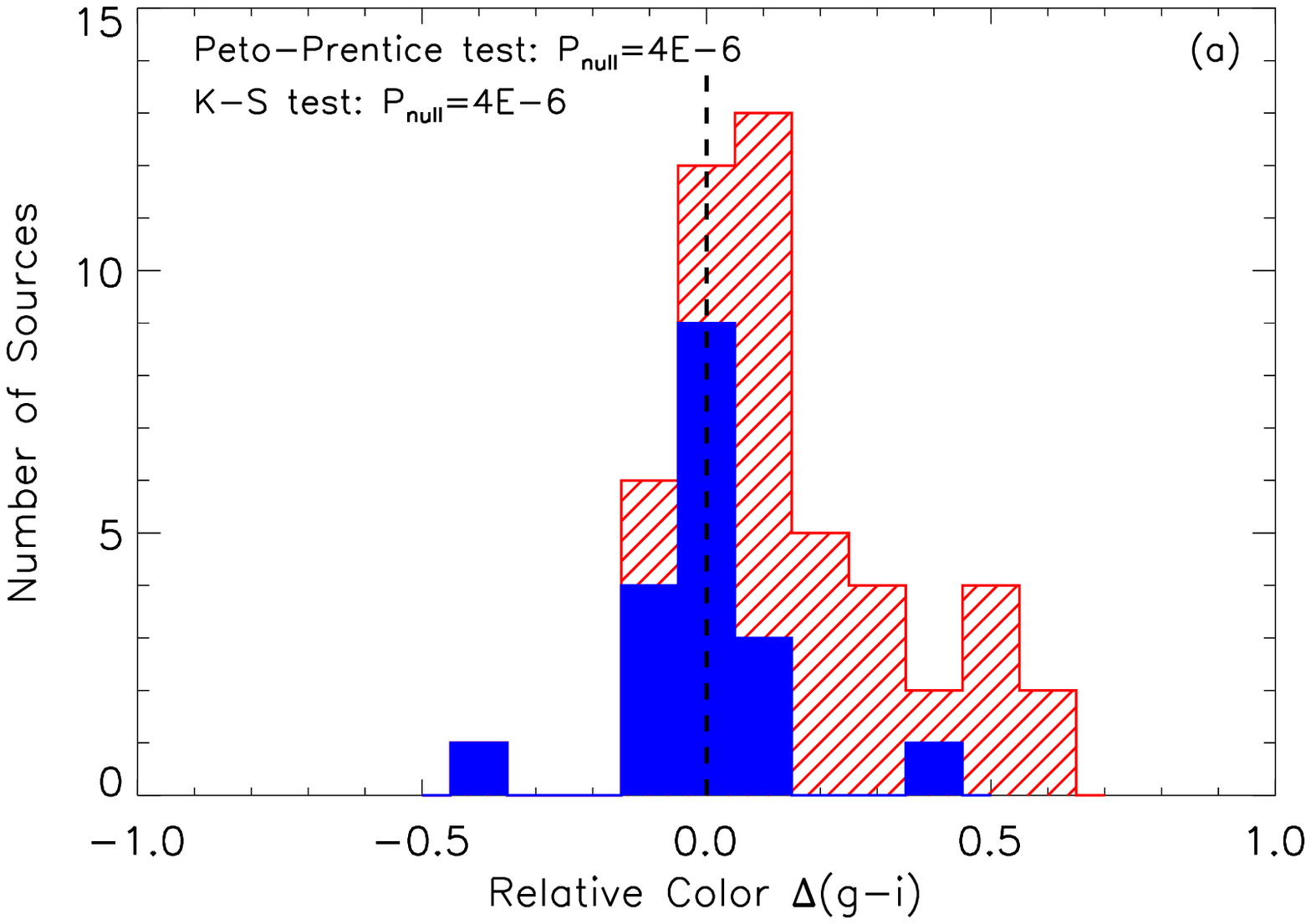}
\includegraphics[scale=0.5]{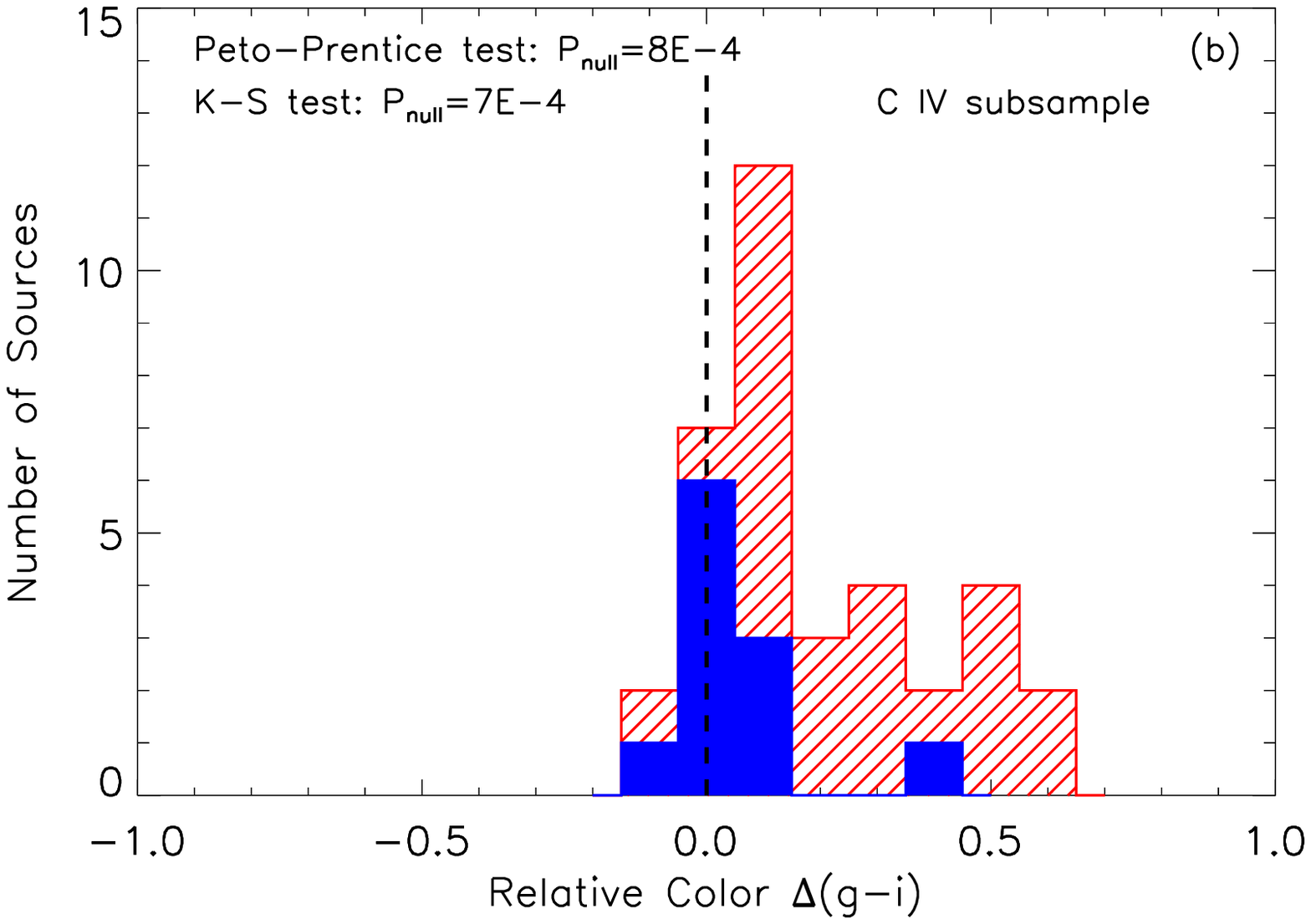}
}
\caption{
Distributions of the relative color, $\Delta(g-i)$, for
the (a) full sample and (b) \iona{C}{iv} subsample.
The hatched red and solid blue shaded histograms represent
X-ray weak and X-ray normal objects, respectively.
The vertical dashed line indicates a relative color of zero.
The probabilities of the two distributions
being drawn from the same parent population based on the Peto-Prentice test
and K-S test are listed.
The X-ray weak objects
are in general redder than the X-ray normal objects and typical SDSS quasars.
}
\label{fig-color}
\end{figure*}

We consider the UV \iona{Fe}{ii} REW and $\Delta(g-i)$ color as 
the most significant diagnostics of X-ray weak PHL~1811 analogs and WLQs. 
The distributions of the 
\iona{Fe}{ii} REWs for the \hbox{X-ray} weak and X-ray normal populations are 
shown in Figure~\ref{fig-fe2rew}, with the results from 
the Peto-Prentice test listed.
The \iona{Fe}{ii} REW measured 
from the SDSS quasar composite spectrum is also plotted for comparison.
The X-ray weak objects have in fact comparable UV \iona{Fe}{ii} emission
to typical SDSS quasars, while the X-ray normal objects have weaker 
than average \iona{Fe}{ii} emission. 
This result is consistent with the findings from the composite spectra 
(Section~\ref{sec-compspec}).
We also compared the distributions of the 
\iona{Fe}{ii} to \iona{C}{iv} REW ratio for the 
{24 X-ray weak and 11 X-ray normal objects with
both \iona{Fe}{ii} REW and \iona{C}{iv} REW constraints}
(Figure~\ref{fig-fe2c4ratio}), and no significant
difference was found (Peto-Prentice test $P_{\rm null}=0.5$).

The distributions of the $\Delta(g-i)$ colors are shown in Figure~\ref{fig-color}, with the results from
the Peto-Prentice test and Kolmogorov-Smirnov (K-S; applicable when
there are not censored data) test listed.
The median $\Delta(g-i)$ color for the X-ray weak population is 0.17,
and it is 0.01 for the X-ray normal population.
In the \iona{C}{iv}
subsample, these values are 0.23 and 0.03, respectively.\footnote{These 
median color
values differ slightly from those we measured from the composite spectra
(Section~\ref{sec-compspec}),
as the $g-i$ color of the composite spectra 
(converted to the observed frame using the 
median redshift) differs from the 
$g-i$ colors of individual objects with different redshifts.}
These color offsets are modest, and as pointed out 
in Section~2.4 of W11, the PHL~1811 analogs and WLQs
are still within the inclusion area for the SDSS color selection of quasars
\citep[e.g.,][]{Richards2002}.

\begin{figure*}
\centerline{
\includegraphics[scale=0.5]{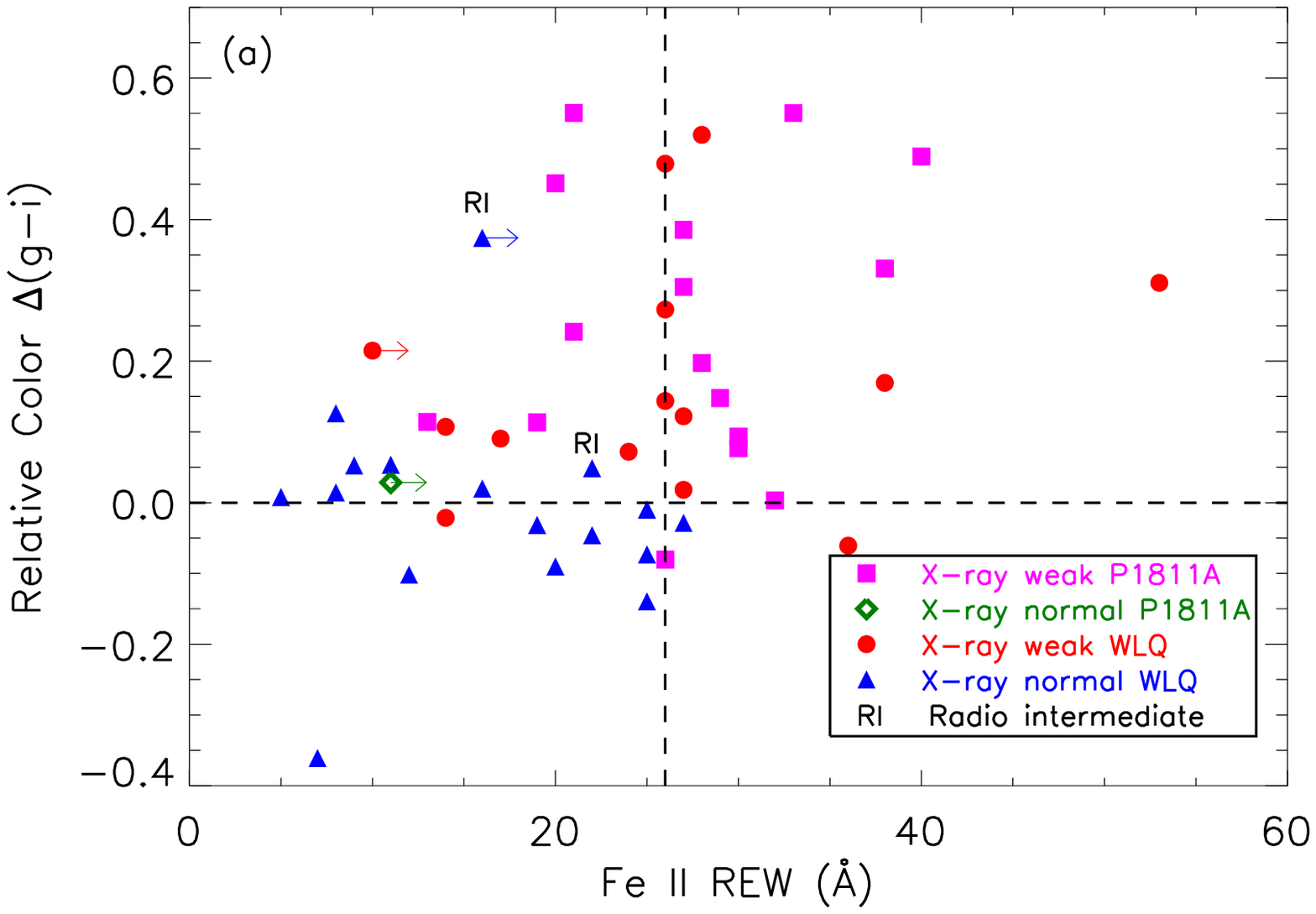}
\includegraphics[scale=0.5]{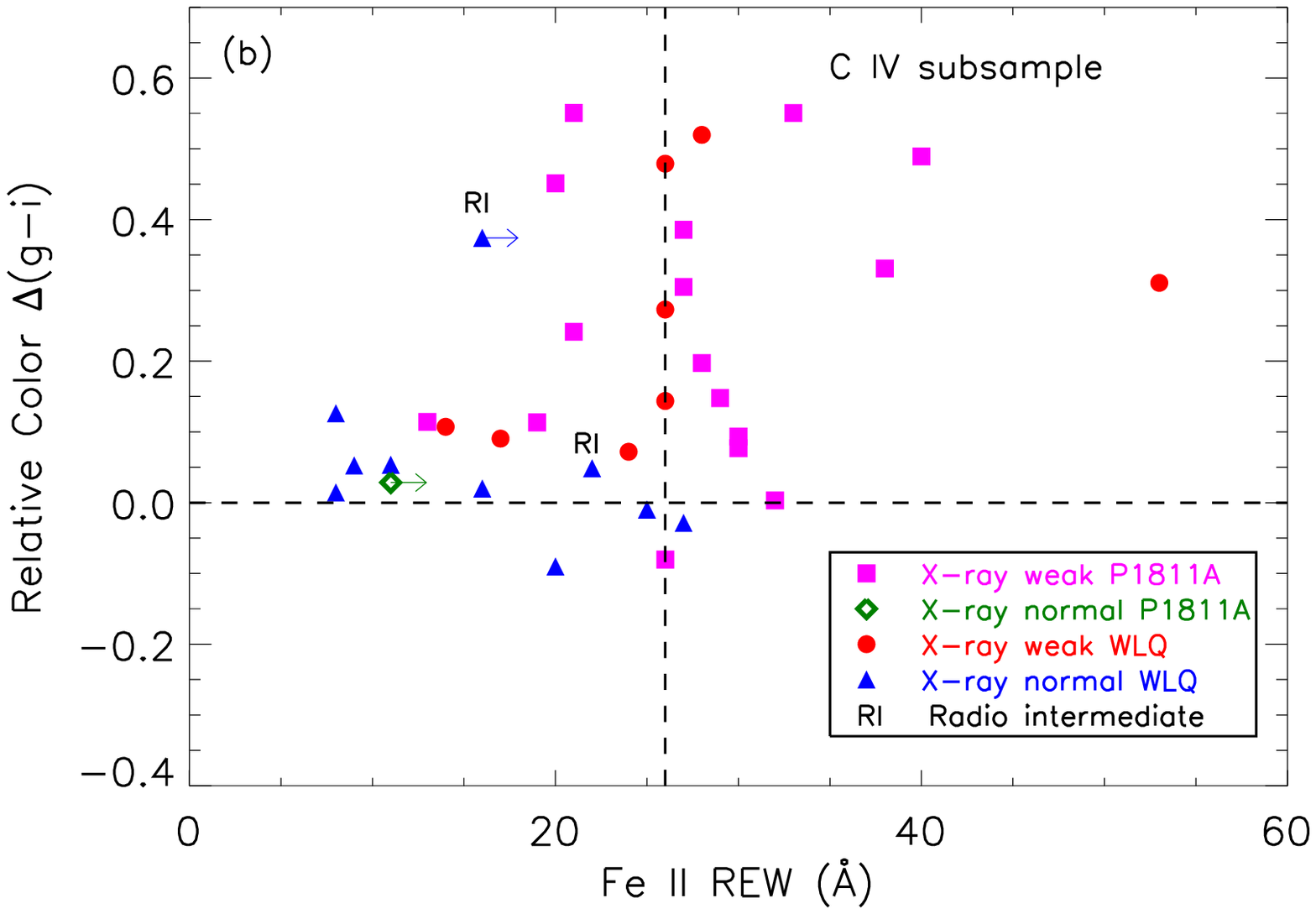}
}
\caption{
Relative color vs.\ \iona{Fe}{ii} REW for
the (a) full sample and (b) \iona{C}{iv} subsample, including the
X-ray weak PHL~1811
analogs (magenta squares), X-ray normal PHL~1811
analog (green diamond), X-ray weak WLQs (red circles), and
X-ray normal WLQs (blue triangles). The X-ray normal PHL~1811 analog
has an
estimated \iona{Fe}{ii} REW of
$13\pm2$~\AA\ based on the fractional coverage of \iona{Fe}{ii} (Table 2).
The dashed lines represent the median SDSS color and
the \iona{Fe}{ii} REW in the SDSS composite spectrum.
We also mark two objects, J0844+1245
and J1156+1848, that are considered
radio intermediate ($R>10$) in the \citet{Shen2011} SDSS quasar catalog using
a slightly different definition of $R$ from the one used here
(Section~\ref{sec-radio}).
The WLQs and PHL~1811
analogs that have redder colors and/or larger \iona{Fe}{ii} REWs
than the average SDSS values are more likely
to be X-ray weak quasars.
}
\label{fig-colorfe2}
\end{figure*}

Figure~\ref{fig-colorfe2}
displays how the X-ray weak and X-ray normal sample objects are separated
in the relative color vs.\ \iona{Fe}{ii} REW plot.
The normal X-ray emission from J1537+2716, 
the only X-ray normal PHL~1811 analog, can be explained 
by its small $\Delta(g-i)$ and \iona{Fe}{ii} REW (estimated to be
$13\pm2$~\AA\ based on the fractional coverage; Table 2) values.
There is also one obvious outlier in Figure~\ref{fig-colorfe2}, 
J0844+1245, that has an unusually 
red color, $\Delta(g-i)=0.37$, yet is X-ray normal.
This source has a relatively high radio loudness parameter and it is 
possible that its X-ray emission 
has some contribution from jet-linked emission
(see more details in Section~\ref{sec-radio}).

\begin{figure*}
\centerline{
\includegraphics[scale=0.5]{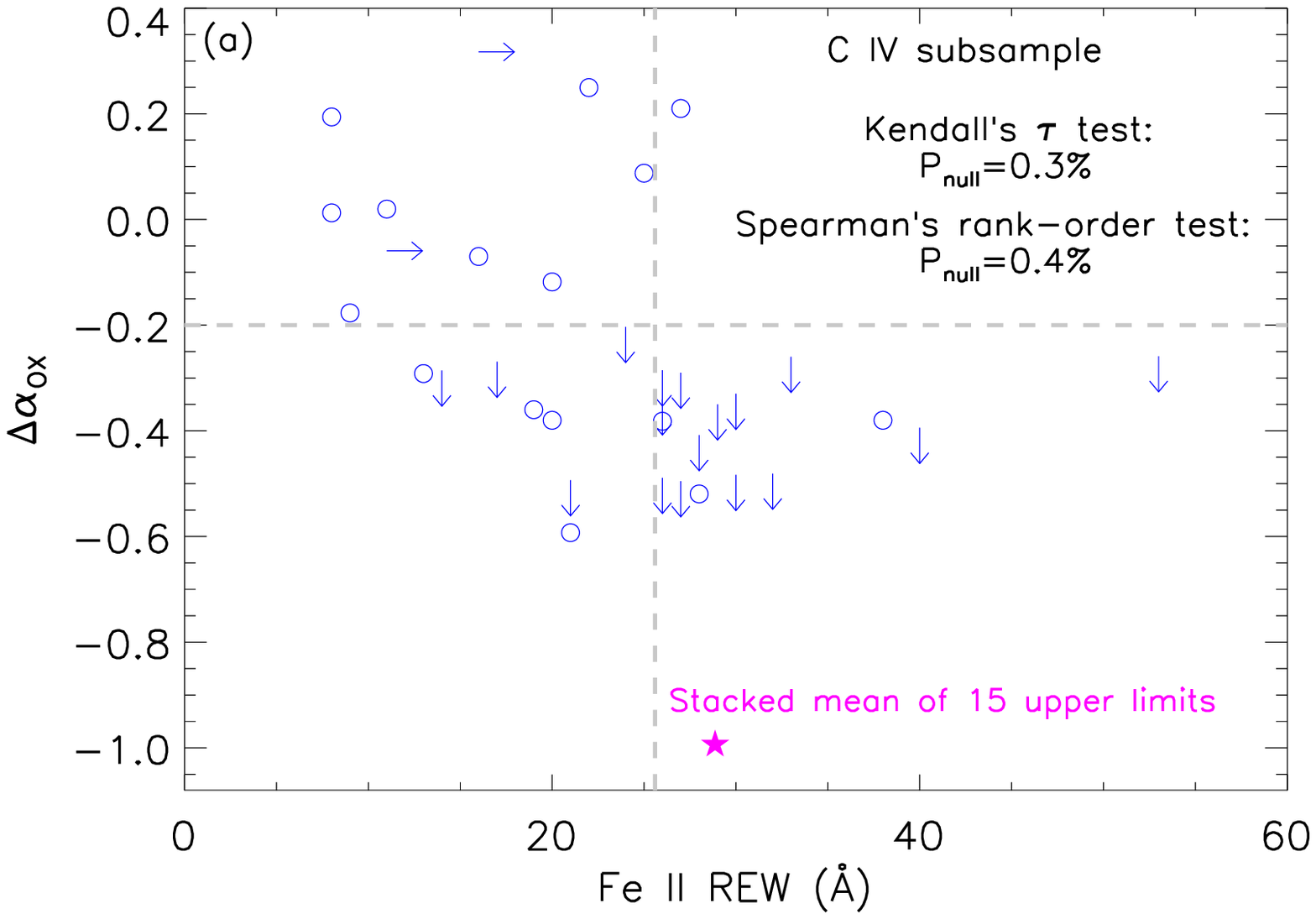}
\includegraphics[scale=0.5]{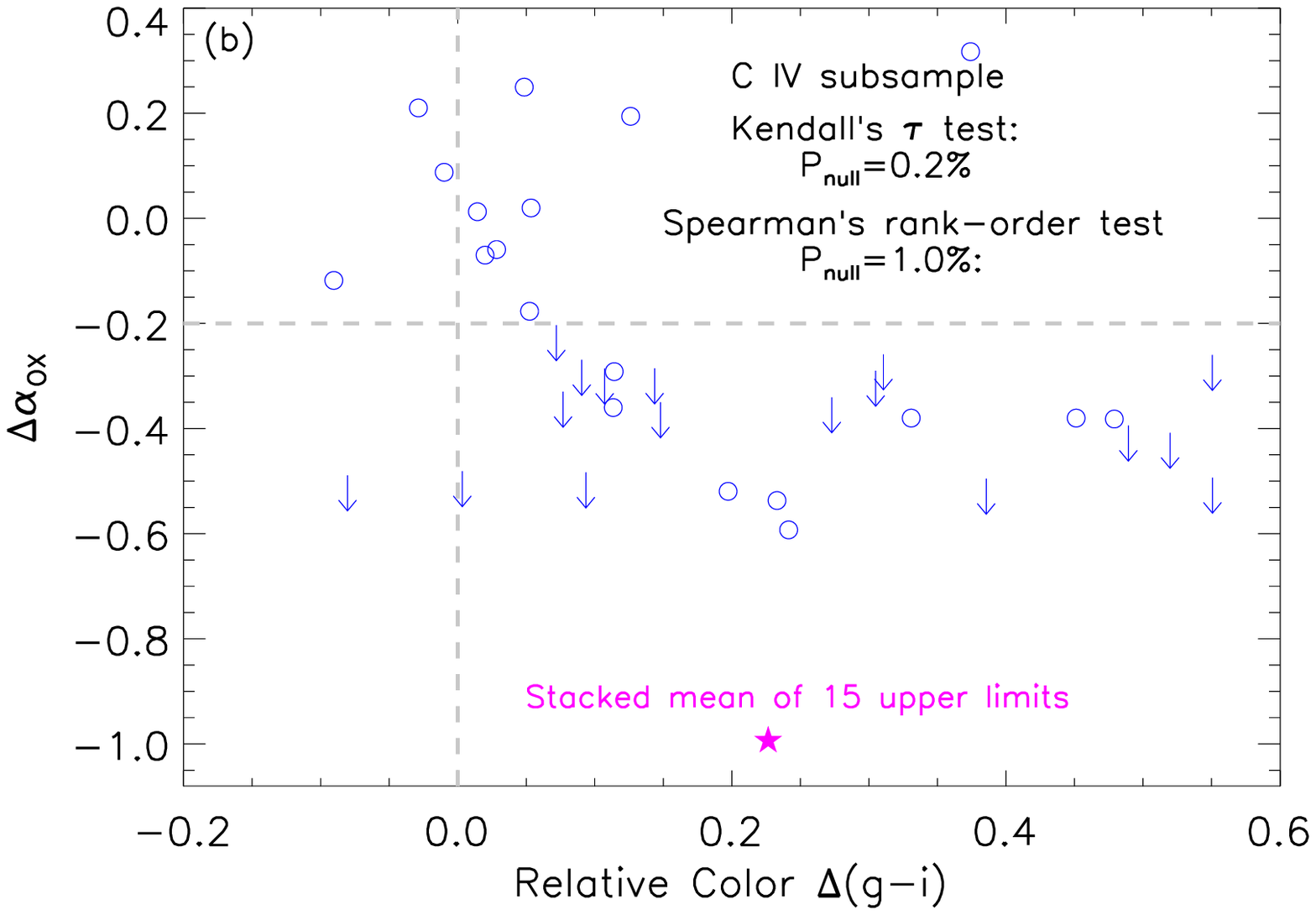}
}
\caption{
$\Delta\alpha_{\rm OX}$ vs.\ (a) \iona{Fe}{ii} REW and (b) relative color 
for the \iona{C}{iv} subsample.
The horizontal gray dashed lines represent the 
division between X-ray weak and X-ray normal quasars in this study.
The vertical gray dashed lines display the \iona{Fe}{ii} REW in the 
SDSS composite spectrum and the median SDSS color, respectively.
For the undetected 15 objects with \chandra\ data (the other two 
undetected sources have \xmm\ observations) in this figure, 
we show their stacked $\Delta\alpha_{\rm OX}$ value and the mean
\iona{Fe}{ii} REW (or relative color) as the magenta star.
We performed both the Kendall's $\tau$ and Spearman's rank-order
tests in the ASURV package for correlation analyses.
These tests work with data having upper and/or lower limits,
and the resulting small null-hypothesis probabilities (0.2\%--1\%)
suggest significant correlations between $\Delta\alpha_{\rm OX}$
and \iona{Fe}{ii} REW or relative color.
}
\label{fig-fe2colaox}
\end{figure*}

The X-ray weakness of our sample objects 
is measured by the $\Delta\alpha_{\rm OX}$ parameter.
Due to the significant fraction of sources not detected in the X-rays, 
it is not feasible to perform correlation analysis between
$\Delta\alpha_{\rm OX}$ and \iona{Fe}{ii} REW or $\Delta(g-i)$.
Instead, we utilized the Kendall's $\tau$ and Spearman's rank-order
tests in the ASURV package to check whether such a correlation exists. 
These tests were performed on the 
\iona{C}{iv}
subsample, and the distributions of the quantities are plotted
in Figure~\ref{fig-fe2colaox}.
The resulting small null-hypothesis probabilities (0.2\%--1\%)
suggest a likely correlation between $\Delta\alpha_{\rm OX}$
and \iona{Fe}{ii} REW or relative color, and the overall trend is such that a
larger \iona{Fe}{ii} REW or $\Delta(g-i)$ corresponds to a smaller
$\Delta\alpha_{\rm OX}$. 
We also show the stacked data points in Figure~\ref{fig-fe2colaox} 
for the 15 \chandra\ undetected sources (excluding the two \xmm\ undetected
sources), which have a mean $\Delta\alpha_{\rm OX}$ of $-0.99$
following the stacking procedure in Section~\ref{sec-stacking}.
Deeper X-ray observations, converting the $\Delta\alpha_{\rm OX}$ upper
limits into detections (factors of $\approx10$ increase in the exposure times
are needed given the stacked \hbox{X-ray} flux level), 
or a larger sample is required to 
quantify the possible correlations.

It is also of interest to probe the correlation between the spectral
properties and $\alpha_{\rm OX}$, which is an indicator of the \hbox{X-ray} 
flux level.
In the \iona{C}{iv}
subsample, the objects have a narrow distribution in 
$L_{\rm 2500~{\textup{\AA}}}$ (within about an order of magnitude) and thus
a narrow range of expected $\alpha_{\rm OX}$ values (within $\sim0.14$)
from the \citet{Just2007}
$\alpha_{\rm OX}$--$L_{\rm 2500~{\textup{\AA}}}$ relation.
Therefore, replacing $\Delta\alpha_{\rm OX}$ with $\alpha_{\rm OX}$
in Figure~\ref{fig-fe2colaox} does not change the distributions significantly,
and similar trends still exist in the sense that a
larger \iona{Fe}{ii} REW or $\Delta(g-i)$ corresponds to a smaller
$\alpha_{\rm OX}$.

\subsection{Selection Bias in the X-ray Weak Quasar Diagnostics?} \label{sec-bias}

We found that X-ray weak PHL~1811 analogs and WLQs generally have 
larger \iona{Fe}{ii} REWs, as well as larger \iona{C}{iv} blueshifts,
\iona{C}{iv} FWHMs, 
and UV line (\iona{C}{iv}, the $\lambda1900$ complex,
\iona{Fe}{iii}, and \iona{Mg}{ii}) REWs at a less significant level,
than the X-ray normal population. However, 
PHL~1811 analogs and WLQs were mixed to form a 
large sample for statistical analysis,
which could in fact introduce 
a selection effect due to the different selection criteria for these
two types of objects. As detailed in Section~2, the PHL~1811 analogs 
were selected with a less stringent requirement on the \iona{C}{iv} REW,
and no requirements on the REWs of the $\lambda1900$ complex and \iona{Mg}{ii}, but with extra requirements 
on the \iona{C}{iv} blueshift and 
UV \iona{Fe}{ii} and \iona{Fe}{iii} strength. The fraction of 
\hbox{X-ray} weak quasars is significantly higher among the PHL~1811 analogs
than the WLQs, and 
X-ray weak PHL~1811 analogs also constitute a substantial fraction ($>50\%$) 
of the X-ray weak population analyzed here. A consequence is that 
all the above criteria would stand out as spectral diagnostics of
X-ray weak quasars in our
analysis while only one or some of these might really be 
responsible for the weak X-ray emission. 

To avoid this possible selection bias, we performed the tests in
Section~\ref{sec-weakdiag} using only the WLQs in the \iona{C}{iv} 
subsample, which has
eight X-ray weak and 10 X-ray normal objects. Not all of
these objects have measurements of all the emission-line properties.
Only three properties in Table~6 differ at the more 
than 1.5$\sigma$ significance 
level between the X-ray weak and X-ray normal populations:
\iona{Fe}{ii} REW (2.0$\sigma$), \iona{Mg}{ii} REW (1.6$\sigma$), and
$\Delta(g-i)$ (2.7$\sigma$). The less-significant test results could either be
due to the selection bias being removed or simply the smaller sample size.
The latter effect is evident in the $\Delta(g-i)$ results (3.7$\sigma$ to
2.7$\sigma$), which are not affected by the selection bias.
Nevertheless, \iona{Fe}{ii} REW and $\Delta(g-i)$ remain
the most robust diagnostics of X-ray weak quasars.

It is also possible that the relatively 
larger REWs of the UV lines (e.g., \iona{C}{iv})
of the X-ray weak population are real instead of being a selection effect, 
if they are intrinsically connected with the relatively 
larger \iona{Fe}{ii} REW,
e.g., in 
the shielding-gas scenario discussed in Section~\ref{sec-unif} below.
With our currently limited sample of 
X-ray observed WLQs, it is difficult to assess whether there is a 
selection bias 
in the diagnostic analysis.

\section{DISCUSSION}

\subsection{Unification of PHL~1811 Analogs and WLQs with the Shielding-Gas Scenario}  \label{sec-shieldingscenario}


The overall similarities between the X-ray weak
PHL~1811 analogs and WLQs, including their IR--UV SEDs
(Figures~\ref{fig-sed1811} and \ref{fig-sedwlq}), composite SDSS spectra
(Figure~\ref{fig-compspec}), and degrees of
X-ray weakness (Figure~\ref{fig-aox}),
suggest that PHL~1811 analogs and WLQs are physically similar types of object.
PHL~1811 analogs are empirically a subset of WLQs (with small differences
caused by the differing selection approaches; Sections~\ref{sec-sel11}
and \ref{sec-selwlq}), and 
with the additional requirements of PHL~1811-like emission-line properties,
we preferentially selected X-ray weak WLQs 
as the PHL~1811 analogs.
Given our analysis of the X-ray weak quasar diagnostics in Section~\ref{sec-weakdiag}, the
criterion of strong UV \iona{Fe}{ii} emission most likely influenced
this outcome, while the criteria of large \iona{C}{iv} blueshift and strong
UV \iona{Fe}{iii} emission might also play a role.

Based on the properties of the small pilot sample of PHL~1811 analogs,
W11 proposed a simple shielding-gas scenario to unify PHL~1811 analogs
(X-ray weak WLQs) and \hbox{X-ray} normal WLQs. In this model,
the shielding gas
has a sufficiently high covering factor to shield
all or most of the BELR from the ionizing continuum, resulting
in the observed weak UV emission lines.
Whether we observe an \hbox{X-ray} 
weak or \hbox{X-ray} normal quasar is then an
orientation effect, depending upon whether our line of sight
intersects the X-ray absorbing shielding gas. X-ray normal WLQs are
observed at small inclination angles, while X-ray weak WLQs and PHL~1811
analogs are observed at
larger inclination angles.

With our extended sample of PHL~1811 analogs and WLQs, the spectral 
stacking results
suggest that the X-ray weak objects are generally absorbed 
instead of being intrinsically \hbox{X-ray} weak (Section~\ref{sec-stacking}).
The presence of the significant fraction of X-ray normal 
WLQs ($\approx50\%$) also indicates that the weak UV line emission
in PHL~1811 analogs and WLQs cannot universally be attributed to
intrinsic \hbox{X-ray} weakness.
The average X-ray weakness factor for our
\hbox{X-ray} weak PHL~1811 analogs or WLQs in the \iona{C}{iv} subsample
is $\approx40$,
corresponding to 
$N_{\rm H}\approx9\times10^{23}$~cm$^{-2}$ (Section~\ref{sec-aox}).
The spectral analysis of J1521+5202 also revealed strong X-ray absorption
with $N_{\rm H}=(1.26_{-0.46}^{+0.54})\times10^{23}$~cm$^{-2}$ 
(Section~\ref{sec-J1521}).
We note that the actual $N_{\rm H}$ values could be much larger if 
the observed \hbox{X-ray} emission is dominated by reflected/scattered \hbox{X-rays}
rather than direct transmission \citep[e.g.,][]{Murphy2009};
a substantial contribution from reflected/scattered X-rays is expected
for such levels of \hbox{X-ray} weakness based on observations of, e.g., 
local Seyfert 2 galaxies.
These findings provide further support for the 
W11 shielding-gas scenario, where 
a soft ionizing continuum due to small-scale absorption is the key 
for creating the BELR line properties of 
PHL~1811 analogs and WLQs.\footnote{
The \iona{He}{ii}~$\lambda1640$ line has been suggested to be
a good indicator of the
strength of the ionizing continuum, with a smaller REW implying
a weaker/softer continuum \citep[e.g.,][]{Leighly2004,Baskin2013}. 
We measured the \iona{He}{ii} REWs for our sample
objects the same way as in \citet{Baskin2013}, and the results do
show weaker than average \iona{He}{ii} REWs for our PHL~1811 analogs and 
WLQs.}

It has been suggested that the FWHMs of low-ionization lines
(e.g., H$\beta$, \iona{Mg}{ii}) have an orientation dependence due to a
flattened BELR geometry \citep[e.g.,][]{Wills1986,Runnoe2013,Shen2014},
with the FWHMs generally being larger at
larger inclination angles (although there is considerable 
object-to-object scatter). There are 23 X-ray weak objects and
17 X-ray normal objects in our sample having \iona{Mg}{ii} FWHM
measurements in the \citet{Shen2011} catalog.\footnote{Upon
visual inspection, we consider that 
it is necessary to subtract a narrow \iona{Mg}{ii}
line component for J1629+2532, and the 
FWHM value of the broad component was adopted.
For the other objects, the FWHM values of 
the whole \iona{Mg}{ii} profile are adopted.}
The X-ray weak objects
have relatively larger 
\iona{Mg}{ii} FWHMs
than the \hbox{X-ray} normal objects (mean  
FWHM values $5400\pm600$~km~s$^{-1}$ vs.\ 
$4100\pm600$~km~s$^{-1}$ and median values 
$5300$~km~s$^{-1}$ vs.\ 
$2900$~km~s$^{-1}$), consistent with the shielding-gas
scenario where the X-ray weak objects are viewed at larger 
inclination angles. However, a K-S test between the two
sets of \iona{Mg}{ii} FWHMs does not indicate a highly
significant difference, so we consider this result 
only suggestive at present.
 
One possible consequence
of the shielding-gas scenario is that the 
$\Delta\alpha_{\rm OX}$ distribution of PHL~1811 analogs and WLQs would
appear bimodal, at least to first approximation, 
as the sources are either heavily X-ray absorbed or X-ray normal.
Due to the large fraction of undetected sources in the 
\hbox{X-ray} weak population, the $\Delta\alpha_{\rm OX}$ distributions for 
our samples (Figure~\ref{fig-aoxhist}) cannot reveal clear bimodality.
However, the very large 
average X-ray weakness factors ($\ga100$; Section~\ref{sec-aox})
for these undetected sources do suggest that the $\Delta\alpha_{\rm OX}$
values cannot have a continuous distribution and bimodality is 
plausible. Deeper X-ray observations are required to constrain better
the $\Delta\alpha_{\rm OX}$ distribution.

Based on the X-ray and multiwavelength properties of PHL~1811 analogs and
WLQs, some basic physical requirements on the shielding gas can be obtained:
\begin{enumerate}

\item
The shielding gas
has a large column density of X-ray absorption (at least $10^{23}$~cm$^{-2}$
and likely much larger).

\item
It lacks accompanying \iona{C}{iv} BALs or mini-BALs at least 
along our line of sight.

\item
It lies closer to the SMBH than the BELR, so that it can screen
the nuclear EUV and \hbox{X-ray} emission from reaching the BELR.
Furthermore, it should have a large
covering factor to the BELR.

\item
The ``waste heat'' from the absorbed high-energy emission 
(that otherwise would have reached the BELR)
is likely 
re-emitted in the unobserved EUV as the continuum IR--UV SED appears normal.
Such re-emission in the EUV is indicative of a small-scale absorber on a scale 
of $\approx10R_{\rm s}$ 
($R_{\rm s}=2GM_{\rm BH}/c^2$ is the Schwarzschild radius).

\end{enumerate}
Given these properties, the shielding gas might naturally be understood
as 
a geometrically thick inner accretion disk (e.g., a slim
disk; see Section~\ref{sec-phy} below). This would require much of the
BELR gas to be in an equatorial configuration, as supported by
observations \citep[e.g.,][]{Shen2014}.

\subsection{A Geometrically Thick Disk Scenario for the X-ray Absorbing Shielding Gas} \label{sec-phy}

Although the X-ray weak and X-ray normal PHL~1811 analogs and WLQs can be
unified under the W11 shielding-gas scenario,
the physical 
nature of this shielding gas, that can produce such quasars with 
extreme emission-line and \hbox{X-ray} properties, remains mysterious.
Based on its physical requirements enumerated above 
(Section~\ref{sec-shieldingscenario}), we propose that a geometrically thick inner accretion disk may naturally serve as the shielding gas.
{We discuss below such a scenario in the context of a puffed-up
disk due to rapid or even super-Eddington accretion. However, we first note
that the basic idea of our model stands 
irrespective of the exact physical processes
leading to the geometrical thickness of the disk.}

Given the small fraction of PHL~1811 analogs and WLQs among SDSS quasars, 
a geometrically thick accretion disk
with a sufficiently large scale height to shield the BELR
almost fully
should be a rare phenomenon.
Under our selection criteria for the PHL~1811 analogs, 66 candidates
were identified from \hbox{$\approx7200$} SDSS quasars satisfying
the redshift, magnitude,
radio-loudness, and non-BAL requirements (Section~\ref{sec-sel11}).
The corresponding fraction is $\approx0.9\%$, consistent with the estimation
of $\la1.2\%$ in W11. The fraction of WLQs among SDSS quasars
is likely larger by a factor of $\approx2$ due to the addition of
the X-ray normal population. The rarity of these quasars suggests
a link to some extreme physical property.
One relevant physical
quantity is the Eddington ratio, as suggested by earlier
studies \citep[e.g.,][]{Leighly2007b,Leighly2007,Shemmer2010}.

PHL~1811 itself has an estimated Eddington ratio of \hbox{$\approx1.6$} 
\citep{Leighly2007}.
PHL~1811, along with several PHL~1811 analogs and WLQs with rest-frame
optical spectra (J1521+5202, 2QZ~J2154$-$3056, and the two high-redshift
WLQs in \citealt{Shemmer2010}), 
have weak or undetected [\iona{O}{iii}]~$\lambda5007$ narrow emission
lines, suggestive of high Eddington ratios \citep[e.g.,][]{Boroson1992,Shen2014}.
Moreover, several 
studies have found that as $L_{\rm Bol}/L_{\rm Edd}$ increases,
the \iona{C}{iv} REW generally decreases and the \iona{C}{iv} blueshift
also increases \citep[e.g.,][]{Bachev2004,Baskin2004,Richards2011,Shen2014,Sulentic2014,Shemmer2015}. 
In these correlations, there is only limited sampling in the 
super-Eddington or low ($<10$~\AA) \iona{C}{iv} REW regime, but 
the overall trends suggest that the Eddington ratio grows as one
moves from typical quasars 
toward WLQs in the \iona{C}{iv} REW vs.\ blueshift space (Figure~\ref{fig-vc4}a).
By the nature of our selection of the PHL~1811 analogs and WLQs, 
demanding that the 
BELR does not produce normal emission lines, we may have
recovered effectively a population of quasars with (extremely) high 
Eddington ratios.
It is difficult to measure $L_{\rm Bol}/L_{\rm Edd}$ directly for our 
exceptional objects, as the SMBH masses estimated from the line-based 
virial method are 
likely highly uncertain and perhaps systematically
in error (Section~\ref{sec-ledd}).
However, based on the empirical $\Gamma\textrm{--}L_{\rm Bol}/L_{\rm Edd}$ 
relations, our joint spectral
analysis of the \hbox{X-ray} normal subsample in Sections~\ref{sec-jointfit} 
and \ref{sec-ledd}
does indicate a high Eddington ratio ($L_{\rm Bol}/L_{\rm Edd}\approx1$) 
in general for our quasars.\footnote{{Narrow-line Seyfert 1
galaxies (NLS1s) also 
generally have steep X-ray spectra and high Eddington ratios.
PHL~1811 itself is considered a NLS1 \citep{Leighly2001}, and 
NIR spectroscopy of a limited sample of WLQs shows that they 
have in general narrow H$\beta$ and strong optical
\iona{Fe}{ii} emission (see Section 5.1 of
\citealt{Plotkin2015}), similar to NLS1s. However,
as the line widths and REWs have likely dependences on luminosity, 
our PHL~1811 analogs and WLQs are not directly comparable to 
local, less-luminous NLS1s.}}

\begin{figure}
\centerline{
\includegraphics[scale=0.35]{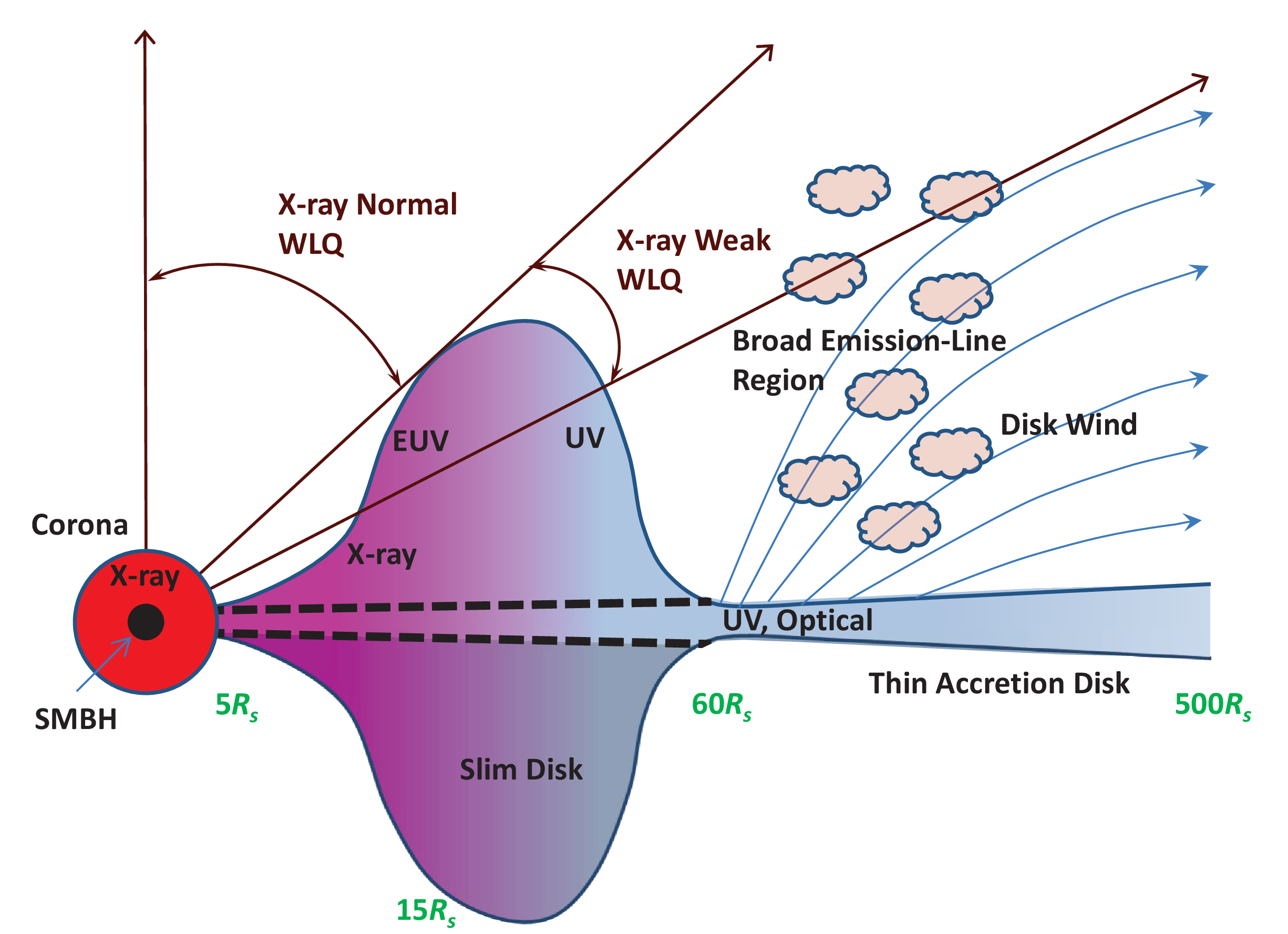}
}
\caption{A schematic diagram of the geometrically thick 
disk scenario for PHL~1811 analogs
and WLQs. 
When the accretion rate is very high, 
the inner region of the accretion disk is puffed up significantly,
which serves as the shielding gas in
the W11 scenario. It blocks the nuclear ionizing continuum from reaching 
the BELR, resulting in the observed weak line emission. The dashed
lines illustrate a standard thin accretion disk, where the BELR is 
illuminated normally. The inner region of the disk can also
absorb the coronal \hbox{X-ray} 
emission, resulting in an X-ray weak
PHL~1811 analog/WLQ when the disk inclination angle is moderate
or an X-ray normal
object when the inclination angle is small.
A standard accretion disk wind is launched from the outer region
of the disk. The BELR contains both a static component and a wind
component, at least for the \iona{C}{iv} line \citep[e.g.,][]{Richards2011}.
We annotate the approximate scale of the disk in several places,
in units of the Schwarzschild radius ($R_s$).
}
\label{fig-cartoon}
\end{figure}

A high Eddington ratio is naturally connected to our proposed
geometrically thick accretion disk.
When the Eddington ratio is high ($L_{\rm Bol}/L_{\rm Edd}\ga0.3$),
optically thick advection becomes important and
a slim accretion disk {\citep[e.g.,][and references therein]{Abramowicz1988,Mineshige2000,Wang2003,Ohsuga2011,Straub2011,Netzer2014,Wang2014b}}
is a more appropriate solution than the standard 
\citet{Shakura1973} thin accretion disk.
The slim disk has a geometrically
thick inner region which, in the case of PHL~1811 analogs and WLQs, 
might act as the shielding gas that blocks the
nuclear ionizing continuum from reaching
the BELR \citep[e.g.,][]{Leighly2004,Wang2014}.
{Recent three-dimensional global MHD simulations of super-Eddington
accretion disks with accurate and self-consistent radiative transfer
(\citealt{Jiang2014}; Y.-F. Jiang 2015,
private communication) or MHD simulations of super-Eddington
accretion disks accounting for general relativity and 
magnetic pressure
\citep{Sadowski2014}
also show this geometrically and optically thick inner region.}
An extremely high Eddington ratio is probably required for the 
inner disk to be puffed up sufficiently to reach the large necessary 
covering factor 
to the BELR. This thick 
inner disk might also block a significant portion
of the ionizing radiation from reaching the narrow line region, leading 
to the observed weak [\iona{O}{iii}] emission \citep[e.g.,][]{Boroson1992}.
A schematic illustration of this scenario is shown in Figure~\ref{fig-cartoon}.
{In order for the puffed-up disk to block the nuclear
high-energy emission from reaching the BELR, the disk corona must
be fairly compact. Studies of the rapid X-ray variability
\citep[e.g.,][]{Shemmer2014,Uttley2014} and 
X-ray microlensing 
\citep[e.g.,][]{Dai2010,Morgan2012} of AGNs
have constrained the sizes of 
the X-ray emitting regions to $\approx5R_{\rm s}$ in these
systems, consistent with the requirements of our model.}

The expected SED from a super-Eddington
accretion disk is uncertain.
In the optical--UV, it likely has a power-law shape similar to a
standard thin accretion disk, but in the FUV 
to soft \hbox{X-ray} range ($\approx10$--100~eV), it is
probably flatter \citep[e.g.,][]{Wang2003}. The nuclear high-energy 
emission absorbed
by the geometrically thick inner disk (the shielding gas)
will likely enhance its
EUV (e.g., $\approx50$~eV) emission, and thus the source will appear to have a typical quasar
continuum SED from the IR to UV (Section~\ref{sec-sed}).
The column density through the puffed-up inner accretion disk 
is high ($N_{\rm H}\gg10^{24}$~cm$^{-2}$), much larger than
the average $N_{\rm H}$ of $\approx9\times10^{23}$~cm$^{-2}$ roughly
estimated for our \hbox{X-ray} weak objects (Section~\ref{sec-shieldingscenario}).
Therefore, the observed X-ray emission from these objects is likely 
dominated by
indirect, reflected/scattered \hbox{X-rays} instead of 
transmitted emission through the disk. The nuclear X-ray emission 
could be Compton reflected by the accretion disk and/or scattered by an ionized 
scattering medium on a larger scale than the inner disk.
A broad range of \hbox{X-ray} weakness factors (e.g., Figure~\ref{fig-aoxhist})
might be expected
as the reflection/scattering efficiency varies for individual
objects, but the overall levels of \hbox{X-ray} weakness should 
be high for reflection/scattering dominated spectra so that bimodality
in the $\Delta\alpha_{\rm OX}$ distribution is still likely expected 
(see Section~\ref{sec-shieldingscenario}).
If the observed X-ray emission is dominated by reflection,
a strong
Fe~K$\alpha$ emission line with a REW of order 1--2~keV
is usually expected
\citep[e.g.,][]{Ghisellini1994,Matt1996}, which cannot be constrained with
our limited data.

If our line of sight is close to the edge of the inner bulge of 
the puffed-up disk in some objects, 
a small change in the covering factor of the inner disk
due to, e.g., accretion-disk instability
could result in transitions between \hbox{X-ray} weak and \hbox{X-ray} 
normal states on
timescales of years.
Such a phenomenon may have been
observed in the $z=0.396$ quasar PHL~1092. It has similar
emission-line properties to PHL~1811 (\citealt{Leighly2007b}; W11), and it 
varied between an X-ray normal quasar and an X-ray weak quasar  
over a timescale of years with a maximum variability factor of $\approx260$
in its 2~keV flux; meanwhile, its UV emission lines did not show such 
a drastic
change \citep{Miniutti2009,Miniutti2012} as would be expected if the 
covering-factor change is small. 
Further X-ray monitoring observations of PHL~1811 analogs 
and WLQs are required to
constrain the frequency and duration of such transitions, which 
may be a useful probe 
of the geometrically thick disk scenario.

\subsection{Broader Implications of Geometrically Thick Disks for 
Quasar Emission Lines}

Based on our geometrically thick disk scenario for PHL~1811 analogs and WLQs,
these extreme and rare quasars are likely not a
distinct population, but are instead extreme members of the continuous
population of quasars.
{The shielding effect from a puffed-up inner disk 
likely exists beyond these extreme objects, and it 
might be applicable at a milder level
to quasars with lower Eddington ratios; in either the slim-disk
model or the \citet{Jiang2014} simulations,
the disk is 
unlikely as thin and flat as what a standard disk model describes
as long as $L_{\rm Bol}/L_{\rm Edd}\ga0.3$.
When the Eddington ratio
is smaller than those of PHL~1811 analogs and WLQs,
the 
radius and scale height of the puffed-up disk decreases, 
and its covering factor to the BELR also decreases, leading to
a larger \iona{C}{iv} REW than those of PHL~1811 analogs and WLQs.} 

{Given a continuously varying covering factor
of the inner disk that is $L_{\rm Bol}/L_{\rm Edd}$ dependent,
a substantial fraction of the quasar \iona{C}{iv} REW distribution 
(spanning about two orders of magnitude; Figure~\ref{fig-vc4}a) 
could be governed by the Eddington ratio.
Additional factors shaping this distribution include
anisotropic continuum emission, anisotropic line emission, 
gas metallicity, \iona{C}{iv} BELR geometry 
(setting the fraction of ionizing radiation that is captured),
and potential
self-shadowing of the BELR \citep[e.g.,][]{Korista1998,Korista2004,Goad2012,Baskin2014,Wang2014}.}
Furthermore, shielding of the nuclear X-ray emission can prevent
the disk wind from being overionized and help the launching of the wind
\citep[e.g.,][]{Murray1995,Proga2000}. A larger covering factor
of the geometrically thick disk (larger $L_{\rm Bol}/L_{\rm Edd}$) could thus 
produce a stronger outflowing wind and result in a larger 
\iona{C}{iv} blueshift. The distribution of quasar \iona{C}{iv}
blueshifts (Figure~\ref{fig-vc4}a) could thus also be partly explained by 
the disk shielding scenario.

\subsection{X-ray Weak Quasar Diagnostics and the
Shielding-Gas Scenario} \label{sec-unif}

We investigated the X-ray weak quasar diagnostics discovered in 
Section~\ref{sec-weakdiag}, and found that they can be plausibly
fitted into the W11 shielding-gas scenario, where
the X-ray weak objects are viewed at larger inclination angles 
than the X-ray normal objects.

The redder $\Delta(g-i)$ color of the X-ray weak population
can be explained by mild excess intrinsic reddening ($A_V\approx0.05$~mag; 
Section~\ref{sec-compspec}) if
the dust tends to reside in the equatorial plane of the quasar (e.g.,
in a dust-driven outer wind scenario as discussed by \citealt{Elvis2012b})
which leads to more extinction at larger inclination angles.
Another possible scenario is that the accretion-disk emission is not
isotropic (e.g., from a puffed-up slim disk; see Section~\ref{sec-phy}),
and less UV emission is received at larger inclination angles, 
producing an
intrinsically redder continuum.

The larger REWs of the UV lines (\iona{C}{iv},
the $\lambda1900$ complex,
\iona{Fe}{iii}, \iona{Fe}{ii}, and \iona{Mg}{ii})
in the X-ray weak population compared to those in the X-ray normal population 
may be a consequence of the aspect dependent effects of accretion-disk
emission \citep[e.g.,][]{Netzer1990,Netzer2014,Wang2014}.
The continuum flux from a standard accretion disk scales approximately
as
$\cos i(1+2\cos i)$,
where $i$ is the inclination angle of the disk \citep{Netzer2014}.
{In the case of a slim disk, the disk continuum emission
is significantly anisotropic (e.g., see 
Figure~4 of \citealt{Wang2014}).}
The emission-line flux, on the other hand, is
presumably from a BELR structure which overall emits more
isotropically. The resulting line REW thus depends on the
inclination angle and may be
increased by a factor of a few moving from a small angle for an
X-ray normal WLQ to a moderate angle for an X-ray weak WLQ.
In this scenario, the ratios of the emission-line REWs
for the X-ray weak and \hbox{X-ray} normal populations
should not differ significantly. This interpretation
is supported by the distributions
of the \iona{Fe}{ii} to \iona{C}{iv} REW ratios for the two populations
in our sample (Figure~\ref{fig-fe2c4ratio}), which show no significant
difference (Peto-Prentice test $P_{\rm null}=0.5$).

The theoretical angular dependence of disk continuum emission
is slightly stronger toward
longer wavelength in the UV (e.g., Figure 32 of \citealt{Netzer1990}).
The \iona{Fe}{ii} emission is at a longer wavelength than the other
studied UV lines except \iona{Mg}{ii}, and it is also relatively
stronger than the other weak UV lines in WLQs
(leading to more reliable measurements).
These two factors combined
probably cause the \iona{Fe}{ii} REW to be the best X-ray weak quasar
diagnostic
among all the REWs in our analysis (\iona{C}{iv}, the $\lambda1900$ complex,
\iona{Fe}{iii}, \iona{Fe}{ii}, and \iona{Mg}{ii}).

The other two less-significant diagnostics of \hbox{X-ray weak} quasars,
the \iona{C}{iv} blueshift (1.7$\sigma$) and FWHM (2.2$\sigma$), 
can be generally incorporated in
the W11 scenario considering that the \iona{C}{iv} line has an
accelerating disk-wind origin so that the line may have both
a larger blueshift and a larger velocity broadening 
if our line of sight is more aligned with directions along which
the wind is strongly accelerated. A large \iona{C}{iv} blueshift was one of 
the selection criteria for our PHL~1811 analogs, but it appears 
less efficient than the requirement of large
UV \iona{Fe}{ii} REW for selecting X-ray weak
quasars; the orientation-angle dependence of the line-of-sight 
wind velocity is 
probably less significant than the orientation-angle dependence of the line 
REWs.

\section{SUMMARY AND FUTURE WORK}

\subsection{Summary of Main Results}

In this paper, we have presented X-ray and multiwavelength 
investigations of the nature of 
a large sample of PHL~1811 analogs and WLQs. 
The key points from this work are the following:

\begin{enumerate}
\item
We obtained \chandra\ exploratory (1.5--9.5~ks) observations 
of 10 PHL~1811 analogs and 22 WLQs. We also acquired a 40~ks \chandra\
observation of J1521+5202, a PHL~1811 analog that is 
one of the most luminous SDSS quasars. 
We measured their X-ray photometric properties, and performed
basic spectral analysis for J1521+5202 which suggests
strong intrinsic
X-ray absorption.
Including the previous samples in W11 and W12, we
constructed a large sample of 18 PHL~1811 analogs and 33 WLQs
at $z=0.5$--2.9 that have X-ray observations. See Sections~2 and 3.

\item
Out of the 18 PHL~1811 analogs, 17 (94\%)
are \hbox{X-ray} weak; 
out of the 33 WLQs, 16 (48\%) are \hbox{X-ray} weak.
The selection criteria of PHL~1811-like emission-line properties
worked effectively for finding X-ray weak quasars, and we can accommodate
the one X-ray normal PHL~1811 analog based on its SDSS spectral properties.
The average X-ray weakness factor 
for the
X-ray weak PHL~1811 analogs or WLQs in the \iona{C}{iv} subsample
is $\approx40$.
The $\Delta\alpha_{\rm OX}$ distributions of the
PHL~1811 analogs and WLQs are significantly different from that of
typical quasars. See Section~4.1.

\item
We constructed IR--X-ray continuum SEDs for the PHL~1811 analogs
and WLQs. 
Both the X-ray weak and X-ray normal groups
have IR--UV continuum SEDs similar to those of typical quasars.
See Section~4.2. 

\item
The stacked effective power-law 
photon indices for the X-ray weak subsamples
are relatively hard ($\Gamma_{\rm eff}\approx0.8$--1.5). These, together with the
strong X-ray absorption found in J1521+5202, suggest
that the X-ray weak
PHL~1811 analogs and WLQs on average are X-ray absorbed, although the
possibility of intrinsic X-ray weakness cannot be completely excluded
for some objects. See Section~5.1.

\item
We performed joint spectral fitting for the 18 X-ray normal
PHL~1811 analogs and WLQs. The best-fit hard X-ray photon indices are
$\Gamma=2.18\pm0.09$ for the 18 objects and $\Gamma=2.26\pm0.11$
for the 10 objects in the \iona{C}{iv} subsample, suggesting
a high Eddington ratio in general for these X-ray normal 
objects (and also the 
X-ray weak objects as they can be unified).
See Sections~5.2 and 5.3.

\item
We compared composite SDSS spectra for the \hbox{X-ray} weak and X-ray normal
PHL~1811 analogs and
WLQs, and investigated their
optical--UV spectral properties as diagnostics for identifying 
X-ray weak quasars.
Statistically, the X-ray weak PHL~1811 analogs and WLQs
have significantly ($>3\sigma$) larger UV \iona{Fe}{ii} REWs 
and redder $\Delta(g-i)$ colors than
the X-ray normal population. They also have in general 
larger \iona{C}{iv} blueshifts and FWHMs, and 
larger UV line (\iona{C}{iv}, the $\lambda1900$ complex,
\iona{Fe}{iii}, and \iona{Mg}{ii}) REWs 
at a less significant level, than
the X-ray normal population.
The normal \hbox{X-ray} emission from J1537+2716,
the only \hbox{X-ray} normal PHL~1811 analog, can be understood
given its small $\Delta(g-i)$ and \iona{Fe}{ii} REW.
See Sections~5.4--5.6.

\item
The PHL~1811 analogs empirically appear to be a subset of WLQs in
general, and these
two groups of objects can be unified under the W11 shielding-gas scenario.
Due to the additional requirements of PHL~1811-like emission-line properties,
we preferentially selected \hbox{X-ray} weak WLQs
as the PHL~1811 analogs.
The \hbox{X-ray} absorption found in J1521+5202 and our stacking results
provide further support for the
W11 shielding-gas scenario.
See Section~6.1.

\item
With the requirement of small \iona{C}{iv} REWs 
for PHL~1811 analogs and WLQs,
we may have
selected effectively a population of quasars with
geometrically thick inner accretion disks that can
block the ionizing continuum from reaching the 
BELR and act naturally as the shielding gas of the W11 scenario.
These quasars probably have unusually high
Eddington ratios so that the inner disk is significantly puffed up
(e.g., a slim disk). See Section~6.2.

\item
PHL~1811 analogs and WLQs are likely not a
distinct population, but are extreme members of the continuous
population of quasars.
Shielding of the BELR by a geometrically thick 
disk is thus perhaps generally applicable to
quasars with lower Eddington ratios, governing substantial fractions
of
the observed broad distributions of quasar \iona{C}{iv} REWs and blueshifts
shown in Figure~\ref{fig-vc4}a.
See Section~6.3.

\end{enumerate}

\subsection{Future Work}

Further investigations are needed to test the ideas
in this study. 
We suggested a geometrically thick disk scenario to explain the nature
of PHL~1811 analogs and WLQs and to unify these objects.
If these quasars are indeed systems with unusually high Eddington ratios,
we would expect 
a larger fraction of them at higher redshifts,
as the quasar Eddington ratio generally grows as redshift
increases \citep[e.g.,][]{Netzer2007,Shen2012b}. 
WLQs do appear to be more common at higher redshifts from
studies of limited WLQ samples at $z\approx3$--6 
\citep{Diamond2009,Banados2014}.
To test further this notion,
a large sample of WLQs selected systematically 
across a broad range of redshift is required.

However, there is no simple way to define universal selection criteria
for WLQs at all redshifts based solely on SDSS spectroscopy due to
the limited spectral coverage (e.g., Section~3.2 of W12).
Broader spectral coverage, such as NIR spectra for high-redshift
WLQs and UV spectra for low-redshift WLQs, are needed to study the 
correlations of weak emission lines and thus define consistent
selection criteria for WLQs at different redshifts. Currently, only a small
sample of WLQs have such broad spectral coverage (e.g., \citealt{Shemmer2010};
\citealt{Plotkin2015}), and NIR/UV
spectroscopic observations of more WLQs are needed. 
Moreover, given the unified nature of PHL~1811 analogs and WLQs, the 
WLQ selection criteria
(REW $\la5$~\AA\ for all emission features)
should be relaxed to
\iona{C}{iv} REW $\la10$~\AA, which will 
allow the selection of a WLQ sample
that is more consistent with the selection of PHL~1811 analogs.

If the distributions of quasar \iona{C}{iv} REWs and blueshifts
(Figure~\ref{fig-vc4}a) are indeed partly the result of
the varying covering factor of a
geometrically thick disk (which depends on the Eddington ratio) as we suggest in Section~6.3,
we would also expect to see more quasars in general having smaller
\iona{C}{iv} REWs and larger \iona{C}{iv} blueshifts 
at higher redshifts due to the redshift evolution of the quasar Eddington 
ratio. Caution should be applied when performing such analyses as the 
\iona{C}{iv} REWs and blueshifts might have other luminosity and/or
redshift dependences (e.g., the Baldwin effect).

{The most-luminous quasar at $z>6$ discovered recently is
also a WLQ \citep{Wu2015}, consistent with our expectation that the WLQ fraction
rises with redshift. It probably has a high or even super-Eddington
accretion rate, as we propose for WLQs (Section~6); a luminosity
exceeding the Eddington limit may help explain its very high luminosity.
This quasar has a \iona{Mg}{ii}-based virial mass of
$\approx1.2\times10^{10}~M_{\sun}$ \citep{Wu2015}. However, as we
discussed in Section~5.3, virial-mass estimates for these extreme quasars
are highly uncertain and perhaps systematically
in error (e.g., \iona{Mg}{ii}-based virial masses
are on average $\approx3$ times larger than H$\beta$-based virial masses
for five of our objects); this factor could alleviate the challenge
of growing such a massive black hole in the early universe.
A \chandra\ observation has been scheduled for this WLQ (PI: X.~Fan).
Our X-ray results on WLQs suggest there is a $\approx50\%$
chance that this quasar is X-ray weak.
This chance is likely higher based on this object's 
relatively red continuum, with a spectral index of $\alpha_\lambda = -1.43$ 
\citep{Wu2015} as compared to the average spectral index 
of $\alpha_\lambda=-1.72$ for SDSS quasars (e.g., Section~3.2 
of \citealt{Krawczyk2014}).
Quantification that its UV \iona{Fe}{ii} emission is strong 
would increase this chance further (see Section~5.5).}

Due to the substantial uncertainties in the SMBH mass estimates, it is 
challenging to measure quantitatively the Eddington ratios of PHL~1811 analogs
and WLQs. However, the Eddington ratio appears to be the primary driver of 
quasar Eigenvector 1 \citep[e.g.,][]{Boroson1992,Netzer2007,Shen2014}, which
is dominated by the optical
\iona{Fe}{ii}/H$\beta$ ratio, [\iona{O}{iii}] REW, and
H$\beta$ FWHM. NIR spectroscopy of a large sample of WLQs could provide 
measurements of 
these emission-line properties in the rest-frame optical. 
If they indeed have very high Eddington
ratios, clustering is expected toward
high \iona{Fe}{ii}/H$\beta$, low [\iona{O}{iii}]
REW, and low H$\beta$ FWHM, relative to quasars with comparable luminosities; there are already hints of such clustering
based on a small sample of five WLQs (see Section 5.1 of
\citealt{Plotkin2015}).
{Moreover, the \citet{Jiang2014} and \citet{Sadowski2014} 
simulations of super-Eddington
accretion disks predict a radiation-driven outflow along the rotation axis.
Signatures of this outflow may be sought in our X-ray normal WLQs 
as high-ionization absorption lines/edges at X-ray and UV 
wavelengths (e.g., \iona{O}{vi} absorption, as discussed in Section~4.4 of W11).
}

To quantify the possible correlations between
$\Delta\alpha_{\rm OX}$
and \iona{Fe}{ii} REW or $\Delta(g-i)$ (Figure~\ref{fig-fe2colaox}) and
probe the underlying physics, deeper X-ray observations of the undetected
objects are required. Given the stacked \hbox{X-ray} flux level,
most of the undetected sources should be detectable
with factors of $\approx10$ increase in the exposure times.
Such observations will also help test whether the 
$\Delta\alpha_{\rm OX}$ distribution is bimodal, as might
be expected in the 
shielding-gas scenario (Section~\ref{sec-shieldingscenario}).
Moreover, long-term X-ray monitoring observations of the PHL~1811 analogs
and WLQs will be useful, as they can constrain
the frequency and duration of the X-ray
state transitions as in PHL~1092, and thus
provide insights into the geometrically thick disk scenario.

Finally, it remains somewhat perplexing that at least many of our
PHL~1811 analogs appear to be X-ray absorbed, while PHL~1811 itself
appears intrinsically X-ray weak; Occam's razor would initially 
favor a single explanation for the \hbox{X-ray} weakness of all these
objects selected to have similar UV emission-line properties. It is 
worth noting that our PHL~1811 analogs are generally being observed in the 
$\approx1.5$--24~keV rest-frame band, while the X-ray properties 
of PHL~1811 itself have only been effectively probed up to $\approx8$~keV.
Perhaps PHL~1811 itself has a highly absorbed component that has yet
to be recognized. Although PHL~1811 has not been detected in the 
{\it Swift}-BAT all-sky survey in the 14--24~keV band (M.~Koss 2014,
private communication), a \nustar\ observation could probe more 
sensitively for a highly absorbed component. 

~\\

We thank M.~Eracleous,
Y.-F.~Jiang, K.~Korista, H.~Netzer, A.~E.~Scott, and J.~M.~Wang 
for helpful discussions.
We thank the referee for carefully
reviewing the manuscript and providing helpful comments.
We acknowledge financial support from
Chandra X-ray Center grant GO3-14100X (B.L., W.N.B.),
NASA ADP grant NNX10AC99G (B.L., W.N.B.), ACIS team contract
SV4-74018 (B.L., W.N.B.), the V.M. Willaman Endowment (B.L., W.N.B.), 
NSERC (P.B.H.), and Smithsonian Astrophysics Observatory contract SV2-82024
(G.P.G.).

Funding for the SDSS and SDSS-II has been provided by the Alfred 
P. Sloan Foundation, the Participating Institutions, the National 
Science Foundation, the U.S.\ Department of Energy, the National 
Aeronautics and Space Administration, the Japanese Monbukagakusho, 
the Max Planck Society, and the Higher Education Funding Council 
for England. The SDSS Web Site is \url{http://www.sdss.org/}.

\begin{deluxetable}{lccccccccc}


\tablewidth{0pt}
\tablecaption{New \chandra\ Observations and X-ray Photometric Properties}
\tablehead{
\colhead{Object Name}                   &
\colhead{Redshift}                  &
\colhead{Observation}                  &
\colhead{Observation}                  &
\colhead{Exposure}                   &
\colhead{Soft Band}                   &
\colhead{Hard Band}                   &
\colhead{Band}                   &
\colhead{$\Gamma_{\rm eff}$}                   &
\colhead{Comment}                   \\
\colhead{(J2000)}   &
\colhead{}   &
\colhead{ID}   &
\colhead{Start Date}   &
\colhead{Time (ks)}   &
\colhead{(0.5--2~keV)}   &
\colhead{(2--8~keV)}   &
\colhead{Ratio}                   &
\colhead{}                   &
\colhead{}   \\
\colhead{(1)}         &
\colhead{(2)}         &
\colhead{(3)}         &
\colhead{(4)}         &
\colhead{(5)}         &
\colhead{(6)}         &
\colhead{(7)}         &
\colhead{(8)}         &
\colhead{(9)}         &
\colhead{(10)}  \\
\noalign{\smallskip}\hline\noalign{\smallskip}
\multicolumn{10}{c}{PHL 1811 Analogs} 
}

\startdata
$       014733.58+000323.2$&$ 2.031$&$  14949$&    2013 Nov 15&$ 6.2$&$                       <2.4$&$          3.2_{-1.8}^{+3.2}$&$                   >1.19$&$                    <0.6$&                      ...\\
$       082508.75+115536.3$&$ 1.998$&$  14951$&    2013 Jun 10&$ 5.1$&$                       <2.4$&$                       <2.5$&$                     ...$&$                     ...$&                      ...\\
$       090809.13+444138.8$&$ 1.742$&$  14950$&    2013 Sep 23&$ 6.8$&$                       <2.4$&$                       <2.5$&$                     ...$&$                     ...$&                      ...\\
$       094808.39+161414.1$&$ 1.805$&$  14947$&    2013 Jun 10&$ 3.7$&$          2.1_{-1.3}^{+2.7}$&$                       <2.5$&$                   <1.42$&$                    >0.4$&                      ...\\
\vspace{2mm}
$       113342.67+114206.2$&$ 2.052$&$  14952$&    2014 Feb 08&$ 9.3$&$                       <2.4$&$          2.1_{-1.4}^{+2.9}$&$                   >0.70$&$                    <1.0$&                      ...\\
$       143525.31+400112.2$&$ 2.267$&$  14954$&    2013 Sep 04&$ 6.9$&$          7.3_{-2.7}^{+3.9}$&$          2.1_{-1.4}^{+2.9}$&$    0.29_{-0.22}^{+0.43}$&$       1.8_{-0.8}^{+1.3}$&                      ...\\
$       152156.48+520238.5$&$ 2.238$&$  15334$&    2013 Oct 22&$37.1$&$         43.5_{-6.7}^{+7.8}$&$         47.9_{-7.3}^{+8.5}$&$    1.10_{-0.24}^{+0.28}$&$       0.6_{-0.2}^{+0.2}$&  W11, archival 4~ks obs.\\
$       153412.68+503405.3$&$ 2.122$&$  14956$&    2013 Dec 31&$ 6.1$&$                       <4.1$&$                       <4.2$&$                     ...$&$                     ...$&                      ...\\
$       153714.26+271611.6$&$ 2.457$&$  14955$&    2013 Dec 22&$ 6.0$&$         44.4_{-6.8}^{+7.9}$&$         22.0_{-4.9}^{+6.1}$&$    0.49_{-0.13}^{+0.16}$&$       1.3_{-0.3}^{+0.3}$&                      ...\\
\vspace{2mm}
$       153913.47+395423.4$&$ 1.934$&$  14948$&    2013 Dec 13&$ 5.3$&$                       <2.4$&$                       <2.5$&$                     ...$&$                     ...$&                      ...\\
$       222256.11-094636.2$&$ 2.913$&$  14953$&    2013 Sep 27&$ 9.4$&$          2.0_{-1.3}^{+2.7}$&$                       <4.2$&$                   <2.87$&$                   >-0.2$&                      ...\\
\noalign{\smallskip}\hline\noalign{\smallskip}
\multicolumn{10}{c}{Weak-line Quasars} \\
\noalign{\smallskip}\hline\noalign{\smallskip}
$       014333.65-391700.1$&$ 1.800$&$  15358$&    2012 Nov 21&$ 3.0$&$         15.6_{-4.0}^{+5.2}$&$          7.7_{-2.8}^{+4.2}$&$    0.49_{-0.22}^{+0.31}$&$       1.4_{-0.4}^{+0.5}$&             HE 0141-3932\\
$       082722.73+032755.9$&$ 2.031$&$  15342$&    2013 Mar 23&$ 2.5$&$                       <2.4$&$                       <2.5$&$                     ...$&$                     ...$&                      ...\\
$       084424.24+124546.5$&$ 2.492$&$  15337$&    2013 Apr 01&$ 1.5$&$         51.2_{-7.3}^{+8.4}$&$         11.0_{-3.4}^{+4.7}$&$    0.21_{-0.07}^{+0.10}$&$       2.1_{-0.3}^{+0.4}$&                      ...\\
$       090843.25+285229.8$&$ 0.933$&$  15352$&    2013 Feb 25&$ 2.0$&$         40.6_{-6.5}^{+7.6}$&$          7.7_{-2.8}^{+4.2}$&$    0.19_{-0.08}^{+0.11}$&$       2.2_{-0.4}^{+0.5}$&                      ...\\
\vspace{2mm}
$       100517.54+331202.8$&$ 1.802$&$  15351$&    2013 Mar 01&$ 3.0$&$                       <4.1$&$                       <2.5$&$                     ...$&$                     ...$&                      ...\\
$       113413.48+001042.0$&$ 1.487$&$  15343$&    2013 Feb 19&$ 2.5$&$                       <4.1$&$                       <2.5$&$                     ...$&$                     ...$&                      ...\\
$       115637.02+184856.5$&$ 1.993$&$  15341$&    2013 Mar 20&$ 2.0$&$         43.9_{-6.7}^{+7.9}$&$         11.0_{-3.4}^{+4.7}$&$    0.25_{-0.09}^{+0.12}$&$       2.0_{-0.3}^{+0.4}$&                      ...\\
$       130312.89+321911.4$&$ 0.638$&$  15350$&    2012 Nov 19&$ 2.0$&$                       <2.4$&$                       <2.5$&$                     ...$&$                     ...$&                      ...\\
$       131059.77+560140.2$&$ 1.285$&$  15344$&    2013 Apr 10&$ 2.0$&$         18.7_{-4.4}^{+5.5}$&$          5.5_{-2.4}^{+3.7}$&$    0.29_{-0.14}^{+0.22}$&$       1.8_{-0.5}^{+0.6}$&                      ...\\
\vspace{2mm}
$       132130.21+481719.1$&$ 1.409$&$  15347$&    2013 Feb 19&$ 2.5$&$                       <2.4$&$                       <2.5$&$                     ...$&$                     ...$&                      ...\\
$       132809.59+545452.7$&$ 2.116$&$  15338$&    2013 Jul 15&$ 2.0$&$          6.2_{-2.5}^{+3.7}$&$          2.2_{-1.4}^{+2.9}$&$    0.35_{-0.26}^{+0.52}$&$       1.7_{-0.8}^{+1.2}$&                      ...\\
$       133222.62+034739.9$\tablenotemark{a}&$ 1.447$&$  15340$&    2013 Feb 13&$ 2.0$&$          2.1_{-1.3}^{+2.7}$&$          6.6_{-2.6}^{+3.9}$&$    3.18_{-2.21}^{+6.15}$&$      -0.3_{-0.9}^{+1.1}$&              Lensed, BAL\\
$       134601.28+585820.2$&$ 1.664$&$  15336$&    2013 Apr 28&$ 1.5$&$                       <2.4$&$                       <2.5$&$                     ...$&$                     ...$&                      ...\\
$       140710.26+241853.6$&$ 1.668$&$  15345$&    2012 Nov 19&$ 2.5$&$                       <2.4$&$                       <2.5$&$                     ...$&$                     ...$&                      ...\\
\vspace{2mm}
$       141141.96+140233.9$&$ 1.753$&$  15353$&    2012 Dec 16&$ 3.4$&$         48.9_{-7.1}^{+8.2}$&$         13.2_{-3.8}^{+5.0}$&$    0.27_{-0.09}^{+0.11}$&$       1.9_{-0.3}^{+0.3}$&                      ...\\
$       141730.92+073320.7$&$ 1.710$&$  15349$&    2012 Dec 05&$ 2.5$&$         21.9_{-4.7}^{+5.9}$&$          5.5_{-2.4}^{+3.7}$&$    0.25_{-0.12}^{+0.18}$&$       2.0_{-0.5}^{+0.6}$&                      ...\\
$       142943.64+385932.2$&$ 0.928$&$  15335$&    2012 Dec 08&$ 1.5$&$         14.6_{-3.8}^{+5.0}$&$          4.4_{-2.1}^{+3.5}$&$    0.30_{-0.16}^{+0.26}$&$       1.8_{-0.6}^{+0.7}$&                      ...\\
$       144741.76-020339.1$&$ 1.431$&$  15355$&    2013 Jan 13&$ 2.0$&$          6.2_{-2.5}^{+3.7}$&$                       <4.3$&$                   <0.70$&$                    >1.0$&                      ...\\
$       162933.60+253200.6$&$ 1.339$&$  15356$&    2012 Dec 07&$ 2.9$&$         18.7_{-4.4}^{+5.5}$&$         17.6_{-4.4}^{+5.6}$&$    0.94_{-0.32}^{+0.41}$&$       0.8_{-0.3}^{+0.4}$&                      ...\\
\vspace{2mm}
$       164302.03+441422.1$&$ 1.650$&$  15348$&    2013 May 30&$ 3.0$&$         40.6_{-6.5}^{+7.6}$&$          8.8_{-3.0}^{+4.3}$&$    0.22_{-0.08}^{+0.11}$&$       2.1_{-0.4}^{+0.4}$&                      ...\\
$       172858.16+603512.7$&$ 1.807$&$  15346$&    2013 Jan 05&$ 2.5$&$         15.6_{-4.0}^{+5.2}$&$                       <4.3$&$                   <0.29$&$                    >1.8$&                      ...\\
$       212416.05-074129.8$&$ 1.402$&$  15339$&    2012 Nov 20&$ 1.5$&$         13.5_{-3.7}^{+4.9}$&$                       <4.3$&$                   <0.35$&$                    >1.7$&                      ...\\
$       215454.35-305654.3$&$ 0.494$&$  15354$&    2012 Dec 13&$ 2.5$&$                       <2.4$&$                       <2.5$&$                     ...$&$                     ...$&     2QZ J215454.3-305654\\
\enddata
\tablecomments{
Cols. (1) and (2): object name in the J2000 equatorial coordinate format 
and redshift.
Cols. (3)--(5): \chandra\ observation ID, observation start date, and background-flare cleaned effective exposure time in the 0.5--8~keV band.
Cols. (6)--(7): Aperture-corrected source 
counts in the soft (0.5--2~keV) and hard (2--8~keV) bands. An 
upper limit at a 90\%
confidence level
is given if the source is not detected.
Col. (8): ratio between the soft-band and hard-band counts. An entry
of ``...'' indicates that
the source is undetected in both bands.
Col.~(9): {0.5--8~keV effective power-law photon 
index}, derived from the band ratio 
assuming a power-law spectrum modified with Galactic absorption. 
An entry of ``...'' indicates that it cannot be
constrained.
Col. (10): comments on special objects.}
\tablenotetext{a}{J1332+0347 is included in this table but not
in the figures or our analyses (see Section~\ref{sec-selwlq}).}
\label{tbl-xbasic}
\end{deluxetable}

\begin{deluxetable}{lccccccccc}


\tablewidth{0pt}
\tablecaption{Quasar UV Emission-Line Measurements from the SDSS Spectra}
\tablehead{
\colhead{Object Name}                   &
\colhead{MJD}                  &
\colhead{\iona{C}{iv} Blueshift}                  &
\colhead{\iona{C}{iv} FWHM}                   &
\colhead{REW}                   &
\colhead{REW}                   &
\colhead{REW}                   &
\colhead{REW}                   &
\colhead{REW}                   &
\colhead{REW}                   \\
\colhead{(J2000)}   &
\colhead{}   &
\colhead{(km~s$^{-1}$)}   &
\colhead{(km~s$^{-1}$)}   &
\colhead{\iona{C}{iv}}   &
\colhead{\iona{Si}{iv}}   &
\colhead{$\lambda1900$~\AA}   &
\colhead{\iona{Fe}{ii}}   &
\colhead{\iona{Fe}{iii}}   &
\colhead{\iona{Mg}{ii}}   \\
\colhead{(1)}         &
\colhead{(2)}         &
\colhead{(3)}         &
\colhead{(4)}         &
\colhead{(5)}         &
\colhead{(6)}         &
\colhead{(7)}         &
\colhead{(8)}         &
\colhead{(9)}         &
\colhead{(10)}  \\
\noalign{\smallskip}\hline\noalign{\smallskip}
\multicolumn{10}{c}{PHL 1811 Analogs}
}

\startdata
$       014733.58+000323.2$&$  51793      $&$      -6170\pm 440$&$       7380\pm 560$&$        4.5\pm 0.3$&$              <3.0$&$        3.8\pm 0.3$&$           21\pm 2$&$        6.2\pm 0.3$&$       12.2\pm 0.6$\\
$       082508.75+115536.3$&$  54149      $&$      -4600\pm 340$&$       8100\pm 640$&$        5.8\pm 0.3$&$        4.0\pm 0.3$&$        8.2\pm 0.4$&$           26\pm 2$&$        5.4\pm 0.4$&$       13.1\pm 0.8$\\
$       090809.13+444138.8$&$  52312      $&$      -4250\pm 640$&$       5650\pm 460$&$        4.1\pm 0.5$&$               ...$&$        7.9\pm 0.5$&$           32\pm 3$&$        4.8\pm 0.8$&$       17.1\pm 1.1$\\
$       094808.39+161414.1$&$  54095      $&$      -3700\pm 380$&$       8290\pm 660$&$        5.9\pm 0.3$&$               ...$&$        3.7\pm 0.3$&$           28\pm 2$&$        4.7\pm 0.3$&$       16.5\pm 0.5$\\
\vspace{2mm}
$       113342.67+114206.2$&$  53055      $&$     -2460\pm 1740$&$       2900\pm 240$&$        2.6\pm 0.8$&$        1.4\pm 1.2$&$        4.9\pm 0.9$&$           27\pm 2$&$        3.4\pm 0.7$&$       10.4\pm 1.9$\\
$       143525.31+400112.2$&$  52797      $&$      -2050\pm 440$&$       6160\pm 500$&$        7.4\pm 0.6$&$        3.2\pm 0.5$&$        8.3\pm 0.8$&$           13\pm 1$&$        1.9\pm 0.5$&$               ...$\\
$       152156.48+520238.5$&$  52376      $&$      -9300\pm 610$&$      11700\pm 800$&$        9.1\pm 0.6$&$        2.7\pm 0.3$&$        8.2\pm 0.6$&$           21\pm 2$&$        7.1\pm 0.6$&$       14.9\pm 0.6$\\
$       153412.68+503405.3$&$  52401      $&$       -970\pm 320$&$       7610\pm 580$&$        8.7\pm 0.5$&$        3.1\pm 0.3$&$       15.8\pm 0.5$&$           40\pm 4$&$        7.8\pm 0.4$&$       25.6\pm 1.0$\\
$       153714.26+271611.6$&$  54180      $&$      -1310\pm 220$&$       6560\pm 520$&$        6.0\pm 0.3$&$        3.9\pm 0.2$&$        8.0\pm 0.3$&$               >11$ ($13\pm2$)&$        1.9\pm 0.3$&$               ...$\\
\vspace{2mm}
$       153913.47+395423.4$&$  53171      $&$      -3930\pm 600$&$      10570\pm 740$&$        6.9\pm 0.5$&$        2.2\pm 0.4$&$       11.2\pm 0.6$&$           30\pm 3$&$        5.4\pm 0.4$&$       19.9\pm 0.8$\\
$       222256.11-094636.2$&$  52206      $&$     -2570\pm 1300$&$       5030\pm 440$&$        3.7\pm 0.5$&$        5.2\pm 0.9$&$        8.7\pm 0.7$&$               ...$&$              <3.0$&$               ...$\\
\noalign{\smallskip}\hline\noalign{\smallskip}
\multicolumn{10}{c}{Weak-line Quasars} \\
\noalign{\smallskip}\hline\noalign{\smallskip}
$       014333.65-391700.1$&$  ...        $&$               ...$&$               ...$&$               ...$&$               ...$&$               ...$&$               ...$&$               ...$&$               ...$\\
$       082722.73+032755.9$&$  52642      $&$      -1460\pm 520$&$       2560\pm 200$&$        4.0\pm 0.8$&$              <1.1$&$        2.5\pm 1.1$&$           53\pm 5$&$        6.0\pm 0.8$&$       22.2\pm 2.1$\\
$       084424.24+124546.5$&$  53801      $&$      -1450\pm 500$&$       5870\pm 480$&$        5.5\pm 0.5$&$        2.1\pm 0.4$&$        3.1\pm 0.4$&$               >16$ ($26\pm2$)&$        3.9\pm 0.5$&$               ...$\\
$       090843.25+285229.8$&$  53330      $&$               ...$&$               ...$&$               ...$&$               ...$&$               ...$&$           19\pm 1$&$              <1.6$&$        4.9\pm 0.8$\\
\vspace{2mm}
$       100517.54+331202.8$&$  53378      $&$     -2420\pm 1040$&$       4020\pm 340$&$        5.8\pm 1.0$&$               ...$&$        3.4\pm 1.0$&$           24\pm 2$&$        2.6\pm 1.3$&$       12.5\pm 2.1$\\
$       113413.48+001042.0$&$  51630,51658$&$               ...$&$               ...$&$               ...$&$               ...$&$        5.1\pm 0.5$&$           38\pm 3$&$        3.2\pm 0.5$&$       12.9\pm 0.6$\\
$       115637.02+184856.5$&$  54180      $&$      -1790\pm 900$&$       3180\pm 260$&$        1.2\pm 0.5$&$        1.5\pm 0.4$&$        5.0\pm 0.8$&$           22\pm 2$&$        1.8\pm 0.6$&$        8.5\pm 1.5$\\
$       130312.89+321911.4$&$  53819      $&$               ...$&$               ...$&$               ...$&$               ...$&$               ...$&$               >10$&$               ...$&$        7.3\pm 0.9$\\
$       131059.77+560140.2$&$  52791      $&$               ...$&$               ...$&$               ...$&$               ...$&$        4.5\pm 0.8$&$           12\pm 1$&$        1.2\pm 0.5$&$        7.3\pm 1.0$\\
\vspace{2mm}
$       132130.21+481719.1$&$  52759      $&$               ...$&$               ...$&$               ...$&$               ...$&$        5.6\pm 1.0$&$           27\pm 2$&$        3.0\pm 0.9$&$        9.3\pm 1.0$\\
$       132809.59+545452.7$&$  52724      $&$      -2560\pm 340$&$       3970\pm 300$&$        4.4\pm 0.4$&$        3.5\pm 0.3$&$        3.7\pm 0.5$&$           20\pm 2$&$        3.5\pm 0.5$&$       18.3\pm 2.2$\\
$       133222.62+034739.9$\tablenotemark{a}&$  52374      $&$               ...$&$               ...$&$               ...$&$               ...$&$        4.4\pm 1.1$&$            8\pm 1$&$              <1.5$&$        9.3\pm 0.9$\\
$       134601.28+585820.2$&$  52425,52466$&$      -4570\pm 760$&$       5580\pm 440$&$        2.9\pm 0.3$&$               ...$&$        4.8\pm 0.3$&$           14\pm 1$&$        1.7\pm 0.3$&$       12.1\pm 0.5$\\
$       140710.26+241853.6$&$  53770      $&$       -780\pm 600$&$       1170\pm 140$&$        1.5\pm 0.6$&$               ...$&$        5.4\pm 0.8$&$           26\pm 2$&$        4.1\pm 0.9$&$       10.0\pm 1.2$\\
\vspace{2mm}
$       141141.96+140233.9$&$  53442      $&$      -1850\pm 740$&$       2420\pm 180$&$        2.4\pm 0.9$&$               ...$&$        6.6\pm 0.9$&$           27\pm 2$&$        5.5\pm 1.2$&$        6.0\pm 1.6$\\
$       141730.92+073320.7$&$  53499      $&$      -5470\pm 760$&$       3240\pm 260$&$        2.0\pm 0.7$&$               ...$&$        4.2\pm 0.8$&$           25\pm 2$&$        2.8\pm 0.8$&$       11.2\pm 1.2$\\
$       142943.64+385932.2$&$  52797      $&$               ...$&$               ...$&$               ...$&$               ...$&$               ...$&$           22\pm 2$&$        2.3\pm 0.5$&$       13.7\pm 0.8$\\
$       144741.76-020339.1$&$  52411      $&$               ...$&$               ...$&$               ...$&$               ...$&$        6.7\pm 0.7$&$            7\pm 1$&$        2.8\pm 0.7$&$       12.4\pm 1.1$\\
$       162933.60+253200.6$&$  53226      $&$               ...$&$               ...$&$               ...$&$               ...$&$        9.5\pm 1.0$&$            5\pm 1$&$        3.2\pm 0.9$&$       10.7\pm 1.1$\\
\vspace{2mm}
$       164302.03+441422.1$&$  52051      $&$               ...$&$               ...$&$              <2.0$&$               ...$&$              <2.0$&$            8\pm 1$&$        9.8\pm 1.6$&$        3.9\pm 1.1$\\
$       172858.16+603512.7$&$  51792      $&$      -2630\pm 720$&$       3380\pm 340$&$        3.1\pm 0.7$&$        2.6\pm 1.1$&$        6.2\pm 0.8$&$           11\pm 1$&$              <1.6$&$        5.9\pm 1.6$\\
$       212416.05-074129.8$&$  52178,52200$&$               ...$&$               ...$&$               ...$&$               ...$&$        2.6\pm 0.5$&$           25\pm 2$&$        1.4\pm 0.4$&$       12.7\pm 0.5$\\
$       215454.35-305654.3$&$  ...        $&$               ...$&$               ...$&$               ...$&$               ...$&$               ...$&$               ...$&$               ...$&$               ...$\\
\enddata
\tablecomments{
Col. (1): object name. 
Col. (2): modified Julian date of the SDSS observation;
an entry of ``...'' indicates that no SDSS spectrum is available.
In a few cases, 
two spectra are available, and the average spectrum was used for the 
measurements. 
Cols. (3) and (4): blueshift and full width at half maximum (FWHM) of the 
\iona{C}{iv} line.
Cols. (5)--(10): REWs (in units of \AA) of the \iona{C}{iv}~$\lambda1549$, 
\iona{Si}{iv}~$\lambda1397$, $\lambda1900$ complex, \iona{Fe}{ii} (2250--2650~\AA),
\iona{Fe}{iii} UV48 $\lambda2080$, and \iona{Mg}{ii}~$\lambda2799$
emission features. The \iona{Si}{iv}~$\lambda1397$ line is a blend of 
\iona{Si}{iv} and \iona{O}{iv]}, and the $\lambda1900$ emission is
dominated by \iona{C}{iii]}~$\lambda1909$ with additional contributions from
\iona{[Ne}{iii]}~$\lambda1814$, \iona{Si}{ii}~$\lambda1816$, 
\iona{Al}{iii}~$\lambda1857$, \iona{Si}{iii]}~$\lambda1892$,
and several \iona{Fe}{iii} multiplets
(see Table~2 of \citealt{Vandenberk2001} for details).
The UV \iona{Fe}{ii} REW was measured between 
2250\AA\ and 2650\AA; for three objects that do not have 
full spectral coverage of this range, a lower limit was derived (along
with an estimated value in parentheses if the covered fraction is $>60\%$).
For the other properties, 
the measurements were performed the same manner as in W11.
The uncertainties have been multiplied by the factors 
given in the notes of W11 Table 3, and the upper limits are at 
a 3$\sigma$ confidence level.}
\tablenotetext{a}{J1332+0347 is included in this table but not
in the figures or our analyses (see Section~\ref{sec-selwlq}).}
\label{tbl-uvline}
\end{deluxetable}

\begin{turnpage}
\begin{deluxetable}{lccccccccccccc}
\tabletypesize{\scriptsize}

\tablewidth{0pt}
\tablecaption{X-ray and Optical Properties}
\tablehead{
\colhead{Object Name}                   &
\colhead{$M_{i}$}                   &
\colhead{$N_{\rm H,Gal}$}                   &
\colhead{Count Rate}                   &
\colhead{$F_{\rm X}$}                   &
\colhead{$f_{\rm 2~keV}$}                   &
\colhead{$\log L_{\rm X}$}                   &
\colhead{$f_{\rm 2500~\textup{\AA}}$}                   &
\colhead{$\log L_{\rm 2500~\textup{\AA}}$}                   &
\colhead{$\alpha_{\rm OX}$}                   &
\colhead{$\Delta\alpha_{\rm OX}(\sigma)$}                   &
\colhead{$f_{\rm weak}$}                  &
\colhead{$\Delta(g-i)$}  &
\colhead{$R$}  \\
\colhead{(J2000)}   &
\colhead{}   &
\colhead{} & 
\colhead{(0.5--2~keV)}   &
\colhead{(0.5--2~keV)}   &
\colhead{}   &
\colhead{(2--10 keV)}   &
\colhead{}   &
\colhead{}   &
\colhead{}   &
\colhead{}   &
\colhead{}   &
\colhead{}   &
\colhead{}   \\
\colhead{(1)}         &
\colhead{(2)}         &
\colhead{(3)}         &
\colhead{(4)}         &
\colhead{(5)}         &
\colhead{(6)}         &
\colhead{(7)}         &
\colhead{(8)}         &
\colhead{(9)}         &
\colhead{(10)}         &
\colhead{(11)}         &
\colhead{(12)}         &
\colhead{(13)}       &
\colhead{(14)}           \\
\noalign{\smallskip}\hline\noalign{\smallskip}
\multicolumn{14}{c}{PHL 1811 Analogs}
}

\startdata
$       014733.58+000323.2$&$ -27.91$&$   2.90$&$                    <0.39$&$   <0.16$&$   <0.68$&$  <43.77$&$    3.68$&$   31.58$&$ <-2.20$&$<-0.49(3.32)$&$ >18.40$&$   0.55$&$   <1.0$\\
$       082508.75+115536.3$&$ -28.31$&$   3.67$&$                    <0.47$&$   <0.19$&$   <0.81$&$  <43.83$&$    4.62$&$   31.66$&$ <-2.21$&$<-0.48(3.30)$&$ >17.95$&$  -0.08$&$   <1.0$\\
$       090809.13+444138.8$&$ -27.26$&$   1.66$&$                    <0.35$&$   <0.14$&$   <0.56$&$  <43.56$&$    2.09$&$   31.21$&$ <-2.14$&$<-0.47(3.25)$&$ >17.28$&$   0.00$&$   <1.9$\\
$       094808.39+161414.1$&$ -28.43$&$   3.34$&$     0.55_{-0.36}^{+0.74}$&$    0.22$&$    0.90$&$   43.80$&$    6.59$&$   31.74$&$  -2.25$&$ -0.51(3.51)$&$  21.67$&$   0.20$&$   <0.6$\\
\vspace{2mm}
$       113342.67+114206.2$&$ -27.52$&$   3.32$&$                    <0.26$&$   <0.10$&$   <0.45$&$  <43.60$&$    1.99$&$   31.32$&$ <-2.17$&$<-0.49(3.34)$&$ >18.61$&$   0.39$&$   <2.3$\\
$       143525.31+400112.2$&$ -27.84$&$   1.00$&$     1.05_{-0.39}^{+0.57}$&$    0.42$&$    1.89$&$   44.32$&$    3.32$&$   31.62$&$  -2.01$&$ -0.29(2.00)$&$   5.75$&$   0.11$&$   <1.4$\\
$       152156.48+520238.5$&$ -30.22$&$   1.58$&$     1.17_{-0.18}^{+0.21}$&$    0.46$&$    0.98$&$   44.51$&$   19.86$&$   32.39$&$  -2.42$&$ -0.59(4.51)$&$  34.73$&$   0.24$&$   <0.2$\\
$       153412.68+503405.3$&$ -28.70$&$   1.56$&$                    <0.67$&$   <0.27$&$   <1.20$&$  <44.05$&$    3.69$&$   31.61$&$ <-2.11$&$<-0.39(2.64)$&$ >10.13$&$   0.49$&$   <1.2$\\
$       153714.26+271611.6$&$ -28.67$&$   3.06$&$     7.45_{-1.14}^{+1.32}$&$    2.92$&$   10.23$&$   45.31$&$    5.75$&$   31.92$&$  -1.82$&$ -0.06(0.45)$&$   1.43$&$   0.03$&$    2.6$\\
\vspace{2mm}
$       153913.47+395423.4$&$ -27.99$&$   1.70$&$                    <0.45$&$   <0.18$&$   <0.76$&$  <43.78$&$    3.86$&$   31.56$&$ <-2.19$&$<-0.48(3.26)$&$ >17.39$&$   0.09$&$   <1.2$\\
$       222256.11-094636.2$&$ -28.55$&$   4.34$&$     0.22_{-0.14}^{+0.29}$&$    0.09$&$    0.47$&$   43.88$&$    4.33$&$   31.92$&$  -2.29$&$ -0.52(4.01)$&$  23.29$&$   0.23$&$   <1.3$\\
\noalign{\smallskip}\hline\noalign{\smallskip}
\multicolumn{14}{c}{Weak-line Quasars} \\
\noalign{\smallskip}\hline\noalign{\smallskip}
$       014333.65-391700.1$&$ -29.22$&$   1.71$&$     5.28_{-1.35}^{+1.74}$&$    2.07$&$    6.75$&$   44.88$&$   15.28$&$   32.10$&$  -2.06$&$ -0.27(2.03)$&$   4.93$&$    ...$&$   <0.5$\\
$       082722.73+032755.9$&$ -27.27$&$   3.73$&$                    <0.97$&$   <0.39$&$   <1.68$&$  <44.16$&$    1.68$&$   31.24$&$ <-1.92$&$<-0.25(1.72)$&$  >4.50$&$   0.31$&$   <2.6$\\
$       084424.24+124546.5$&$ -28.17$&$   3.81$&$    33.58_{-4.77}^{+5.50}$&$   13.53$&$   74.79$&$   45.90$&$    3.81$&$   31.75$&$  -1.42$&$  0.32(2.17)$&$   0.15$&$   0.37$&$    8.0$\\
$       090843.25+285229.8$&$ -25.10$&$   2.46$&$    20.28_{-3.23}^{+3.80}$&$    8.30$&$   23.71$&$   44.55$&$    1.48$&$   30.54$&$  -1.46$&$  0.11(0.57)$&$   0.51$&$  -0.03$&$   <2.0$\\
\vspace{2mm}
$       100517.54+331202.8$&$ -26.70$&$   1.45$&$                    <1.38$&$   <0.55$&$   <2.22$&$  <44.19$&$    1.38$&$   31.06$&$ <-1.84$&$<-0.20(1.35)$&$  >3.26$&$   0.07$&$   <2.8$\\
$       113413.48+001042.0$&$ -26.48$&$   2.63$&$                    <1.64$&$   <0.66$&$   <2.39$&$  <44.07$&$    1.57$&$   30.96$&$ <-1.85$&$<-0.22(1.11)$&$  >3.75$&$   0.17$&$   <2.4$\\
$       115637.02+184856.5$&$ -27.30$&$   2.83$&$    21.91_{-3.36}^{+3.92}$&$    8.78$&$   38.74$&$   45.49$&$    2.04$&$   31.31$&$  -1.43$&$  0.25(1.71)$&$   0.22$&$   0.05$&$    7.4$\\
$       130312.89+321911.4$&$ -24.20$&$   1.17$&$                    <1.20$&$   <0.48$&$   <1.20$&$  <43.03$&$    1.45$&$   30.19$&$ <-1.95$&$<-0.43(2.17)$&$ >13.09$&$   0.21$&$   <1.9$\\
$       131059.77+560140.2$&$ -26.09$&$   1.51$&$     9.36_{-2.19}^{+2.77}$&$    3.73$&$   12.42$&$   44.69$&$    1.64$&$   30.86$&$  -1.58$&$  0.03(0.17)$&$   0.82$&$  -0.10$&$   <2.3$\\
\vspace{2mm}
$       132130.21+481719.1$&$ -26.20$&$   1.13$&$                    <0.97$&$   <0.39$&$   <1.36$&$  <43.78$&$    1.38$&$   30.86$&$ <-1.92$&$<-0.31(1.54)$&$  >6.26$&$   0.12$&$   <2.6$\\
$       132809.59+545452.7$&$ -27.72$&$   1.16$&$     3.12_{-1.24}^{+1.86}$&$    1.23$&$    4.89$&$   44.75$&$    2.70$&$   31.47$&$  -1.82$&$ -0.12(0.81)$&$   2.03$&$  -0.09$&$   <1.7$\\
$       133222.62+034739.9$\tablenotemark{a}&$ -26.70$&$   1.98$&$     1.03_{-0.67}^{+1.37}$&$    0.41$&$    0.63$&$   44.47$&$    1.75$&$   30.98$&$  -2.09$&$ -0.46(2.31)$&$  15.58$&$   0.34$&$   <2.2$\\
$       134601.28+585820.2$&$ -27.38$&$   1.63$&$                    <1.57$&$   <0.63$&$   <2.43$&$  <44.17$&$    3.22$&$   31.36$&$ <-1.97$&$<-0.28(1.92)$&$  >5.37$&$   0.11$&$   <1.3$\\
$       140710.26+241853.6$&$ -26.70$&$   1.65$&$                    <0.97$&$   <0.39$&$   <1.50$&$  <43.96$&$    1.50$&$   31.03$&$ <-1.92$&$<-0.28(1.41)$&$  >5.36$&$   0.14$&$   <2.8$\\
\vspace{2mm}
$       141141.96+140233.9$&$ -26.56$&$   1.43$&$    14.42_{-2.10}^{+2.43}$&$    5.75$&$   22.89$&$   45.18$&$    1.13$&$   30.95$&$  -1.42$&$  0.21(1.06)$&$   0.28$&$  -0.03$&$   <3.6$\\
$       141730.92+073320.7$&$ -26.64$&$   2.12$&$     8.81_{-1.91}^{+2.37}$&$    3.53$&$   14.14$&$   44.93$&$    1.64$&$   31.09$&$  -1.56$&$  0.09(0.60)$&$   0.59$&$  -0.01$&$   <3.0$\\
$       142943.64+385932.2$&$ -26.01$&$   0.95$&$     9.56_{-2.53}^{+3.30}$&$    3.81$&$   11.02$&$   44.37$&$    3.57$&$   30.91$&$  -1.73$&$ -0.11(0.55)$&$   1.91$&$  -0.05$&$   <0.9$\\
$       144741.76-020339.1$&$ -26.00$&$   4.53$&$     3.11_{-1.24}^{+1.86}$&$    1.25$&$    4.46$&$   44.31$&$    1.69$&$   30.96$&$  -1.76$&$ -0.13(0.64)$&$   2.15$&$  -0.36$&$   <2.3$\\
$       162933.60+253200.6$&$ -25.35$&$   3.58$&$     6.39_{-1.49}^{+1.89}$&$    2.51$&$    6.44$&$   44.87$&$    1.05$&$   30.70$&$  -1.62$&$ -0.02(0.12)$&$   1.16$&$   0.01$&$   <3.4$\\
\vspace{2mm}
$       164302.03+441422.1$&$ -26.56$&$   1.52$&$    13.73_{-2.19}^{+2.57}$&$    5.52$&$   22.46$&$   45.06$&$    1.20$&$   30.93$&$  -1.43$&$  0.19(0.98)$&$   0.31$&$   0.13$&$   <3.1$\\
$       172858.16+603512.7$&$ -26.88$&$   3.32$&$     6.29_{-1.61}^{+2.08}$&$    2.52$&$   10.21$&$   44.86$&$    1.86$&$   31.19$&$  -1.64$&$  0.03(0.18)$&$   0.85$&$   0.05$&$   <2.4$\\
$       212416.05-074129.8$&$ -26.64$&$   5.58$&$     8.89_{-2.43}^{+3.21}$&$    3.60$&$   12.68$&$   44.74$&$    2.56$&$   31.12$&$  -1.65$&$ -0.00(0.00)$&$   1.00$&$  -0.14$&$   <1.5$\\
$       215454.35-305654.3$&$ -23.19$&$   1.84$&$                    <0.97$&$   <0.39$&$   <0.90$&$  <42.68$&$    0.61$&$   29.59$&$ <-1.85$&$<-0.42(2.53)$&$ >12.25$&$    ...$&$   <8.3$\\
\enddata
\tablecomments{
Col. (1): object name.
Col. (2): absolute $i$-band magnitudes.
Col. (3): Galactic neutral hydrogen column density \citep{Dickey1990}.
Col. (4): observed 0.5--2~keV \chandra\ count rate in units of 10$^{-3}$~s$^{-1}$.
Col. (5): Galactic absorption-corrected observed-frame
0.5--2~keV flux in units of
$10^{-14}$~\flux.
Col. (6): rest-frame 2~keV flux density in units of $10^{-32}$~\mflux.
Col. (7): logarithm of the rest-frame 2--10~keV luminosity in units
of \lum, derived from
the observed 0.5--2~keV flux.
Col. (8): flux density at rest-frame 2500~\AA\ in units of $10^{-27}$~\mflux.
Col. (9): logarithm of the rest-frame 2500~\AA\ luminosity density in units
of \mlum.
Col. (10): measured $\alpha_{\rm OX}$ parameter.
Col. (11): difference between the measured $\alpha_{\rm OX}$ and the
expected $\alpha_{\rm OX}$ from the \citet{Just2007}
\hbox{$\alpha_{\rm OX}$--$L_{\rm 2500~{\textup{\AA}}}$} relation. 
The statistical significance of this difference,
measured in units of the $\alpha_{\rm OX}$ rms scatter in Table~5 of 
\citet{Steffen2006}, is given in the parenthesis.
Col. (12): factor of X-ray weakness in accordance with $\Delta\alpha_{\rm OX}$.
Col. (13): relative SDSS $g-i$ color.
Col. (14): radio-loudness parameter, {defined as 
$R=f_{5~{\rm GHz}}/f_{\rm 4400~{\textup{\AA}}}$
\citep[e.g.,][]{Kellermann1989}, where $f_{5~{\rm GHz}}$ and
$f_{\rm 4400~{\textup{\AA}}}$ are the flux densities at
rest-frame 5~GHz and 4400~\AA, respectively.}
}
\tablenotetext{a}{J1332+0347 is included in this table but not
in the figures or our analyses (see Section~\ref{sec-selwlq}).}
\label{tbl-aox}
\end{deluxetable}
\end{turnpage}

\begin{deluxetable}{ccccccccc}
\tabletypesize{\footnotesize}

\tablecaption{Stacked X-ray Properties for the X-ray Weak Subsamples}
\tablehead{
\colhead{Subsample} &
\colhead{Number of} &
\colhead{Mean} &
\colhead{Total Stacked} &
\colhead{Soft-Band} &
\colhead{Hard-Band} &
\colhead{$\Gamma_{\rm eff}$}  &
\colhead{$\alpha_{\rm OX}$}                   &
\colhead{$\Delta\alpha_{\rm OX}(\sigma)$}                         \\
\colhead{} &
\colhead{Sources} &
\colhead{Redshift} &
\colhead{Exposure (ks)} &
\colhead{Counts} &
\colhead{Counts} &
\colhead{}   &                   
\colhead{}   &
\colhead{}
}
\startdata
P1811A\tablenotemark{a} & 15 &2.16 & 106.6& $26.5^{+6.4}_{-5.3}$&
$16.1^{+5.6}_{-4.4}$& $1.16^{+0.37}_{-0.32}$&$-2.33$&$-0.61(4.16)$\\
P1811A undet\tablenotemark{b} & 7 &2.08 & 46.8& $2.9^{+3.0}_{-1.7}$&
$2.7^{+3.2}_{-1.8}$& $0.79^{+1.63}_{-0.93}$&$-2.57$&$-0.87(5.94)$\\
P1811A SB undet\tablenotemark{c} & 9 &2.07 & 62.3& $2.8^{+3.0}_{-1.7}$&
\vspace{2mm}
$7.9^{+4.3}_{-3.0}$& $-0.23^{+0.94}_{-0.85}$&$-2.74$&$-1.04(7.09)$\\
WLQ\tablenotemark{d} & 15 &1.53 & 61.7& $31.8^{+6.9}_{-5.8}$&
$12.2^{+5.0}_{-3.8}$& $1.58^{+0.40}_{-0.35}$&$-2.15$&$-0.47(3.22)$\\
WLQ undet\tablenotemark{e} & 10 & 1.54 &31.4 &$ 1.8^{+2.7}_{-1.3}$&
\vspace{2mm}
$<2.5$& $>0.44$&$-2.44$&$-0.78(5.35)$\\
\iona{C}{iv} WLQ\tablenotemark{f} & 7 &1.89 & 36.8& $5.0^{+3.5}_{-2.2}$&
$<3.8$& $>1.1$&$-2.30$&$-0.62(4.27)$\\
\iona{C}{iv} WLQ undet\tablenotemark{g} & 6 & 1.89 &21.9 &$ <3.9$&
\vspace{2mm}
$<2.5$& ...&$<-2.26$&$<-0.59(4.02)$\\
All\tablenotemark{h} & 30 &1.85 & 168.3& $58.4^{+9.0}_{-7.8}$&
$28.3^{+7.0}_{-5.8}$& $1.37^{+0.25}_{-0.23}$&$-2.24$&$-0.54(3.68)$\\
\enddata
\tablenotetext{a}{X-ray weak PHL~1811 analogs with \chandra\ observations
except J1521+5202 (nine in the \chandra\ Cycle 14 sample and six in W11).}
\tablenotetext{b}{Undetected PHL~1811 analogs with \chandra\ 
observations
(four in the \chandra\ Cycle 14 sample and three in W11).}
\tablenotetext{c}{Soft-band undetected PHL~1811 analogs with \chandra\ 
observations
(six in the \chandra\ Cycle 14 sample and three in W11). The are two
objects that are not detected in the soft band but are detected in the hard band.
This stacked $\Delta\alpha_{\rm OX}$ value was used in Figure~\ref{fig-aoxhist}a.}
\tablenotetext{d}{X-ray weak WLQs with \chandra\ observations (nine in
the \chandra\ Cycle 14 sample and six in W12).}
\tablenotetext{e}{Undetected WLQs with \chandra\ observations
(eight in the \chandra\ Cycle 14 sample and two in W12). This stacked $\Delta\alpha_{\rm OX}$ value was used in Figure~\ref{fig-aoxhist}b.}
\tablenotetext{f}{X-ray weak WLQs with \chandra\ observations 
in the \iona{C}{iv} subsample (four in
the \chandra\ Cycle 14 sample and three in W12).}
\tablenotetext{g}{Undetected WLQs with \chandra\ observations
in the \iona{C}{iv} subsample 
(four in the \chandra\ Cycle 14 sample and two in W12).}
\tablenotetext{h}{All X-ray weak PHL~1811 analogs and WLQs with \chandra\ observations except J1521+5202.}
\label{tbl-stack}
\end{deluxetable}

\begin{deluxetable}{lccccccccc}


\tablewidth{0pt}
\tablecaption{Bolometric Luminosity, SMBH Mass, and 
Eddington Ratio Estimates}
\tablehead{
\colhead{Object Name}                   &
\colhead{$\log L_{\rm Bol}$}                  &
\colhead{$\log M_{\rm BH}$}                  &
\colhead{$L_{\rm Bol}/L_{\rm Edd}$}                   \\
\colhead{(J2000)}   &
\colhead{(erg s$^{-1}$)}                   &
\colhead{($M_{\sun}$)}                   &
\colhead{}   \\
\noalign{\smallskip}\hline\noalign{\smallskip}
\multicolumn{4}{c}{\chandra\ Cycle 14 PHL 1811 Analogs}
}

\startdata
$       014733.58+000323.2$&$    46.8$&$     ...$&$     ...$\\
$       082508.75+115536.3$&$    47.5$&$     ...$&$     ...$\\
$       090809.13+444138.8$&$    46.8$&$     9.8$&$    0.07$\\
$       094808.39+161414.1$&$    47.5$&$     9.9$&$    0.38$\\
\vspace{2mm}
$       113342.67+114206.2$&$    46.8$&$     ...$&$     ...$\\
$       143525.31+400112.2$&$    46.8$&$     ...$&$     ...$\\
$       152156.48+520238.5$&$    48.2$&$    9.8*$&$    2.09$\\
$       153412.68+503405.3$&$    46.8$&$     ...$&$     ...$\\
$       153714.26+271611.6$&$    46.8$&$     ...$&$     ...$\\
$       153913.47+395423.4$&$    47.3$&$     ...$&$     ...$\\
\enddata
\tablecomments{The SMBH masses are the \iona{Mg}{ii}-based 
single-epoch virial masses from \citet{Shen2011}, except
for those objects marked with a ``*'', which are the
H$\beta$-based virial masses from W11 or \citet{Plotkin2015}.
{We caution that the SMBH mass and bolometric luminosity estimates have 
substantial uncertainties (see
Sections~\ref{sec-ledd} and \ref{sec-sed}), and thus so do the Eddington ratios.}
\\
Only a portion of this table is shown here to 
demonstrate its form and content. A machine-readable 
version of the full table is available.
}
\end{deluxetable}

\begin{deluxetable}{ccccccccc}
\tablecaption{Peto-Prentice Test Results for the 
Spectral Properties of X-ray Weak and X-ray Normal Samples}
\tablehead{
\colhead{Spectral} &
\multicolumn{4}{c}{Full Sample} &
\multicolumn{4}{c}{\iona{C}{iv} Subsample} \\
\cmidrule(r){2-5} \cmidrule(l){6-9}
\colhead{Property} &
\colhead{$N_{\rm weak}$} &
\colhead{$N_{\rm normal}$} &
\colhead{$\sigma$} &
\colhead{$P_{\rm null}$} &
\colhead{$N_{\rm weak}$} &
\colhead{$N_{\rm normal}$} &
\colhead{$\sigma$} &
\colhead{$P_{\rm null}$} 
}
\startdata
\iona{C}{iv} REW & 25 & 11 &2.3 &$0.02$ & 25 & 11&2.3 &0.02 \\
\iona{C}{iv} blueshift & 23 & 9 &1.7 &$0.09$ & 23 & 9&1.7 &0.09 \\
\iona{C}{iv} FWHM  & 23 & 9 &2.2 &$0.03$ & 23 & 9&2.2 &0.03 \\
$\lambda1900$ \AA\ REW  & 29 & 16 &1.7 &$0.10$ & 24 & 11 &1.7 &0.10 \\ 
\iona{Fe}{ii} REW  & 30 & 18 &3.8 &$1\times10^{-4}$ & 24 & 11&2.9 &$4\times10^{-3}$ \\
\iona{Fe}{iii} REW  & 29 & 18 &2.5 &$0.01$ & 24 & 11&1.2 &0.21 \\
\iona{Mg}{ii} REW  & 21 & 16 &2.4 &$0.02$ & 15 & 9&2.7 &$8\times10^{-3}$ \\  
$\Delta(g-i)$  & 31 & 18 &4.6 &$4\times10^{-6}$ & 25 & 11&3.7 &$2\times10^{-4}$ \\
\enddata
\tablecomments{
In cases where there are no censored data,
the Peto-Prentice test reduces to Gehan's Wilcoxon test
\citep[e.g.,][]{Lavalley1992}. The tests were performed for both 
the full sample and the \iona{C}{iv} subsample. 
We list for each test the number of X-ray weak objects ($N_{\rm weak}$), 
number of X-ray normal objects ($N_{\rm normal}$), test statistic ($\sigma$),
and probability of the data being drawn from the same parent 
population ($P_{\rm null}$).
}
\label{tbl-pptest}
\end{deluxetable}


\begin{thebibliography}{130}
\expandafter\ifx\csname natexlab\endcsname\relax\def\natexlab#1{#1}\fi

\bibitem[{{Abazajian} {et~al.}(2009){Abazajian}, {Adelman-McCarthy},
  {Ag{\"u}eros}, {et~al.}}]{Abazajian2009}
{Abazajian}, K.~N., {Adelman-McCarthy}, J.~K., {Ag{\"u}eros}, M.~A., {et~al.}
  2009, \apjs, 182, 543

\bibitem[{{Abramowicz} {et~al.}(1988){Abramowicz}, {Czerny}, {Lasota}, \&
  {Szuszkiewicz}}]{Abramowicz1988}
{Abramowicz}, M.~A., {Czerny}, B., {Lasota}, J.~P., \& {Szuszkiewicz}, E. 1988,
  \apj, 332, 646

\bibitem[{{Ade} {et~al.}(2015){Ade}, {Aghanim}, {Arnaud}, {et~al.}}]{Ade2014}
{Ade}, P.~A.~R., {Aghanim}, N., {Arnaud}, M., {et~al.} 2015, \aap, submitted
  (arXiv:1406.7482)

\bibitem[{{Anderson} {et~al.}(2001){Anderson}, {Fan}, {Richards},
  {et~al.}}]{Anderson2001}
{Anderson}, S.~F., {Fan}, X., {Richards}, G.~T., {et~al.} 2001, \aj, 122, 503

\bibitem[{{Arnaud}(1996)}]{Arnaud1996}
{Arnaud}, K.~A. 1996, in ASP Conf. Ser., Vol. 101, Astronomical Data Analysis
  Software and Systems V, ed. G.~H. {Jacoby} \& J.~{Barnes}, 17

\bibitem[{{Ba{\~n}ados} {et~al.}(2014){Ba{\~n}ados}, {Venemans}, {Morganson},
  {et~al.}}]{Banados2014}
{Ba{\~n}ados}, E., {Venemans}, B.~P., {Morganson}, E., {et~al.} 2014, \aj, 148,
  14

\bibitem[{{Bachev} {et~al.}(2004){Bachev}, {Marziani}, {Sulentic},
  {et~al.}}]{Bachev2004}
{Bachev}, R., {Marziani}, P., {Sulentic}, J.~W., {et~al.} 2004, \apj, 617, 171

\bibitem[{{Barvainis} {et~al.}(2005){Barvainis}, {Leh{\'a}r}, {Birkinshaw},
  {Falcke}, \& {Blundell}}]{Barvainis2005}
{Barvainis}, R., {Leh{\'a}r}, J., {Birkinshaw}, M., {Falcke}, H., \&
  {Blundell}, K.~M. 2005, \apj, 618, 108

\bibitem[{{Baskin} \& {Laor}(2004)}]{Baskin2004}
{Baskin}, A., \& {Laor}, A. 2004, \mnras, 350, L31

\bibitem[{{Baskin} \& {Laor}(2005)}]{Baskin2005}
---. 2005, \mnras, 356, 1029

\bibitem[{{Baskin} {et~al.}(2013){Baskin}, {Laor}, \& {Hamann}}]{Baskin2013}
{Baskin}, A., {Laor}, A., \& {Hamann}, F. 2013, \mnras, 432, 1525

\bibitem[{{Baskin} {et~al.}(2014){Baskin}, {Laor}, \& {Stern}}]{Baskin2014}
{Baskin}, A., {Laor}, A., \& {Stern}, J. 2014, \mnras, 438, 604

\bibitem[{{Becker} {et~al.}(1995){Becker}, {White}, \& {Helfand}}]{Becker1995}
{Becker}, R.~H., {White}, R.~L., \& {Helfand}, D.~J. 1995, \apj, 450, 559

\bibitem[{{Boroson} \& {Green}(1992)}]{Boroson1992}
{Boroson}, T.~A., \& {Green}, R.~F. 1992, \apjs, 80, 109

\bibitem[{{Brandt} \& {Alexander}(2015)}]{Brandt2014}
{Brandt}, W.~N., \& {Alexander}, D.~M. 2015, \aapr, 23, 1

\bibitem[{{Brightman} {et~al.}(2013){Brightman}, {Silverman}, {Mainieri},
  {et~al.}}]{Brightman2013}
{Brightman}, M., {Silverman}, J.~D., {Mainieri}, V., {et~al.} 2013, \mnras,
  433, 2485

\bibitem[{{Broos} {et~al.}(2007){Broos}, {Feigelson}, {Townsley},
  {et~al.}}]{Broos2007}
{Broos}, P.~S., {Feigelson}, E.~D., {Townsley}, L.~K., {et~al.} 2007, \apjs,
  169, 353

\bibitem[{{Cardelli} {et~al.}(1989){Cardelli}, {Clayton}, \&
  {Mathis}}]{Cardelli1989}
{Cardelli}, J.~A., {Clayton}, G.~C., \& {Mathis}, J.~S. 1989, \apj, 345, 245

\bibitem[{{Cash}(1979)}]{Cash1979}
{Cash}, W. 1979, \apj, 228, 939

\bibitem[{{Collinge} {et~al.}(2005){Collinge}, {Strauss}, {Hall},
  {et~al.}}]{Collinge2005}
{Collinge}, M.~J., {Strauss}, M.~A., {Hall}, P.~B., {et~al.} 2005, \aj, 129,
  2542

\bibitem[{{Dai} {et~al.}(2010){Dai}, {Kochanek}, {Chartas}, {et~al.}}]{Dai2010}
{Dai}, X., {Kochanek}, C.~S., {Chartas}, G., {et~al.} 2010, \apj, 709, 278

\bibitem[{{Diamond-Stanic} {et~al.}(2009){Diamond-Stanic}, {Fan}, {Brandt},
  {et~al.}}]{Diamond2009}
{Diamond-Stanic}, A.~M., {Fan}, X., {Brandt}, W.~N., {et~al.} 2009, \apj, 699,
  782

\bibitem[{{Dickey} \& {Lockman}(1990)}]{Dickey1990}
{Dickey}, J.~M., \& {Lockman}, F.~J. 1990, \araa, 28, 215

\bibitem[{{Eddington}(1922)}]{Eddington1922}
{Eddington}, A.~S. 1922, \mnras, 82, 432

\bibitem[{{Elvis}(2012)}]{Elvis2012b}
{Elvis}, M. 2012, in Astronomical Society of the Pacific Conference Series,
  Vol. 460, AGN Winds in Charleston, ed. G.~{Chartas}, F.~{Hamann}, \& K.~M.
  {Leighly}, 186

\bibitem[{{Falcke} {et~al.}(1996){Falcke}, {Sherwood}, \&
  {Patnaik}}]{Falcke1996}
{Falcke}, H., {Sherwood}, W., \& {Patnaik}, A.~R. 1996, \apj, 471, 106

\bibitem[{{Fan} {et~al.}(1999){Fan}, {Strauss}, {Gunn}, {et~al.}}]{Fan1999}
{Fan}, X., {Strauss}, M.~A., {Gunn}, J.~E., {et~al.} 1999, \apjl, 526, L57

\bibitem[{{Feigelson} \& {Nelson}(1985)}]{Feigelson1985}
{Feigelson}, E.~D., \& {Nelson}, P.~I. 1985, \apj, 293, 192

\bibitem[{{Freeman} {et~al.}(2002){Freeman}, {Kashyap}, {Rosner}, \&
  {Lamb}}]{Freeman2002}
{Freeman}, P.~E., {Kashyap}, V., {Rosner}, R., \& {Lamb}, D.~Q. 2002, \apjs,
  138, 185

\bibitem[{{Garmire} {et~al.}(2003){Garmire}, {Bautz}, {Ford}, {Nousek}, \&
  {Ricker}}]{Garmire2003}
{Garmire}, G.~P., {Bautz}, M.~W., {Ford}, P.~G., {Nousek}, J.~A., \& {Ricker},
  Jr., G.~R. 2003, Proc. SPIE, 4851, 28

\bibitem[{{Gehrels}(1986)}]{Gehrels1986}
{Gehrels}, N. 1986, \apj, 303, 336

\bibitem[{{Ghisellini} {et~al.}(1994){Ghisellini}, {Haardt}, \&
  {Matt}}]{Ghisellini1994}
{Ghisellini}, G., {Haardt}, F., \& {Matt}, G. 1994, \mnras, 267, 743

\bibitem[{{Gibson} {et~al.}(2008){Gibson}, {Brandt}, \&
  {Schneider}}]{Gibson2008}
{Gibson}, R.~R., {Brandt}, W.~N., \& {Schneider}, D.~P. 2008, \apj, 685, 773

\bibitem[{{Glikman} {et~al.}(2012){Glikman}, {Urrutia}, {Lacy},
  {et~al.}}]{Glikman2012}
{Glikman}, E., {Urrutia}, T., {Lacy}, M., {et~al.} 2012, \apj, 757, 51

\bibitem[{{Goad} {et~al.}(2012){Goad}, {Korista}, \& {Ruff}}]{Goad2012}
{Goad}, M.~R., {Korista}, K.~T., \& {Ruff}, A.~J. 2012, \mnras, 426, 3086

\bibitem[{{Gordon} {et~al.}(2003){Gordon}, {Clayton}, {Misselt}, {Landolt}, \&
  {Wolff}}]{Gordon2003}
{Gordon}, K.~D., {Clayton}, G.~C., {Misselt}, K.~A., {Landolt}, A.~U., \&
  {Wolff}, M.~J. 2003, \apj, 594, 279

\bibitem[{{G{\"u}ver} \& {{\"O}zel}(2009)}]{Guver2009}
{G{\"u}ver}, T., \& {{\"O}zel}, F. 2009, \mnras, 400, 2050

\bibitem[{{Harrison} {et~al.}(2013){Harrison}, {Craig}, {Christensen},
  {et~al.}}]{Harrison2013}
{Harrison}, F.~A., {Craig}, W.~W., {Christensen}, F.~E., {et~al.} 2013, \apj,
  770, 103

\bibitem[{{Hewett} \& {Wild}(2010)}]{Hewett2010}
{Hewett}, P.~C., \& {Wild}, V. 2010, \mnras, 405, 2302

\bibitem[{{Hopkins} {et~al.}(2004){Hopkins}, {Strauss}, {Hall},
  {et~al.}}]{Hopkins2004}
{Hopkins}, P.~F., {Strauss}, M.~A., {Hall}, P.~B., {et~al.} 2004, \aj, 128,
  1112

\bibitem[{{Hryniewicz} {et~al.}(2010){Hryniewicz}, {Czerny}, {Niko{\l}ajuk}, \&
  {Kuraszkiewicz}}]{Hryniewicz2010}
{Hryniewicz}, K., {Czerny}, B., {Niko{\l}ajuk}, M., \& {Kuraszkiewicz}, J.
  2010, \mnras, 404, 2028

\bibitem[{{Jiang} {et~al.}(2014){Jiang}, {Stone}, \& {Davis}}]{Jiang2014}
{Jiang}, Y.-F., {Stone}, J.~M., \& {Davis}, S.~W. 2014, \apj, 796, 106

\bibitem[{{Just} {et~al.}(2007){Just}, {Brandt}, {Shemmer},
  {et~al.}}]{Just2007}
{Just}, D.~W., {Brandt}, W.~N., {Shemmer}, O., {et~al.} 2007, \apj, 665, 1004

\bibitem[{{Kellermann} {et~al.}(1989){Kellermann}, {Sramek}, {Schmidt},
  {Shaffer}, \& {Green}}]{Kellermann1989}
{Kellermann}, K.~I., {Sramek}, R., {Schmidt}, M., {Shaffer}, D.~B., \& {Green},
  R. 1989, \aj, 98, 1195

\bibitem[{{Korista} {et~al.}(1998){Korista}, {Baldwin}, \&
  {Ferland}}]{Korista1998}
{Korista}, K., {Baldwin}, J., \& {Ferland}, G. 1998, \apj, 507, 24

\bibitem[{{Korista} \& {Goad}(2004)}]{Korista2004}
{Korista}, K.~T., \& {Goad}, M.~R. 2004, \apj, 606, 749

\bibitem[{{Kraft} {et~al.}(1991){Kraft}, {Burrows}, \& {Nousek}}]{Kraft1991}
{Kraft}, R.~P., {Burrows}, D.~N., \& {Nousek}, J.~A. 1991, \apj, 374, 344

\bibitem[{{Kratzer} \& {Richards}(2015)}]{Kratzer2015}
{Kratzer}, R.~M., \& {Richards}, G.~T. 2015, \aj, 149, 61

\bibitem[{{Krawczyk} {et~al.}(2015){Krawczyk}, {Richards}, {Gallagher},
  {et~al.}}]{Krawczyk2014}
{Krawczyk}, C.~M., {Richards}, G.~T., {Gallagher}, S.~C., {et~al.} 2015, \apj,
  submitted (arXiv:1412.7039)

\bibitem[{{Krawczyk} {et~al.}(2013){Krawczyk}, {Richards}, {Mehta},
  {et~al.}}]{Krawczyk2013}
{Krawczyk}, C.~M., {Richards}, G.~T., {Mehta}, S.~S., {et~al.} 2013, \apjs,
  206, 4

\bibitem[{{Lane} {et~al.}(2011){Lane}, {Shemmer}, {Diamond-Stanic},
  {et~al.}}]{Lane2011}
{Lane}, R.~A., {Shemmer}, O., {Diamond-Stanic}, A.~M., {et~al.} 2011, \apj,
  743, 163

\bibitem[{{Laor} \& {Davis}(2011)}]{Laor2011}
{Laor}, A., \& {Davis}, S.~W. 2011, \mnras, 417, 681

\bibitem[{{Lavalley} {et~al.}(1992){Lavalley}, {Isobe}, \&
  {Feigelson}}]{Lavalley1992}
{Lavalley}, M., {Isobe}, T., \& {Feigelson}, E. 1992, in Astronomical Society
  of the Pacific Conference Series, Vol.~25, Astronomical Data Analysis
  Software and Systems I, ed. D.~M. {Worrall}, C.~{Biemesderfer}, \&
  J.~{Barnes}, 245

\bibitem[{{Leighly}(2004)}]{Leighly2004}
{Leighly}, K.~M. 2004, \apj, 611, 125

\bibitem[{{Leighly} {et~al.}(2001){Leighly}, {Halpern}, {Helfand}, {Becker}, \&
  {Impey}}]{Leighly2001}
{Leighly}, K.~M., {Halpern}, J.~P., {Helfand}, D.~J., {Becker}, R.~H., \&
  {Impey}, C.~D. 2001, \aj, 121, 2889

\bibitem[{{Leighly} {et~al.}(2007{\natexlab{a}}){Leighly}, {Halpern},
  {Jenkins}, \& {Casebeer}}]{Leighly2007b}
{Leighly}, K.~M., {Halpern}, J.~P., {Jenkins}, E.~B., \& {Casebeer}, D.
  2007{\natexlab{a}}, \apjs, 173, 1

\bibitem[{{Leighly} {et~al.}(2007{\natexlab{b}}){Leighly}, {Halpern},
  {Jenkins}, {et~al.}}]{Leighly2007}
{Leighly}, K.~M., {Halpern}, J.~P., {Jenkins}, E.~B., {et~al.}
  2007{\natexlab{b}}, \apj, 663, 103

\bibitem[{{Londish} {et~al.}(2004){Londish}, {Heidt}, {Boyle}, {Croom}, \&
  {Kedziora-Chudczer}}]{Londish2004}
{Londish}, D., {Heidt}, J., {Boyle}, B.~J., {Croom}, S.~M., \&
  {Kedziora-Chudczer}, L. 2004, \mnras, 352, 903

\bibitem[{{Luo} {et~al.}(2013){Luo}, {Brandt}, {Alexander}, {et~al.}}]{Luo2013}
{Luo}, B., {Brandt}, W.~N., {Alexander}, D.~M., {et~al.} 2013, \apj, 772, 153

\bibitem[{{Luo} {et~al.}(2014){Luo}, {Brandt}, {Alexander}, {et~al.}}]{Luo2014}
---. 2014, \apj, 794, 70

\bibitem[{{Lusso} {et~al.}(2010){Lusso}, {Comastri}, {Vignali},
  {et~al.}}]{Lusso2010}
{Lusso}, E., {Comastri}, A., {Vignali}, C., {et~al.} 2010, \aap, 512, A34

\bibitem[{{Lyons}(1991)}]{Lyons1991}
{Lyons}, L. 1991, Data Analysis for Physical Science Students (Cambridge:
  Cambridge Univ. Press)

\bibitem[{{Martin} {et~al.}(2005){Martin}, {Fanson}, {Schiminovich},
  {et~al.}}]{Martin2005}
{Martin}, D.~C., {Fanson}, J., {Schiminovich}, D., {et~al.} 2005, \apjl, 619,
  L1

\bibitem[{{Matt} {et~al.}(1996){Matt}, {Brandt}, \& {Fabian}}]{Matt1996}
{Matt}, G., {Brandt}, W.~N., \& {Fabian}, A.~C. 1996, \mnras, 280, 823

\bibitem[{{McDowell} {et~al.}(1995){McDowell}, {Canizares}, {Elvis},
  {et~al.}}]{McDowell1995}
{McDowell}, J.~C., {Canizares}, C., {Elvis}, M., {et~al.} 1995, \apj, 450, 585

\bibitem[{{Meusinger} \& {Balafkan}(2014)}]{Meusinger2014}
{Meusinger}, H., \& {Balafkan}, N. 2014, \aap, 568, A114

\bibitem[{{Mineshige} {et~al.}(2000){Mineshige}, {Kawaguchi}, {Takeuchi}, \&
  {Hayashida}}]{Mineshige2000}
{Mineshige}, S., {Kawaguchi}, T., {Takeuchi}, M., \& {Hayashida}, K. 2000,
  \pasj, 52, 499

\bibitem[{{Miniutti} {et~al.}(2012){Miniutti}, {Brandt}, {Schneider},
  {et~al.}}]{Miniutti2012}
{Miniutti}, G., {Brandt}, W.~N., {Schneider}, D.~P., {et~al.} 2012, \mnras,
  425, 1718

\bibitem[{{Miniutti} {et~al.}(2009){Miniutti}, {Fabian}, {Brandt}, {Gallo}, \&
  {Boller}}]{Miniutti2009}
{Miniutti}, G., {Fabian}, A.~C., {Brandt}, W.~N., {Gallo}, L.~C., \& {Boller},
  T. 2009, \mnras, 396, L85

\bibitem[{{Morgan} {et~al.}(2012){Morgan}, {Hainline}, {Chen},
  {et~al.}}]{Morgan2012}
{Morgan}, C.~W., {Hainline}, L.~J., {Chen}, B., {et~al.} 2012, \apj, 756, 52

\bibitem[{{Morokuma} {et~al.}(2007){Morokuma}, {Inada}, {Oguri},
  {et~al.}}]{Morokuma2007}
{Morokuma}, T., {Inada}, N., {Oguri}, M., {et~al.} 2007, \aj, 133, 214

\bibitem[{{Murphy} \& {Yaqoob}(2009)}]{Murphy2009}
{Murphy}, K.~D., \& {Yaqoob}, T. 2009, \mnras, 397, 1549

\bibitem[{{Murray} {et~al.}(1995){Murray}, {Chiang}, {Grossman}, \&
  {Voit}}]{Murray1995}
{Murray}, N., {Chiang}, J., {Grossman}, S.~A., \& {Voit}, G.~M. 1995, \apj,
  451, 498

\bibitem[{{Netzer}(1990)}]{Netzer1990}
{Netzer}, H. 1990, in Active Galactic Nuclei, ed. R.~D. {Blandford},
  H.~{Netzer}, L.~{Woltjer}, T.~J.-L. {Courvoisier}, \& M.~{Mayor}, 57

\bibitem[{{Netzer} \& {Trakhtenbrot}(2007)}]{Netzer2007}
{Netzer}, H., \& {Trakhtenbrot}, B. 2007, \apj, 654, 754

\bibitem[{{Netzer} \& {Trakhtenbrot}(2014)}]{Netzer2014}
---. 2014, \mnras, 438, 672

\bibitem[{{Niko{\l}ajuk} \& {Walter}(2012)}]{Nikolajuk2012}
{Niko{\l}ajuk}, M., \& {Walter}, R. 2012, \mnras, 420, 2518

\bibitem[{{O'Donnell}(1994)}]{Odonnell1994}
{O'Donnell}, J.~E. 1994, \apj, 422, 158

\bibitem[{{Ohsuga} \& {Mineshige}(2011)}]{Ohsuga2011}
{Ohsuga}, K., \& {Mineshige}, S. 2011, \apj, 736, 2

\bibitem[{{Park} {et~al.}(2006){Park}, {Kashyap}, {Siemiginowska},
  {et~al.}}]{Park2006}
{Park}, T., {Kashyap}, V.~L., {Siemiginowska}, A., {et~al.} 2006, \apj, 652,
  610

\bibitem[{{Plotkin} {et~al.}(2010{\natexlab{a}}){Plotkin}, {Anderson},
  {Brandt}, {et~al.}}]{Plotkin2010b}
{Plotkin}, R.~M., {Anderson}, S.~F., {Brandt}, W.~N., {et~al.}
  2010{\natexlab{a}}, \apj, 721, 562

\bibitem[{{Plotkin} {et~al.}(2010{\natexlab{b}}){Plotkin}, {Anderson},
  {Brandt}, {et~al.}}]{Plotkin2010}
---. 2010{\natexlab{b}}, \aj, 139, 390

\bibitem[{{Plotkin} {et~al.}(2008){Plotkin}, {Anderson}, {Hall},
  {et~al.}}]{Plotkin2008}
{Plotkin}, R.~M., {Anderson}, S.~F., {Hall}, P.~B., {et~al.} 2008, \aj, 135,
  2453

\bibitem[{{Plotkin} {et~al.}(2015){Plotkin}, {Shemmer}, {Trakhtenbrot},
  {et~al.}}]{Plotkin2015}
{Plotkin}, R.~M., {Shemmer}, O., {Trakhtenbrot}, B., {et~al.} 2015, \apj,
  Submitted

\bibitem[{{Proga} {et~al.}(2000){Proga}, {Stone}, \& {Kallman}}]{Proga2000}
{Proga}, D., {Stone}, J.~M., \& {Kallman}, T.~R. 2000, \apj, 543, 686

\bibitem[{{Rafiee} \& {Hall}(2011)}]{Rafiee2011}
{Rafiee}, A., \& {Hall}, P.~B. 2011, \apjs, 194, 42

\bibitem[{{Reeves} {et~al.}(2004){Reeves}, {Porquet}, \& {Turner}}]{Reeves2004}
{Reeves}, J.~N., {Porquet}, D., \& {Turner}, T.~J. 2004, \apj, 615, 150

\bibitem[{{Reeves} {et~al.}(1997){Reeves}, {Turner}, {Ohashi}, \&
  {Kii}}]{Reeves1997}
{Reeves}, J.~N., {Turner}, M.~J.~L., {Ohashi}, T., \& {Kii}, T. 1997, \mnras,
  292, 468

\bibitem[{{Reimers} {et~al.}(2005){Reimers}, {Janknecht}, {Fechner},
  {et~al.}}]{Reimers2005}
{Reimers}, D., {Janknecht}, E., {Fechner}, C., {et~al.} 2005, \aap, 435, 17

\bibitem[{{Richards} {et~al.}(2002){Richards}, {Fan}, {Newberg},
  {et~al.}}]{Richards2002}
{Richards}, G.~T., {Fan}, X., {Newberg}, H.~J., {et~al.} 2002, \aj, 123, 2945

\bibitem[{{Richards} {et~al.}(2001){Richards}, {Fan}, {Schneider},
  {et~al.}}]{Richards2001}
{Richards}, G.~T., {Fan}, X., {Schneider}, D.~P., {et~al.} 2001, \aj, 121, 2308

\bibitem[{{Richards} {et~al.}(2011){Richards}, {Kruczek}, {Gallagher},
  {et~al.}}]{Richards2011}
{Richards}, G.~T., {Kruczek}, N.~E., {Gallagher}, S.~C., {et~al.} 2011, \aj,
  141, 167

\bibitem[{{Richards} {et~al.}(2006){Richards}, {Lacy}, {Storrie-Lombardi},
  {et~al.}}]{Richards2006}
{Richards}, G.~T., {Lacy}, M., {Storrie-Lombardi}, L.~J., {et~al.} 2006, \apjs,
  166, 470

\bibitem[{{Risaliti} {et~al.}(2009){Risaliti}, {Young}, \&
  {Elvis}}]{Risaliti2009}
{Risaliti}, G., {Young}, M., \& {Elvis}, M. 2009, \apjl, 700, L6

\bibitem[{{Runnoe} {et~al.}(2013){Runnoe}, {Brotherton}, {Shang}, {Wills}, \&
  {DiPompeo}}]{Runnoe2013}
{Runnoe}, J.~C., {Brotherton}, M.~S., {Shang}, Z., {Wills}, B.~J., \&
  {DiPompeo}, M.~A. 2013, \mnras, 429, 135

\bibitem[{{S{\c a}dowski} {et~al.}(2014){S{\c a}dowski}, {Narayan}, {McKinney},
  \& {Tchekhovskoy}}]{Sadowski2014}
{S{\c a}dowski}, A., {Narayan}, R., {McKinney}, J.~C., \& {Tchekhovskoy}, A.
  2014, \mnras, 439, 503

\bibitem[{{Schneider} {et~al.}(2010){Schneider}, {Richards}, {Hall},
  {et~al.}}]{Schneider2010}
{Schneider}, D.~P., {Richards}, G.~T., {Hall}, P.~B., {et~al.} 2010, \aj, 139,
  2360

\bibitem[{{Scott} {et~al.}(2011){Scott}, {Stewart}, {Mateos},
  {et~al.}}]{Scott2011}
{Scott}, A.~E., {Stewart}, G.~C., {Mateos}, S., {et~al.} 2011, \mnras, 417, 992

\bibitem[{{Shakura} \& {Sunyaev}(1973)}]{Shakura1973}
{Shakura}, N.~I., \& {Sunyaev}, R.~A. 1973, \aap, 24, 337

\bibitem[{{Shemmer} {et~al.}(2009){Shemmer}, {Brandt}, {Anderson},
  {et~al.}}]{Shemmer2009}
{Shemmer}, O., {Brandt}, W.~N., {Anderson}, S.~F., {et~al.} 2009, \apj, 696,
  580

\bibitem[{{Shemmer} {et~al.}(2008){Shemmer}, {Brandt}, {Netzer}, {Maiolino}, \&
  {Kaspi}}]{Shemmer2008}
{Shemmer}, O., {Brandt}, W.~N., {Netzer}, H., {Maiolino}, R., \& {Kaspi}, S.
  2008, \apj, 682, 81

\bibitem[{{Shemmer} {et~al.}(2014){Shemmer}, {Brandt}, {Paolillo},
  {et~al.}}]{Shemmer2014}
{Shemmer}, O., {Brandt}, W.~N., {Paolillo}, M., {et~al.} 2014, \apj, 783, 116

\bibitem[{{Shemmer} {et~al.}(2006){Shemmer}, {Brandt}, {Schneider},
  {et~al.}}]{Shemmer2006}
{Shemmer}, O., {Brandt}, W.~N., {Schneider}, D.~P., {et~al.} 2006, \apj, 644,
  86

\bibitem[{{Shemmer} \& {Lieber}(2015)}]{Shemmer2015}
{Shemmer}, O., \& {Lieber}, S. 2015, \apj, Submitted

\bibitem[{{Shemmer} {et~al.}(2010){Shemmer}, {Trakhtenbrot}, {Anderson},
  {et~al.}}]{Shemmer2010}
{Shemmer}, O., {Trakhtenbrot}, B., {Anderson}, S.~F., {et~al.} 2010, \apjl,
  722, L152

\bibitem[{{Shen}(2013)}]{Shen2013}
{Shen}, Y. 2013, Bulletin of the Astronomical Society of India, 41, 61

\bibitem[{{Shen} \& {Ho}(2014)}]{Shen2014}
{Shen}, Y., \& {Ho}, L.~C. 2014, \nat, 513, 210

\bibitem[{{Shen} \& {Kelly}(2012)}]{Shen2012b}
{Shen}, Y., \& {Kelly}, B.~C. 2012, \apj, 746, 169

\bibitem[{{Shen} {et~al.}(2011){Shen}, {Richards}, {Strauss},
  {et~al.}}]{Shen2011}
{Shen}, Y., {Richards}, G.~T., {Strauss}, M.~A., {et~al.} 2011, \apjs, 194, 45

\bibitem[{{Skrutskie} {et~al.}(2006){Skrutskie}, {Cutri}, {Stiening},
  {et~al.}}]{Skrutskie2006}
{Skrutskie}, M.~F., {Cutri}, R.~M., {Stiening}, R., {et~al.} 2006, \aj, 131,
  1163

\bibitem[{{Steffen} {et~al.}(2006){Steffen}, {Strateva}, {Brandt},
  {et~al.}}]{Steffen2006}
{Steffen}, A.~T., {Strateva}, I., {Brandt}, W.~N., {et~al.} 2006, \aj, 131,
  2826

\bibitem[{{Straub} {et~al.}(2011){Straub}, {Bursa}, {S{\c a}dowski},
  {et~al.}}]{Straub2011}
{Straub}, O., {Bursa}, M., {S{\c a}dowski}, A., {et~al.} 2011, \aap, 533, A67

\bibitem[{{Sulentic} {et~al.}(2014){Sulentic}, {Marziani}, {del Olmo},
  {et~al.}}]{Sulentic2014}
{Sulentic}, J.~W., {Marziani}, P., {del Olmo}, A., {et~al.} 2014, \aap, 570,
  A96

\bibitem[{{Teng} {et~al.}(2014){Teng}, {Brandt}, {Harrison},
  {et~al.}}]{Teng2014}
{Teng}, S.~H., {Brandt}, W.~N., {Harrison}, F.~A., {et~al.} 2014, \apj, 785, 19

\bibitem[{{Trakhtenbrot} \& {Netzer}(2012)}]{Trakhtenbrot2012}
{Trakhtenbrot}, B., \& {Netzer}, H. 2012, \mnras, 427, 3081

\bibitem[{{Turner} \& {Miller}(2009)}]{Turner2009}
{Turner}, T.~J., \& {Miller}, L. 2009, \aapr, 17, 47

\bibitem[{{Uttley} {et~al.}(2014){Uttley}, {Cackett}, {Fabian}, {Kara}, \&
  {Wilkins}}]{Uttley2014}
{Uttley}, P., {Cackett}, E.~M., {Fabian}, A.~C., {Kara}, E., \& {Wilkins},
  D.~R. 2014, \aapr, 22, 72

\bibitem[{{Vanden Berk} {et~al.}(2001){Vanden Berk}, {Richards}, {Bauer},
  {et~al.}}]{Vandenberk2001}
{Vanden Berk}, D.~E., {Richards}, G.~T., {Bauer}, A., {et~al.} 2001, \aj, 122,
  549

\bibitem[{{Wang} {et~al.}(2014{\natexlab{a}}){Wang}, {Du}, {Hu},
  {et~al.}}]{Wang2014b}
{Wang}, J.-M., {Du}, P., {Hu}, C., {et~al.} 2014{\natexlab{a}}, \apj, 793, 108

\bibitem[{{Wang} \& {Netzer}(2003)}]{Wang2003}
{Wang}, J.-M., \& {Netzer}, H. 2003, \aap, 398, 927

\bibitem[{{Wang} {et~al.}(2014{\natexlab{b}}){Wang}, {Qiu}, {Du}, \&
  {Ho}}]{Wang2014}
{Wang}, J.-M., {Qiu}, J., {Du}, P., \& {Ho}, L.~C. 2014{\natexlab{b}}, \apj,
  797, 65

\bibitem[{{Weymann} {et~al.}(1991){Weymann}, {Morris}, {Foltz}, \&
  {Hewett}}]{Weymann1991}
{Weymann}, R.~J., {Morris}, S.~L., {Foltz}, C.~B., \& {Hewett}, P.~C. 1991,
  \apj, 373, 23

\bibitem[{{White} {et~al.}(1997){White}, {Becker}, {Helfand}, \&
  {Gregg}}]{White1997}
{White}, R.~L., {Becker}, R.~H., {Helfand}, D.~J., \& {Gregg}, M.~D. 1997,
  \apj, 475, 479

\bibitem[{{Wills} \& {Browne}(1986)}]{Wills1986}
{Wills}, B.~J., \& {Browne}, I.~W.~A. 1986, \apj, 302, 56

\bibitem[{{Wright} {et~al.}(2010){Wright}, {Eisenhardt}, {Mainzer},
  {et~al.}}]{Wright2010}
{Wright}, E.~L., {Eisenhardt}, P.~R.~M., {Mainzer}, A.~K., {et~al.} 2010, \aj,
  140, 1868

\bibitem[{{Wu} {et~al.}(2012){Wu}, {Brandt}, {Anderson}, {et~al.}}]{Wu2012}
{Wu}, J., {Brandt}, W.~N., {Anderson}, S.~F., {et~al.} 2012, \apj, 747, 10
  (W12)

\bibitem[{{Wu} {et~al.}(2011){Wu}, {Brandt}, {Hall}, {et~al.}}]{Wu2011}
{Wu}, J., {Brandt}, W.~N., {Hall}, P.~B., {et~al.} 2011, \apj, 736, 28 (W11)

\bibitem[{{Wu} {et~al.}(2015){Wu}, {Wang}, {Fan}, {et~al.}}]{Wu2015}
{Wu}, X.-B., {Wang}, F., {Fan}, X., {et~al.} 2015, \nat, 518, 512

\bibitem[{{Xue} {et~al.}(2011){Xue}, {Luo}, {Brandt}, {et~al.}}]{Xue2011}
{Xue}, Y.~Q., {Luo}, B., {Brandt}, W.~N., {et~al.} 2011, \apjs, 195, 10

\bibitem[{{York} {et~al.}(2000){York}, {Adelman}, {Anderson},
  {et~al.}}]{York2000}
{York}, D.~G., {Adelman}, J., {Anderson}, Jr., J.~E., {et~al.} 2000, \aj, 120,
  1579

\end{thebibliography}
\end{document}